\documentclass[fleqn,usenatbib]{mnras}

\usepackage[T1]{fontenc}
\usepackage{ae,aecompl}

\usepackage{graphicx}	
\usepackage{amsmath}	
\usepackage{amssymb}	
\usepackage{mathptmx}
\usepackage{txfonts}
\usepackage{multicol}
\usepackage{dblfloatfix}
\usepackage[position=top, singlelinecheck=off, farskip=3pt]{subfig, caption}
\usepackage{reledmac}
\arrangementX[A]{twocol}
\usepackage{epstopdf}
\usepackage{cleveref}
\usepackage{color, soul}

\newcommand{\mmsn}[1]{$#1_\mathrm{MMSN}$}
\newcommand{\e}[1]{$e_\mathrm{#1}$}
\newcommand{\arm}[1]{$a_\mathrm{#1}$}
\newcommand{\Pb}[1]{$#1\,P_\mathrm{b}$}
\crefname{figure}{Fig.}{Figs.}

\title[Self-Gravity in Circumbinary Systems]{The Role of Disc Self-Gravity in Circumbinary Planet Systems: \\ II. Planet Evolution}

\author[M. M. Mutter et al.]{
Matthew M. Mutter$^{1,2}$\thanks{E-mail: m.m.mutter@qmul.ac.uk\newline},
Arnaud Pierens$^{3,4}$ and
Richard P. Nelson$^{1,2}$
\\
$^{1}$Astronomy Unit, Queen Mary University of London, Mile End Rd, London, E1 4NS, UK\\
$^{2}$Kavli Institute for Theoretical Physics, University of California, Santa Barbara, CA 93106, USA.\\
$^{3}$Universit\'e de Bordeaux, Observatoire Aquitain des Sciences de \v lUnivers, BP89 33271 Floirac Cedex, France\\
$^{4}$Laboratoire \v dAstrophysique de Bordeaux, Univ. Bordeaux, CNRS, B18N, all\'ee Geoffroy Saint-Hilaire, 33615 Pessac, France
}

\date{Accepted 2017 May 5. Received 2017 May 4; in original form 2017 March 18}

\pubyear{2017}

\begin{document}
\let\textlabel\label
\label{firstpage}
\pagerange{\pageref{firstpage}--\pageref{lastpage}}
\maketitle

\begin{abstract}
We present the results of hydrodynamic simulations examining migration and growth of planets embedded in self-gravitating circumbinary discs. The binary star parameters are chosen to mimic those of the Kepler-16, -34 and -35 systems; the aim of this study is to examine the role of disc mass in determining the stopping locations of migrating planets at the edge of the cavity created by the central binary. Disc self-gravity can cause significant shrinkage of the cavity for disc masses in excess of 5--10 $\times$ the minimum mass solar nebula model. Planets forming early in the disc lifetime can migrate through the disc and stall at locations closer to the central star than is normally the case for lower mass discs, resulting in closer agreement between simulated and observed orbital architecture. The presence of a planet orbiting in the cavity of a massive disc can prevent the cavity size from expanding to the size of a lower mass disc. As the disc mass reduces over long time scales, this indicates that circumbinary planet systems retain memory of their initial conditions. Our simulations produce planetary orbits in good agreement with Kepler-16b without the need for self-gravity; Kepler-34 analogue systems produce wide and highly eccentric cavities, and self-gravity improves the agreement between simulations and data. Kepler-35b is more difficult to explain in detail due to it's relatively low mass, which results in the simulated stopping location being at a larger radius than that observed.
\end{abstract}




\section{Introduction}\label{sec:intro}

\begin{figure*}
	\centering
	\includegraphics[width=\textwidth]{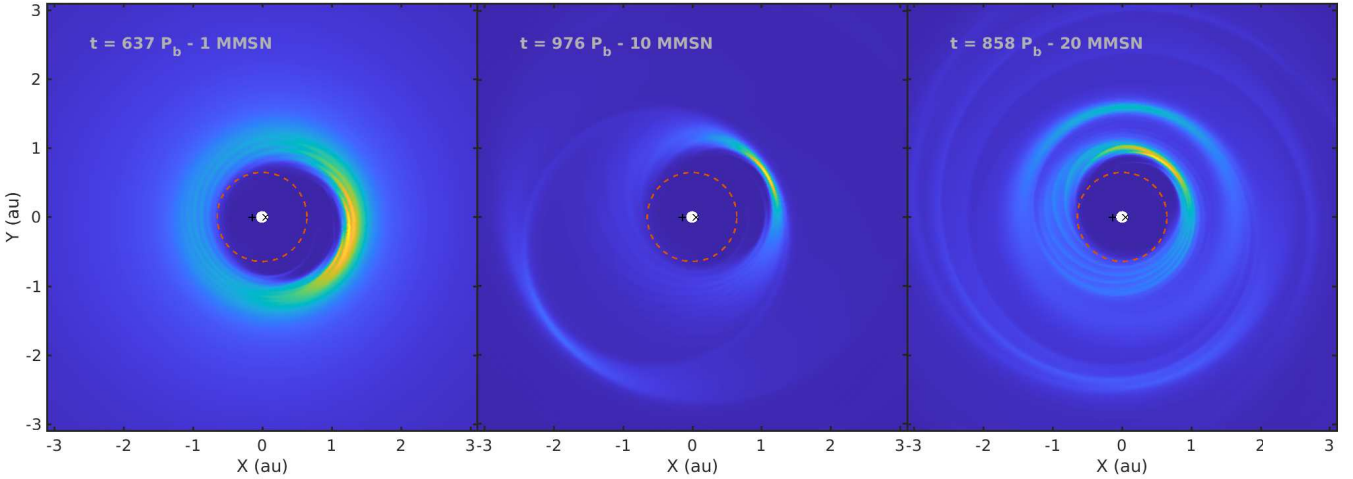}
	\caption{Changed plot 2D surface density image contours from self-gravitating circumbinary disc simulations around the Kepler-16 system from Paper I. These are all taken at a point when the disc has reached a pseudo-steady state. The panel on the left shows a typical low-mass (1 MMSN) self-gravitating disc in this system. The middle shows a \mmsn{10} case, and the right panel shows the most massive \mmsn{20} model. Note the additional eccentric features in the massive discs. The central eccentric cavity common to circumbinary discs can be seen around the binary in all three cases. The positions of the binary stars are marked with the `x' (primary) and `+' (secondary) symbols.}
	\label{fig:sg_disc_surf}
\end{figure*}

With the current tally of Kepler circumbinary planets standing at 11, (Kepler-16b\footnoteA{\citep{Doyle2011}}, Kepler-34b and Kepler-35b\footnoteA{\citep{Welsh2012b}}, Kepler-38b\footnoteA{\citep{Orosz2012}}, Kepler-47b,c\footnoteA{\citep{Orosz2012a}}, and d\footnoteA{\citep{Welsh2015a}}, Kepler-64b\footnoteA{\citep{Kostov2013}}, Kepler-413b\footnoteA{\citep{Kostov2014}}, Kepler-453b\footnoteA{\citep{Welsh2015}} and Kepler-1647b\footnoteA{\citep{Kostov2016}}), this class of object is one of the most interesting outcomes of the now two decade old search for planets around stars other than the Sun. They form part of an exoplanet catalogue that contains planets with very diverse orbital and physical characteristics orbiting within a broad variety of stellar systems.

Interest in circumbinary planets pre-dates their discovery, with their theorised presence prompting a number of studies into their formation and dynamical evolution. One of the key findings from this period, which has been validated by observations in recent years, is the work of \citet{Holman1999}. They found a limit for dynamical stability around short-period binaries. This critical limit depends on the mass and orbital properties of the binary. The majority of the Kepler circumbinary planets lie close to this limit within their respective systems. It is assumed that this, as well as the mutual co-planarity of the planet and binary orbital planes, is a fingerprint of the planets in these systems having formed in a common circumbinary disc.

Two general scenarios of planet formation theory have been used to try and explain the observed positions of the planets in these systems. The first is that the planets formed in-situ from material in close vicinity of the binary. In this case planet-forming material must be brought together under the strong influence of the gravitational field of the binary. Previous studies have shown numerous disruptive effects to the goal of bringing planetesimals together in a manner which results in mass-growth: N-body simulations show excitation of planetesimal eccentricity leading to relative velocities disruptive to accretion and the formation of planetary bodies \citep{Paardekooper2012, Meschiari2012a, Meschiari2012, Lines2014, Bromley2015}; differential pericentre-alignment of eccentric planetesimals of different sizes leads to corrosive collisions \citep{Scholl2007}; and gravitational interactions with asymmetric features in the gas disc, and the global eccentric mode, leading to large impact velocities of planetesimals \citep{Marzari2008, Kley2010}.

The second, and the focus of this series of papers, is that the protoplanetary cores formed in the quiescent exterior of the disc, where the disruptive influence of the binary is negligible, and then moved inwards to their observed position through disc-driven migration -- either Type-I \citep{Ward1997, Tanaka2002} or Type-II \citep{Lin1986, Nelson2000}. Whether this period of migration and mass-growth occurs early or late in the lifetime of these circumbinary discs is unclear. Whilst this scenario solves the problems with in-situ formation we must now answer how the migrating planets stop in the inner disc. Again, it was prior to the discovery of the Kepler circumbinary planets that this question was answered. The influence of the binary exerts a tidal torque on the inner circumbinary disc which sweeps material away from the binary, creating a central cavity that can act as a barrier to migration. The radial extent of this cavity depends on the mass and orbital properties of the binary, as well as disc parameters \citep{Artymowicz1994}. More recent work has shown that the interaction of the binary with this feature leads to an asymmetric, eccentric, precessing disc \citep{Pierens2013, Pelupessy2013, Kley2014, Mutter2017}. The directly-imaged circumbinary disc in the GG Tau system shows an inner cavity \citep{Dutrey1994}.

Paper I in this series, \citep{Mutter2017}, examined the impact of disc self-gravity and disc mass on the evolution and structure of circumbinary discs in analogue Kepler-16, -34 and -35 systems. Self-gravity has already been examined in low-mass circumbinary discs \citep{Marzari2009}, where it was discounted as an unimportant factor in the disc evolution. As pointed out in \citet{Lines2015}, even at low-mass, disc self-gravity can modify the precession frequencies associated with low-frequency global eccentricity modes \citep{Papaloizou2002}. Disc masses ranging from 1 to 20$\times$MMSN (where MMSN refers to the minimum mass solar nebula\citep{Hayashi1981}) were examined in Paper I, where we confirmed these findings for low-mass discs. However we found for more massive discs, corresponding to 10$\times$MMSN and beyond, that self-gravity can dramatically alter the evolution and structure of the disc. In addition to the central eccentric cavity, a series of eccentric modes were found to develop at larger orbital radii in all three of the Kepler circumbinary systems simulated, arising from the tidal field of the binary. Furthermore we see that the radial size of the cavity decreases in more massive discs. Figure \ref{fig:sg_disc_surf} shows the surface density profiles of discs in the Kepler-16 system, for three disc masses: 1, 10, and \mmsn{20} once they have reached a pseudo-steady-state -- i.e. the cavity size evolution has stopped and the eccentricity of the disc oscillates around a constant value. In this paper we present the results of simulations examining a number of migration and accretion scenarios for planet cores, in the full-set of evolved self-gravitating discs from Paper I, including the massive models which show additional eccentric features. While the long term aim of our work on circumbinary discs is to produce simulation outcomes that fit the Kepler data, because of the relative simplicity of our disc models in this study we set ourselves the less ambitious goal of examining how disc mass and self-gravity influence the final orbital elements of planets that form and migrate in circumbinary discs.

The evolution of giant planets in evolved self-gravitating circumbinary discs has not been studied before, however their counterparts in non-self-gravitating discs have been. The interaction of giant migrating planets with circumbinary discs was first studied by \citet{Nelson2003}. This study showed that Jovian-mass planets generally migrated into the central cavity, where they were captured into a $4:1$ mean-motion-resonance with the binary. These giant planets often underwent close-encounters with the binary, with scattering events ejecting them from the system. Lighter, Saturnian-mass planets underwent stable migration to the disc cavity where they then remained in stable orbits \citep{Pierens2008}. Less massive planets undergo Type-I migration until they are halted at the inner cavity edge by a strong positive co-rotation torque, counteracting the Lindblad torque -- see \citet{Pierens2007, Pierens2008, Pierens2008a} for more details. The techniques developed in these works were then applied to a number of the newly discovered Kepler circumbinary systems, in attempts to explain and recreate the orbits of their planets \citep{Pierens2013, Kley2014, Kley2015}. \citet{Pierens2013}, henceforth referred to as PN13, had difficulty recreating both the semi-major axes and eccentricities for the observed planets, with a range of disc parameters, under the assumption of an isothermal equation of state. \citet{Kley2014, Kley2015}, referred to from now on as KH14 and KH15, included a more realistic equation of state and radiation effects, as well as the role of multi-planet migration and interaction. These works also had difficulty in recreating all the observed properties of the Kepler circumbinary planets, although with a little more success than PN13. Using 3D SPH simulations \citet{Dunhill2013} argued the near-circular orbit of Kepler-16b hints that it formed in a massive disc in which the orbit of the planet is heavily damped by the disc. Understanding the physics and parameters which affect the environment in which the planets form and evolve -- the circumbinary disc -- is key to understanding the final, observed states of these intriguing systems.

Our motivation in Paper I and this work, is to probe the early dynamical history of circumbinary discs -- as we increase the disc mass we effectively examine earlier and earlier times in the system's history. We aim to address the questions: Does a high-mass disc leave a fingerprint on the planet population if circumbinary planets form early? Is this erased by the transition to a low-mass disc as the system evolves? Does the epoch when planets form, accrete gas, and migrate affect their final orbital configuration or mass? 

Using the results from Paper I as a starting point, we examine the impact of self-gravity and disc mass on migration and accretion scenarios for protoplanetary cores, in systems intended to mimic the Kepler-16, -34 and -35 circumbinary systems. The scenarios are carried out in evolved self-gravitating discs with masses equivalent to 1, 2, 5, 10 and 20$\times$MMSN, in each of the binary systems. These different disc masses are proxies for different eras in the lifetime of the disc, so we can answer the questions raised above. To simulate the evolution of the system from a high- to low-mass disc state we also carry out simulations where the orbital evolution of the planet is tracked as the disc mass is exponentially dissipated.

The outline for this paper is as follows. Section 2 describes the physical model and initial conditions used in our simulations. Section 3 looks at the results of the orbital migration of protoplanetary cores in the whole range of evolved self-gravitating discs from Paper I. Section 4 examines the orbital evolution of gas accreting cores. Section 5 contains results from our investigation into the impact of disc dissipation on planetary core migration, and final halting position. Our results from Sections 3--5 are summarised and discussed in Section 6.

\section{Numerical Setup}\label{sec:num_setup}
In this section we outline the extensions to our numerical model from Paper I, which deal with the interaction of the planet with the disc. For a full description of the equations pertaining to the evolution of the binary-disc system please refer to Section 2.1 of Paper I. The simulations are conducted in a reference frame based on the centre of mass of the binary system. The stellar orbital elements maintain constant values appropriate to the various Kepler systems that we are studying, and the N-body system comprising the stars and planet is evolved such that the centre of mass of this system is non-accelerating. Given that we are working in a frame centred on the binary centre of mass, an indirect term is required in the equations of motion that accounts for the acceleration of the binary centre of mass, as outlined below.
\subsection{Equations of Motion}\label{sec:eq_of_mot}
\subsubsection{Disc Evolution}\label{sec:disc_evol}
The equations of motion of the binary-disc system are detailed in Section 2.1.1 of Paper I in two-dimensional polar co-ordinates $(R,\,\phi)$ with the origin kept at the center of mass of the binary. 

The first extension we make to the system described in Paper I consisting of a close binary system surrounded by a self-gravitating disc is the addition of a massive, interacting planetary core. The planet is free to interact with the disc, and vice-versa. This results in the potential felt by the disc (Equation 4. in Paper I) having two additional terms, represented by $\Phi_\mathrm{p}$ and $\Phi_\mathrm{i}$:
\begin{equation}	
\Phi = \Phi_\mathrm{SG} + \sum_{k=1}^{2}\Phi_{\mathrm{s},k} + \Phi_\mathrm{p} + \Phi_\mathrm{i}.
\label{eq:pot}
\end{equation}
The first two terms in this equation are those created by the disc itself through self-gravity, and the two binary stars (with indices s). Their form is described in Equations 5 and 7 in Paper I. The form of the potential created by the planet, of mass $m_\mathrm{p}$ is as follows:
\begin{equation}
	\Phi_\mathrm{p} = -\frac{Gm_\mathrm{p}}{\sqrt{R^2 + R^2_\mathrm{p} - 2RR_\mathrm{p}\cos\left(\phi - \phi_\mathrm{p}\right) + \epsilon^2}}.
	\label{eq:pot_p}
\end{equation}	
$\epsilon$ is a softening length used to avoid singularities in the calculation of the planet's potential; it takes a value equal to 0.4$H$ in this work, where $H$ is the disc thickness. Readers of Paper I will note that this is the same prescription as used for the smoothing length used for the calculation of the disc's self-gravitating potential. The term represented by $\Phi_\mathrm{i}$ is the indirect term resulting from the acceleration of the binary centre of mass due to the gravity of the planet \footnote{In principle we should also include an additional term from the disc acting on the centre of mass of the binary. Extensive tests were undertaken in Paper I which demonstrated that including this term did not change the results.}.

\subsubsection{Orbital Evolution}\label{sec:orb_evol}
Table \ref{tab:kepler_params} contains the best-fit observed binary and planetary orbital and mass parameters of the Kepler-16, -34 and -35 circumbinary planetary systems, as quoted in \citet{Doyle2011} and \citet{Welsh2012b}.
One of the eventual goals of this work is to recreate the observed state of the Kepler circumbinary systems. To minimise the initial parameter space of this work, and as noted in Paper I, the binaries' orbital parameters remain fixed throughout our simulations. The orbital evolution of the binary system is therefore independent of the disc and planet system; we hope to revisit the back reaction of massive self-gravitating discs on the binary in a later work. In Paper I we discuss the drawbacks to this approach, where in our most massive systems -- where the disc mass is comparable to the mass of one of the binary stars contained within a notional radius of 30 au -- significant modification of the binary will occur if back-reaction is allowed. See Paper I for a detailed description of these problems, in the initial set-up of our simulations and in the discussion. The equation of motion for the binary stars remains unchanged from Equation 8 in Paper I.
\begin{table}
	\centering
	\caption{Binary and planet parameters.}
	\label{tab:kepler_params}
	\begin{tabular}{cccc}
		\hline
		\hline
		 & Kepler-16 & Kepler-34 & Kepler-35\\
		\hline
		$M_\mathrm{A}\ (M_\odot)$ & 0.690 & 1.048 & 0.888\\
		$M_\mathrm{B}\ (M_\odot)$ & 0.203 & 1.021 & 0.809\\
		$m_\mathrm{p}\ (M_J)$ & 0.333 & 0.220 & 0.127\\
		$q_\mathrm{b} = M_\mathrm{B}/M_\mathrm{A}$ & 0.294 & 0.974 & 0.912\\
		$q_\mathrm{p} = m_\mathrm{p}/M_\star$ & $3.54\times 10^{-4}$ & $1.01\times 10^{-4}$ & $7.13\times 10^{-5}$ \\
		$a_\mathrm{b}$ (au) & 0.224 & 0.228 & 0.176\\
		$a_\mathrm{p}$ (au) & 0.705 & 1.090 & 0.603\\
		\e{b} &	 0.159 & 0.521 & 0.142 \\
		\e{p} &	 0.007 & 0.182 & 0.042\\
		Reference &	\citep{Doyle2011} & \multicolumn{2}{c}{\citep{Welsh2012b}} \\
		\hline
	\end{tabular}
\end{table}
The equation of motion for a planet of mass $m_\mathrm{p}$, interacting with the binary and disc system is as follows:
\begin{equation}
\frac{d^2\mathbf{R}_\mathrm{p}}{dt^2} = -\sum_{k=1}^{2}\frac{GM_{\mathrm{s},k}\left(\mathbf{R}_\mathrm{p}-\mathbf{R}_{\mathrm{s}, k}\right)}{|\mathbf{R}_\mathrm{p}-\mathbf{R}_{\mathrm{s}, k}|^3} + \mathbf{f}_\mathrm{dp} - \mathbf{f}_\mathrm{i},
\label{eq:motion_planet}
\end{equation}
where $\mathbf{f}_\mathrm{dp}$ is the force acting on the planet from the disc, and is given by:
\begin{equation}
\mathbf{f}_\mathrm{dp} = \int_S\frac{\Sigma(\mathbf{R})\mathrm{d}\mathbf{R}}{\sqrt{R^2 + R_\mathrm{p}^2 - 2RR_\mathrm{p}\cos\left(\phi - \phi_\mathrm{p}\right) + \epsilon^2}}.
\label{eq:force_disc}
\end{equation}
The term $\mathbf{f}_\mathrm{i}$ represents the acceleration of the binary centre of mass by the gravity of the planet.

\subsection{Hydrodynamic Model}\label{sec:hydro_model}
The hydrodynamic set-up used in this work follows that in Paper I, and builds on the disc-binary results. The simulation work-load for the results presented here, and in Paper I, was split across two separate numerical codes, after comparing test simulations to verify that the results agreed. These codes were {\small FARGO-ADSG} and {\small GENESIS}. {\small FARGO-ADSG} is an updated version of the widely used {\small FARGO} code \citep{Masset1999}, which includes the calculation of disc self-gravity as well as an adiabatic equation of state \citep{Baruteau2008b,Baruteau2008}. {\small GENESIS} uses an advection scheme based on the \citet{VanLeer1977} monotonic transport algorithm to solve the disc equations, and contains the {\small FARGO} time stepping upgrade, as well as a module to calculate self-gravity. In both codes the binary and planetary orbits are evolved using a fifth-order Runge-Kutta integrator scheme \citep{Press1992}.

These codes were used to run 2D hydrodynamic simulations in the plane of the binary's orbit. The calculations presented here use a grid resolution of $N_R \times N_\phi=550 \times 550$ cells. The radial grid-spacing between $R_\mathrm{in}$ and $R_\mathrm{out}$ is logarithmic, as required by the self-gravity calculations (refer to Table \ref{tab:bc} for the values used). This has the added benefit of having a finer grid in the inner region of the disc closest to the binary. The azimuthal grid is equally-spaced between $[0,\,2\pi]$. All the disc models use a kinematic alpha-prescription to model turbulence in the disc \citep{Shakura1973}, where $\alpha = 1 \times 10^{-3}$, and a constant disc aspect ratio, $H/R = 0.05$, which gives rise to a locally isothermal set-up.

The following computational units are used: the total mass of the binary $M_\star=M_\mathrm{A} + M_\mathrm{B}=1$, the gravitational constant $G=1$ and the radius $R=1$ is equivalent to 1 au. To present the results of simulations we use the binary orbital period, $P_\mathrm{b}=2\pi \sqrt{GM_\star/a^3_\mathrm{b}}$, as the unit of time.

For the migration scenarios where our giant planet cores are allowed to accrete gas from the disc, we follow the prescription of \citet{Kley1999}. Accretion is modelled by removing a fraction of the gas within the planet's Hill sphere, $R_\mathrm{Hill} = a_\mathrm{p}(m_\mathrm{p}/3M_\star)^{1/3}$, from the disc and adding the equivalent mass to that of the planet. The rate at which gas is removed from the Hill sphere is determined by the accretion time-scale, $t_{acc} = f\,t_{dyn}$, where $t_{dyn}$ is the orbital period of the planet, and $f$ is an adjustable factor. For the simulations where there is no accretion this corresponds to $f=0$.
\subsection{Initial Conditions}\label{sec:init_cond}

\begin{table}
	\centering
	\caption{Starting semi-major axes of protoplanetary cores in each of the Kepler-16, -34 and -35 system models.}
	\label{tab:ap_init}
	\begin{tabular}{c|ccc}
		\hline
		\hline
		 & \multicolumn{3}{c}{$a_{\mathrm{p}, 0}$ (au)}\\
		\hline
		 & Kepler-16 & Kepler-34 & Kepler-35\\
		\hline
		\mmsn{1} & 2.0 & 2.5 & 2.5\\
		\mmsn{2} & 2.0 & 2.5 & 2.5\\
		\mmsn{5} & 2.0 & 2.5 & 2.5\\
		\mmsn{10} & 3.0 & 2.5 & 2.0\\
		\mmsn{20} & 3.0 & 2.5 & 2.5\\
		\hline
	\end{tabular}
\end{table}

\begin{figure*}
\centering
\includegraphics[width=\textwidth]{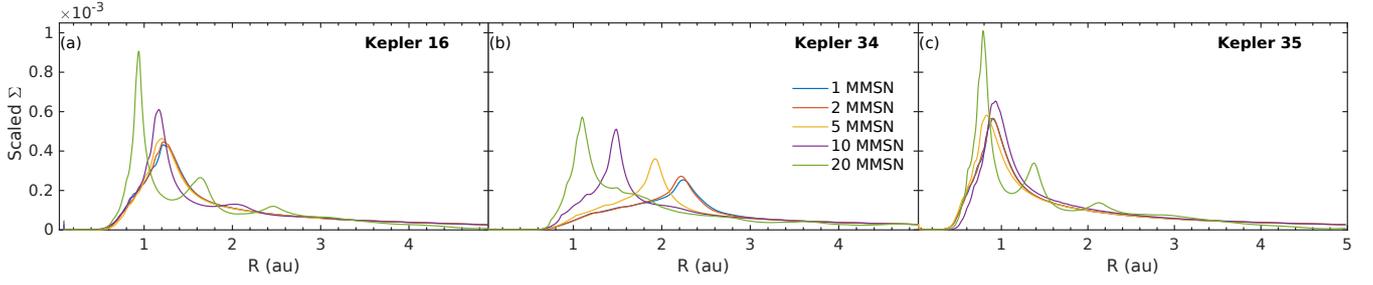}
\caption{Azimuthally averaged surface density profile results of all self-gravitating disc models in the Kepler-16, -34, and -35 systems from Paper I. One can see the central cavity in all disc models, and the density spikes associated with the additional eccentric features in the most massive 10 and \mmsn{20} models. These profiles are calculated once the discs have reached pseudo-steady-state -- \Pb{5000} in the Kepler-16 system, and \Pb{6000} in the Kepler-34 and -35 systems.}
\label{fig:sg_disc_sdprof}
\end{figure*}

\begin{figure*}
\centering
\includegraphics[width=\textwidth]{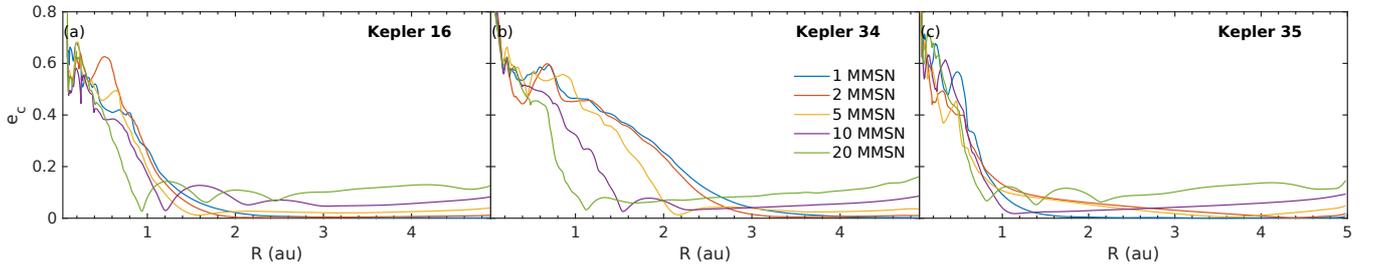}
\caption{Azimuthally averaged cell eccentricity profile results of all self-gravitating disc models in the Kepler-16, -34, and -35 systems from Paper I. The eccentric inner cavity can be seen in all models, with the additional eccentricity bumps associated with the eccentric features in the high-mass discs. Again, these profiles are taken once the models have reached pseudo-steady-state.}
\label{fig:sg_disc_edprof}
\end{figure*}

The initial conditions used to set up the simulations of the disc models used in this work are detailed in Section 2.3 of Paper I. This section will instead focus on the procedure used to initialise the planet cores in the migration, gas accretion and disc dissipation scenarios presented here. Figure \ref{fig:sg_disc_sdprof} summarises the "initial" conditions the planets are inserted into -- the pseudo-steady-state, azimuthally averaged surface density profiles for the 1, 2, 5, 10, and \mmsn{20} models in the Kepler-16 , -34, and -35 systems. These snapshots are taken at $t=$ \Pb{6000}, once the disc has reached a pseudo-equilibrium.

In our first set of simulations we launch protoplanetary cores, on initially circular orbits, in the outer regions of the evolved discs from Paper I and allow them to interact with the discs. The initial mass of the core in each system is chosen so that $q_{\mathrm{p},0}=m_{\mathrm{p},0}/M_\star=6\times 10^{-5}$. If $M_\star=1 M_\odot$ this is equivalent to a $20M_\oplus$ core. We release the cores into the outer region of the disc, where the surface density profile is unperturbed by the binary, and the disc eccentricity is negligible. Referring to the profiles in Fig. \ref{fig:sg_disc_sdprof} this lies at 2 au in the low-mass Kepler-16 and -35 models, and around 2.5 au in the low-mass Kepler-34 systems. The situation in the high-mass systems is a little more complicated due to the additional eccentric features in the outer disc. Starting the planets at an initial starting position beyond 4 au -- exterior to any strong eccentric features -- means the time needed to migrate into the inner disc is too long. However, we speculated in Paper I that the migrating planets could interact with these additional features to produce interesting behaviour, therefore we didn't want to place the planets too close to the binary. In the high-mass discs we used an approach which placed the planet beyond the first additional feature, but not too far out in the outer disc. Table \ref{tab:ap_init} summarises the starting semi-major axes of the cores in all our models. {\small FARGO-ADSG} modifies the initial planet semi-major axes with an initial self-gravity "boost", which increases the starting position by $\approx 10\%$ in the most massive \mmsn{20} models. 

\subsection{Boundary Conditions}\label{sec:bound_cond}
\begin{table}
	\centering
	\caption{Inner and Outer Boundary Conditions}
	\label{tab:bc}
	\begin{tabular}{cccc}
		\hline
		\hline
		 & Kepler-16 & Kepler-34 & Kepler-35\\
		\hline
		$R_\mathrm{in}$ (au) & 0.090 & 0.040 & 0.056\\
		$R_\mathrm{in}$ BC & \multicolumn{3}{c}{Viscous}\\
		$R_\mathrm{out}$ (au) & \multicolumn{3}{c}{5.0}\\
		$R_\mathrm{out}$ BC & \multicolumn{3}{c}{Open}\\
		\hline
	\end{tabular}
\end{table}
In Paper I we carried out a fairly exhaustive investigation into the impact of inner boundary conditions on the structure of circumbinary discs around close binary systems. This was motivated by similar discussions in \citet{Marzari2009}, PN13, KH14 and \citet{Lines2015}, which found discrepancies between different outflow choices, or between the same boundary condition in different systems. For a full description of the investigation and its results we direct the reader to Sections 2.4 and 3 in Paper I. Again, we found no boundary choice which was consistent between the three systems presented here. The choice of inner boundary condition is a way to simulate how much mass flows from the inner disc, out of the boundary and onto the central binary. The Open, Closed or Viscous boundary conditions tested previously all guess at how mass flows through the eccentric cavity, and onto the stars. Our findings from these initial investigations prompted us to develop a way to treat the inner boundary which resulted in a more physically realistic treatment of the material accreting onto the binary. This required us to shrink the size of the inner edge of the disc domain, so the binary is partially embedded in the computational domain rather than sitting entirely interior to the inner boundary.

Decreasing the radius of the inner edge, $R_\mathrm{in}$, increases the computational runtime of the simulations, so a compromise is struck between accuracy and speed. We found this balance came when at least 70\% of the Roche lobe area of the least massive star was contained in the computational grid at all times. The various values for $R_\mathrm{in}$ in the three binary systems simulated here are given in Table \ref{tab:bc}. The outer boundary of the disc is treated the same in each simulation. A value of $R_\mathrm{out}=5$ au is used, with an Open outflow boundary condition. A brief description of the Open and Viscous boundary conditions is given below:
\begin{itemize}
\item Open -- material is allowed to freely leave the disc i.e. outflow. No inflow is allowed. A zero-gradient condition is set in both $\upsilon_R$ and $\Sigma$.
\item Viscous -- this is a limiting condition to stop the inner disc from emptying of gas too quickly. Material in the innermost cells is given a radial velocity, $\upsilon_R=\beta \upsilon_R(R_\mathrm{in})$, where $\upsilon_R(R_\mathrm{in})=-3\nu/2R_\mathrm{in}$, is the viscous drift velocity and $\beta$ is a free factor \citep{Pierens2008}. We follow previous works which use this condition and set $\beta=5$.
\end{itemize}

As we describe in Paper I, the Viscous outflow condition at the inner disc radius tries to model the accretion flow onto the central star(s). It acts as compromise between an unphysical reflecting boundary, and an Open boundary, which empties the inner disc of material at too fast a rate. Our models do not (fully) resolve the circumstellar discs which would form in the Roche lobes of the primary and secondary stars however if we assume that these discs evolve and accrete onto the central objects on the viscous timescale, the Viscous boundary condition `feeds' these discs at a self-consistent rate.

How the azimuthal velocity is treated at the inner radial edge of the disc is also modified. Usually in hydrodynamical codes the viscous stress is maintained by setting $\upsilon_\phi$ to the sub-Keplerian orbital velocity at the locations $R_\mathrm{in}$ and $R_\mathrm{out}$. At the inner boundary the potential created by the binary is extremely non-Keplerian. We therefore set a zero-gradient condition for the azimuthal velocity at this location.

\section{Migration of Protoplanetary Cores}\label{sec:mig_res}
In this section we present the results of simulations examining the migration of protoplanetary cores in evolved self-gravitating discs around the Kepler-16, -34, and -35 binary systems. We insert a non-accreting core, with mass ratio $q_{\mathrm{p},0}=6 \times 10^{-5}$, into each of the 1, 2, 5, 10, and \mmsn{20} discs from Paper I, once the disc has reached a pseudo-steady state. In the Kepler-16 system this is at \Pb{5000}, whilst the discs reach this state after \Pb{6000} in the Kepler-34 and -35 systems. At this point in the simulation the inner eccentric cavity has a stable precession frequency, and in the high-mass discs, models which were shown to exhibit additional eccentric features at any point, have done so. The core mass used in these simulations lies in the regime where Type-I migration is rapid, but is not massive enough to open a gap in the disc i.e. the gap-opening criteria of \citet{Crida2006} is not met.

\subsection{Kepler-16}\label{sec:mig_res_16}
\begin{figure}
\centering
\subfloat[]{\includegraphics[width = 0.47\textwidth]{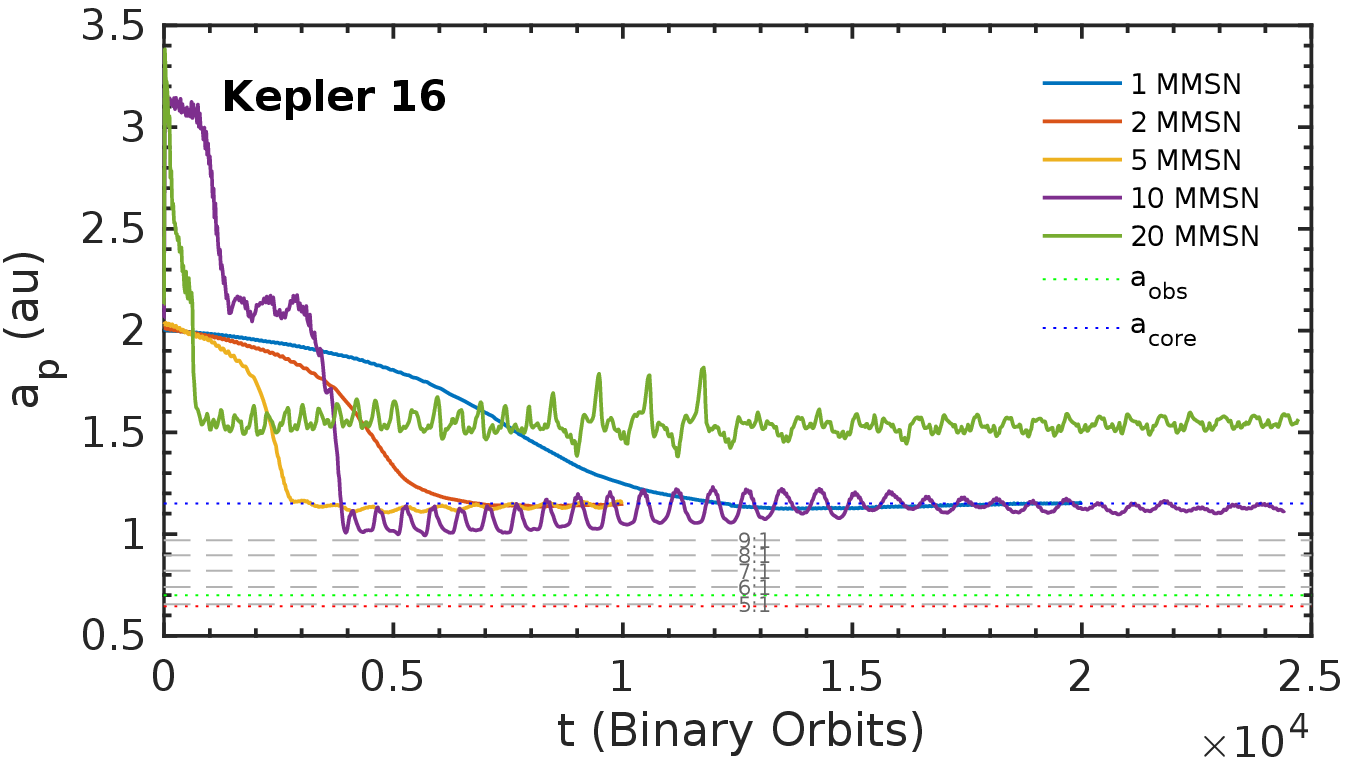}} \\
\vspace{-15pt}
\hspace{-10pt}
\subfloat[]{\includegraphics[width = 0.47\textwidth]{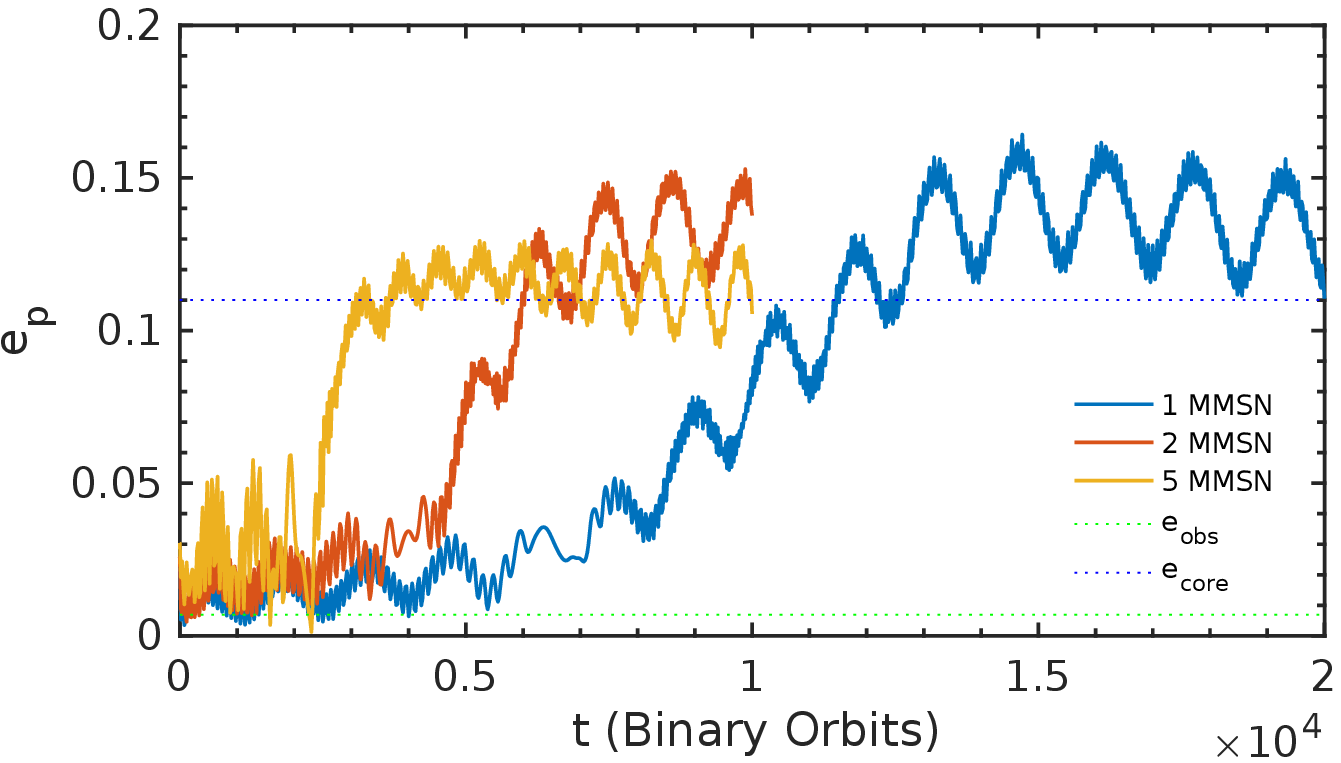}}\\
\vspace{-15pt}
\subfloat[]{\includegraphics[width = 0.47\textwidth]{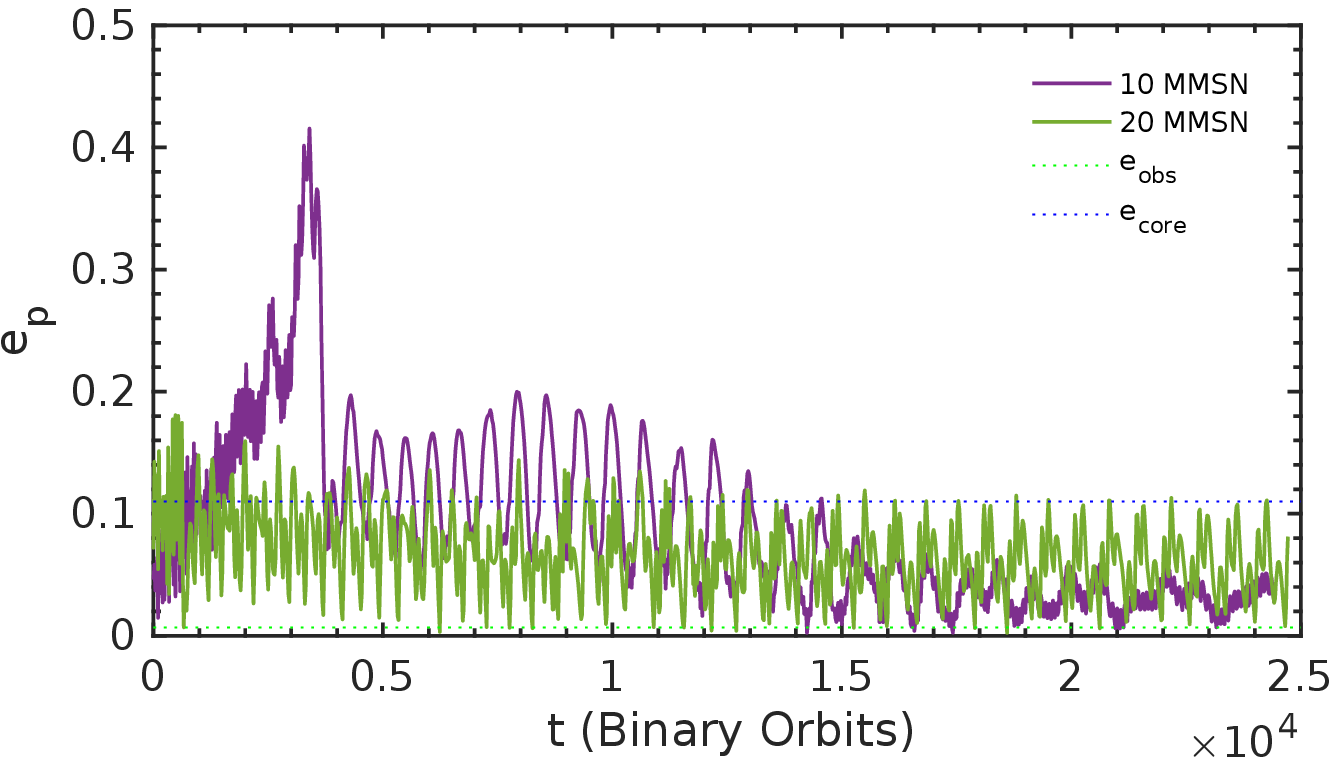}}
\caption{The top panel shows the evolution of $q_{\mathrm{p},0}=6 \times 10^{-5}$ protoplanetary cores' semi-major axes in evolved self-gravitating discs around the Kepler-16 system. The grey dashed lines in this plot show the positions of $n:1$ mean-motion resonances with the binary, which have been shown to be unstable to planetary orbits \citep{Nelson2003, Kostov2013, Kley2014, Kley2015}. The red dotted line shows the location of \arm{crit} from \citet{Holman1999}. The middle and bottom panels show these cores' eccentricity evolution in the low- and high-mass disc models respectively. The green dotted lines in these plots show the values of \arm{p} and \e{p} of the observed planet from \citet{Doyle2011}, and the blue dotted lines show the final values of simulation with the same disc parameters and comparable core mass from PN13.}
\label{fig:mig_res_kep16}
\end{figure}

\begin{figure}
\includegraphics[width = 0.45\textwidth]{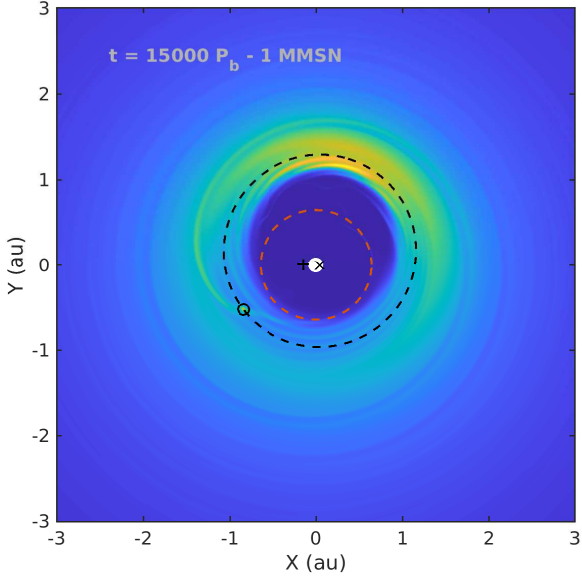}
\caption{A snapshot of the 2D surface density profile in the \mmsn{1} model once the planet has reached a pseudo-steady-state orbit at the edge of the eccentric cavity. The instantaneous orbit of the planet is shown by the black-dashed ellipse, the red-dashed circle shows the location of the critical stability limit. The disc and planet system look nearly identical in the 2 and \mmsn{5} models around the same system. The planet's inwards migration has been halted by a strong positive co-rotation torque balancing the negative Lindblad torque, once the eccentricity of the planet attains a value $\approx$ \e{d}(\arm{p})}.
\label{fig:mig_res_kep16_1sd}
\end{figure}

\begin{figure}
\hspace{-15pt}
\includegraphics[width = 0.5\textwidth]{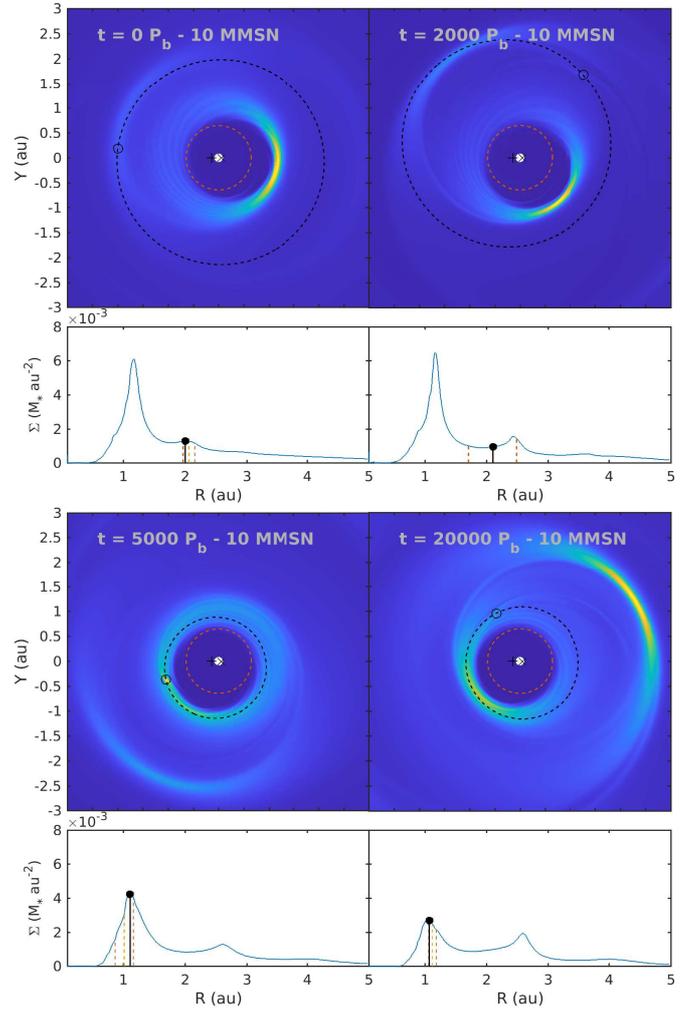}
\vspace{-25pt}
\caption{Evolution of the 2D and azimuthally-averaged 1D surface density profiles of the Kepler-16 \mmsn{10} model, over the course of the protoplanet's migration from its initial starting point, to its final location. The instantaneous orbit of the planet is shown by the black-dashed ellipse in the 2D plots; the red-dashed circle shows the location of the critical stability limit. The black line in the 1D profiles is the core's actual distance from the binary CoM, the orange dashed line is \arm{p} and the inner and outer dashed red lines show the location of $r_{per}=a_\mathrm{p}(1-e_\mathrm{p})$ and $r_{apo}=a_\mathrm{p}(1+e_\mathrm{p})$ respectively.}
\label{fig:mig_res_kep16_10sdgrid}
\end{figure}

\begin{figure}
\centering
\includegraphics[width = 0.45\textwidth]{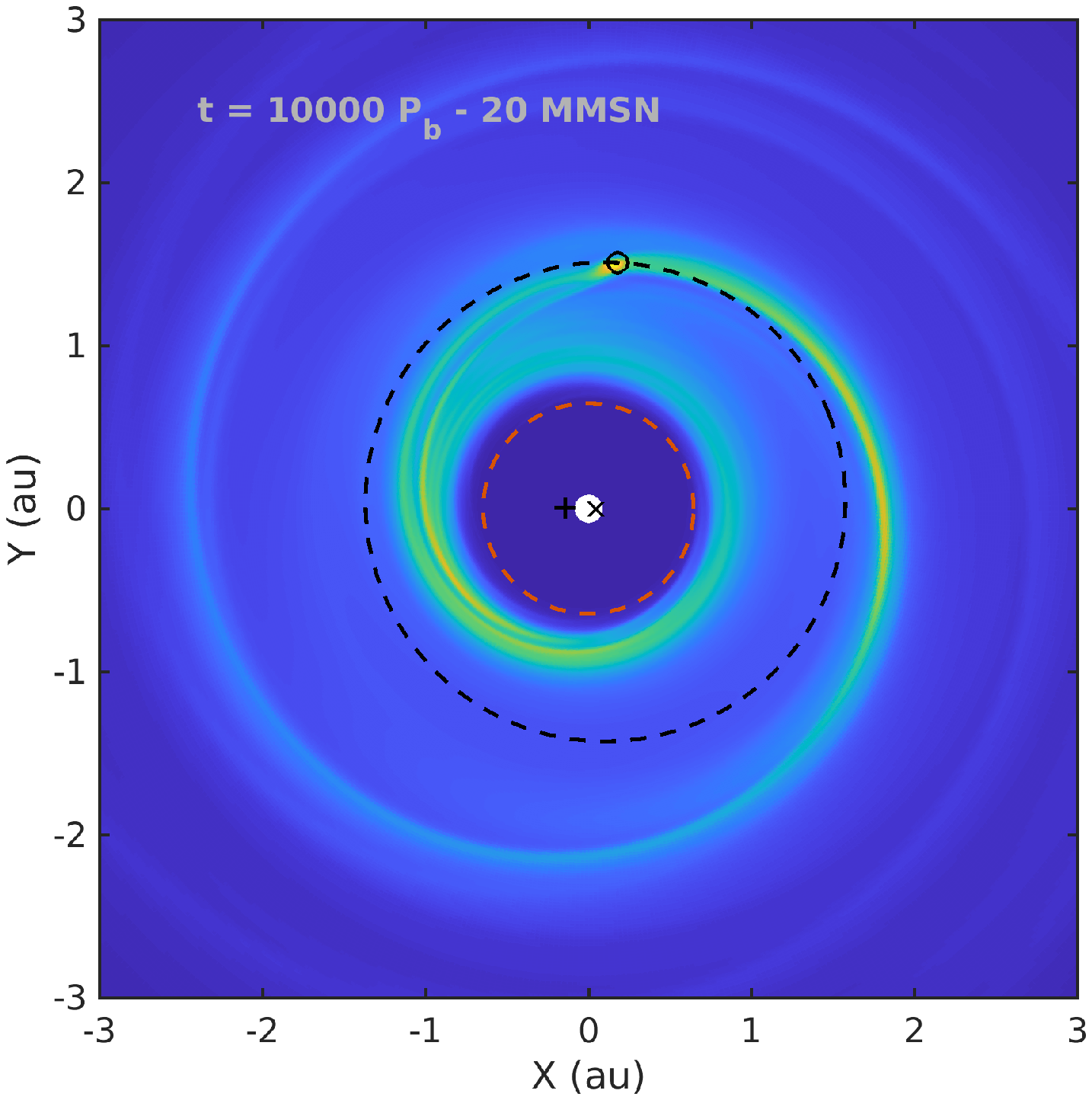}
\caption{A 2D surface density profile of the evolved \mmsn{20} disc model around the Kepler-16 binary. In this snapshot a protoplanetary core has been interacting with the disc for \Pb{1000}. The instantaneous orbit of the core is shown by the black-dashed ellipse. The core can be seen to be orbiting between the inner cavity, and the first eccentric feature. The planet's orbit crosses this feature, and because it has a period of precession ranging in the hundreds of binary orbits (see Paper I), the planet will cross and interact with this feature repeatedly. The eccentric feature is strong enough to trap the planet at this exterior position.}
\label{fig:mig_res_kep16_20sd}
\end{figure}

Our results for the orbital evolution of protoplanetary cores in self-gravitating discs around Kepler-16 are summarised in Fig. \ref{fig:mig_res_kep16}. The upper panel shows the evolution of the cores' semi-major axes, until a pseudo-steady orbit is reached. In addition to the 1--\mmsn{20} models being shown on this plot, several other quantities are plotted. The red dotted line is the semi-empirical critical semi-major axis for stable orbits around Kepler-16 \citep{Holman1999}; the green dashed line is the best-fit observed semi-major axis for Kepler-16b from \citet{Doyle2011}; the blue dotted line (with label a$_\mathrm{core}$) shows the final values for the non-self-gravitating results with comparable disc and core properties from PN13; and the grey dashed lines show the locations of the 5:1--9:1 mean motion resonances with the binary -- locations which have been shown to lead to eccentricity growth, leading to ejections or scattering with the binary \citep{Kostov2014, Kley2015, Kostov2016}. The middle panel shows the evolution of the core eccentricity results for the low-mass (1--\mmsn{5}) disc models, with the high-mass results plotted in the bottom panel for clarity.

We can see that like the disc evolution models in Paper I, the evolution of the protoplanetary cores in the low and high-mass discs can be separated into two distinct regimes of behaviour. In the low-mass discs the cores migrate inwards, albeit with increased rates in the more massive models (the migration rate scales moderately super-linearly with the surface density at the planets' location \citep{Baruteau2008}), from their initial starting position until they reach 1.2 au. This location corresponds well with the surface density peak in the material bounding the tidally truncated inner cavity (Fig. \ref{fig:sg_disc_sdprof}), a result fully expected from previous work. As can be seen in Fig. \ref{fig:mig_res_kep16} the low-mass results agree extremely well with those from PN13, but not with the observed state of Kepler-16b. This result is slightly unexpected as the disc cavity size seen in our models is somewhat smaller than those in PN13 due to our more realistic treatment of the inner disc boundary. 

To explain this we must examine what dictates the halting position of these protoplanetary cores in circumbinary discs. From prior work \citep{Pierens2007, Pierens2013, Kley2014}, we know the stopping behaviour of planets across a range of planetary masses. In the Type I regime, Earth-like planets are stopped by the growth of a strong positive co-rotation torque which counteracts the influence of the negative Lindblad torque \citep{Masset2006, Pierens2007}. These two torques balance each other when the surface density gradient is sufficiently positive. For more massive Saturn-like planets, a different stopping mechanism operates. If the planetary eccentricity is large enough, a torque reversal can be induced -- at apoapse the planet orbits amongst material in the outer disc that is locally travelling faster than itself. When this material overtakes the planet it is focussed by the planet's gravity, leading to a positive torque. The reverse of this occurs at periapse, leading to a negative torque from the inner disc \citep{Pierens2008, Pierens2008a, Pierens2013}. When at a cavity edge, the inner torque is naturally smaller in magnitude than the outer torque, leading to a net positive torque arising from this effect. As can be seen in the first panels of Figs. \ref{fig:sg_disc_sdprof}, \ref{fig:sg_disc_edprof}, and the middle panel of Fig. \ref{fig:mig_res_kep16}, the migration of the protoplanets starts to slow when the planetary eccentricity reaches a significant level, \e{p} $\approx0.13$. This coincides when the local disc cell eccentricity and planetary eccentricity are comparable. These findings lead us to the same conclusion as PN13, that for the protoplanetary core mass used here, it is the torque reversal induced by significant planetary eccentricity which halts migration. Comparing our results with those in PN13, the fact that the planet's stopping location is essentially the same in that study and this one, in spite of the different size of the cavity, arises because of differences in the planetary eccentricity and the structure of the cavity (eccentricity and surface density profile). Figure \ref{fig:mig_res_kep16_1sd} shows the orbit of the core at the exterior edge of the cavity, where the protoplanet's eccentricity is high enough to induce a torque reversal.

In the high-mass disc regime we observe planetary evolution behaviour not seen in the low-mass discs, or previous work on this topic. Whilst the cores still migrate inwards, in the \mmsn{10} model the core briefly halts at 2.2 au. During the period when it is trapped at this location its eccentricity steadily grows from 0.05 to a maximum of 0.4. When the core is then released its eccentricity is quickly damped and it migrates into the inner disc, halting at a location in good agreement with the low-mass results, \arm{p} $=1.1$ au. Unlike the low-mass models the core's eccentricity is slowly damped by the large amount of gas in its vicinity, down to a value in good agreement with the observed value of Kepler-16b. Whilst the core in the \mmsn{20} model doesn't show signs of this trapping immediately, its inwards migration is halted at 1.5 au, a location significantly exterior to the prior results. Examining Figs \ref{fig:mig_res_kep16_10sdgrid} and \ref{fig:mig_res_kep16_20sd}, we can start to explain this behaviour. The discovery of the additional eccentric features in high-mass self-gravitating circumbinary discs prompted us to theorise that they could act as planet traps. In the Type I regime, the positive surface density gradient creates a strong co-rotation torque which could counteract the Lindblad torque or, for more massive planets, the excited eccentricity in these regions could excite the eccentricity of the body sufficiently to induce a torque reversal. Whilst this process requires the planet to have a non-negligible eccentricity, it also requires there to be a surface density gradient across the extremes of the orbit. At apocentre it should find itself in an area of high surface density, and a low surface density at pericentre. In the discs which we obtain in these models, this can be achieved by the planet and disc eccentricities not being exactly equal, or a misalignment between the respective line of nodes. In this case the planet is on a less eccentric orbit than the surrounding disc material.

Despite this prediction, we see two different end results in the 10 and \mmsn{20} models. In the least massive of these cases the core migrates inwards until it reaches the first eccentric feature. At this location it halts. Repeated interaction with this highly eccentric feature lead to the planet's own eccentricity being excited. This can be seen in the second panel of Fig. \ref{fig:mig_res_kep16_10sdgrid}. However the planet's eccentricity becomes so high (\e{p} $=0.4$), the planet's pericentre position decreases until it interacts with the material bounding the inner eccentric cavity. A similar process to the initial trapping then occurs, however the planet's orbit is circularised by the far less extended inner feature. The core's semi-major axis shrinks until the orbit is moderately eccentric, which matches that of the observed Kepler-16b relatively well, at the location of the inner cavity. The core in the \mmsn{20} model can get trapped at the first outer eccentric cavity because the eccentric feature is more tightly localised due to the disc's stronger self-gravity. Therefore the planet's orbit doesn't take it into close proximity of the strong inner eccentric feature, and it remains trapped between the inner and first outer eccentric features.

\subsection{Kepler-34}\label{sec:mig_res_34}
\begin{figure}
\centering
\subfloat[]{\includegraphics[width = 0.47\textwidth]{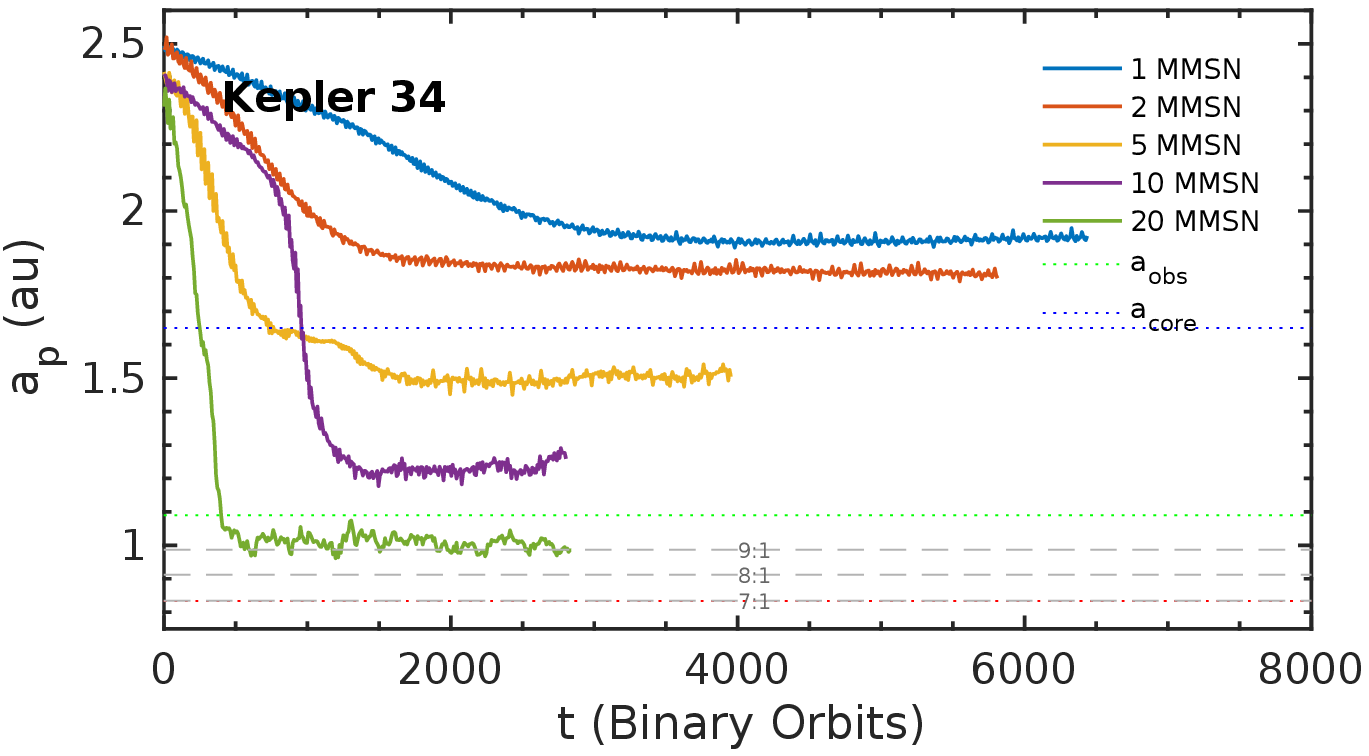}} \\
\vspace{-10pt}
\subfloat[]{\includegraphics[width = 0.47\textwidth]{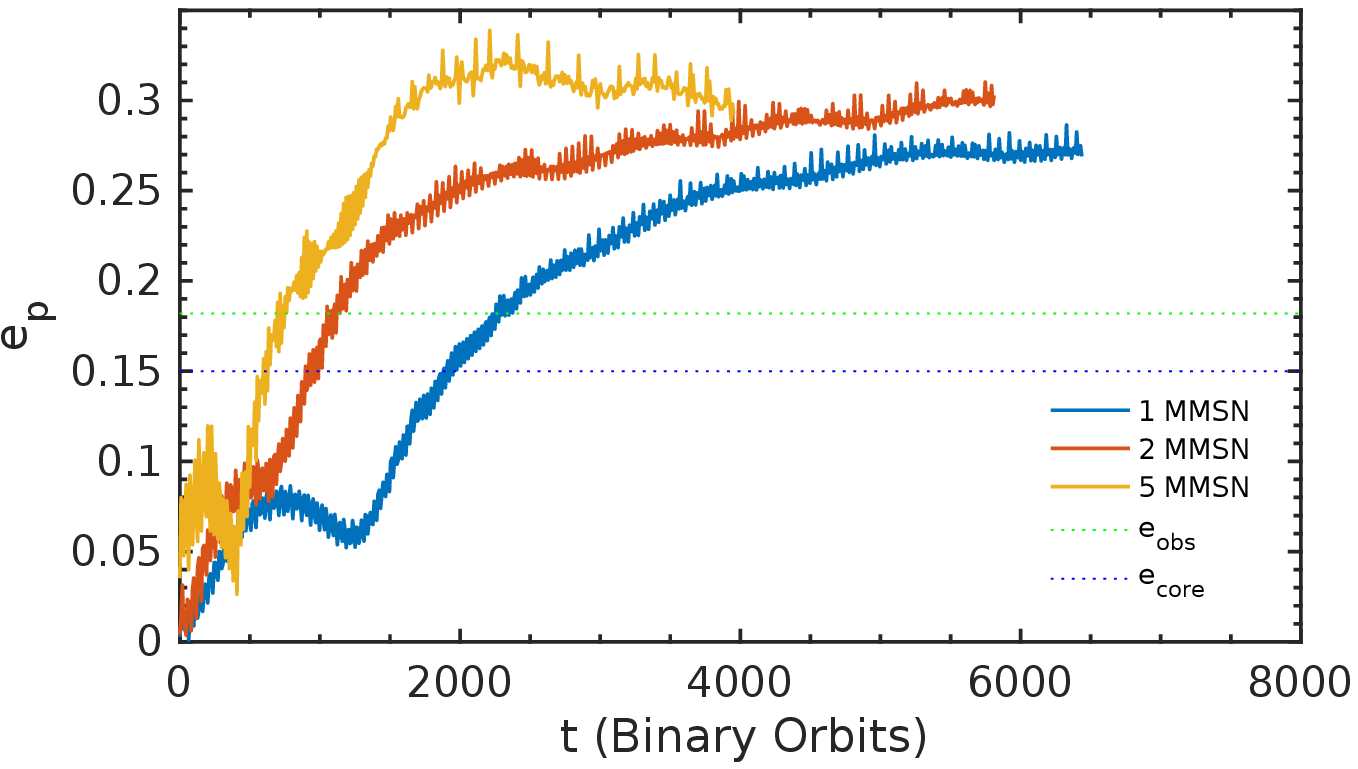}}\\
\vspace{-10pt}
\subfloat[]{\includegraphics[width = 0.47\textwidth]{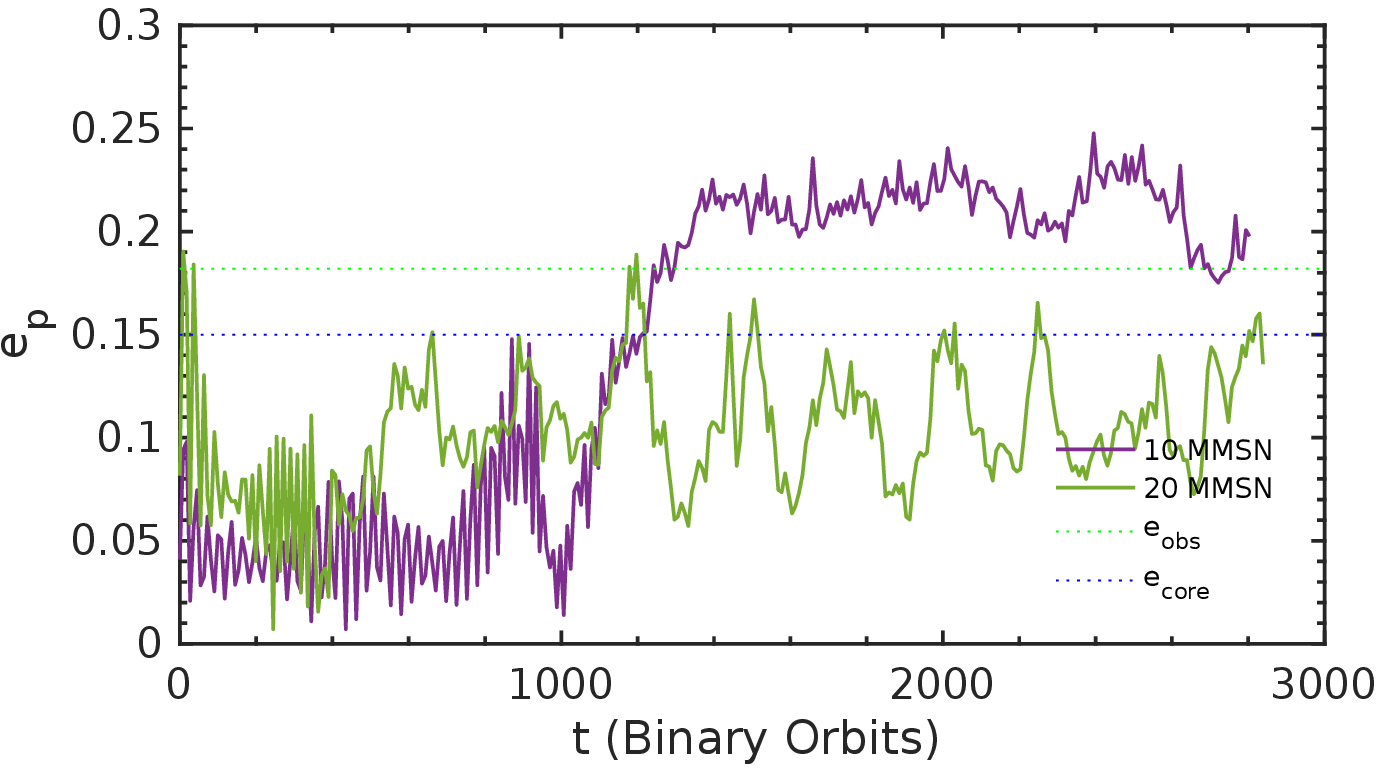}}
\caption{The top panel shows the evolution of $q_{\mathrm{p},0}=6 \times 10^{-5}$ protoplanetary cores' semi-major axes in evolved self-gravitating discs in the Kepler-34 system. The middle and bottom panels shows these cores' eccentricity evolution in the low- and high-mass disc models respectively.}
\label{fig:mig_res_kep34}
\end{figure}

\begin{figure}
\centering
\hspace{-18pt}
\subfloat{\includegraphics[width = 0.25\textwidth]{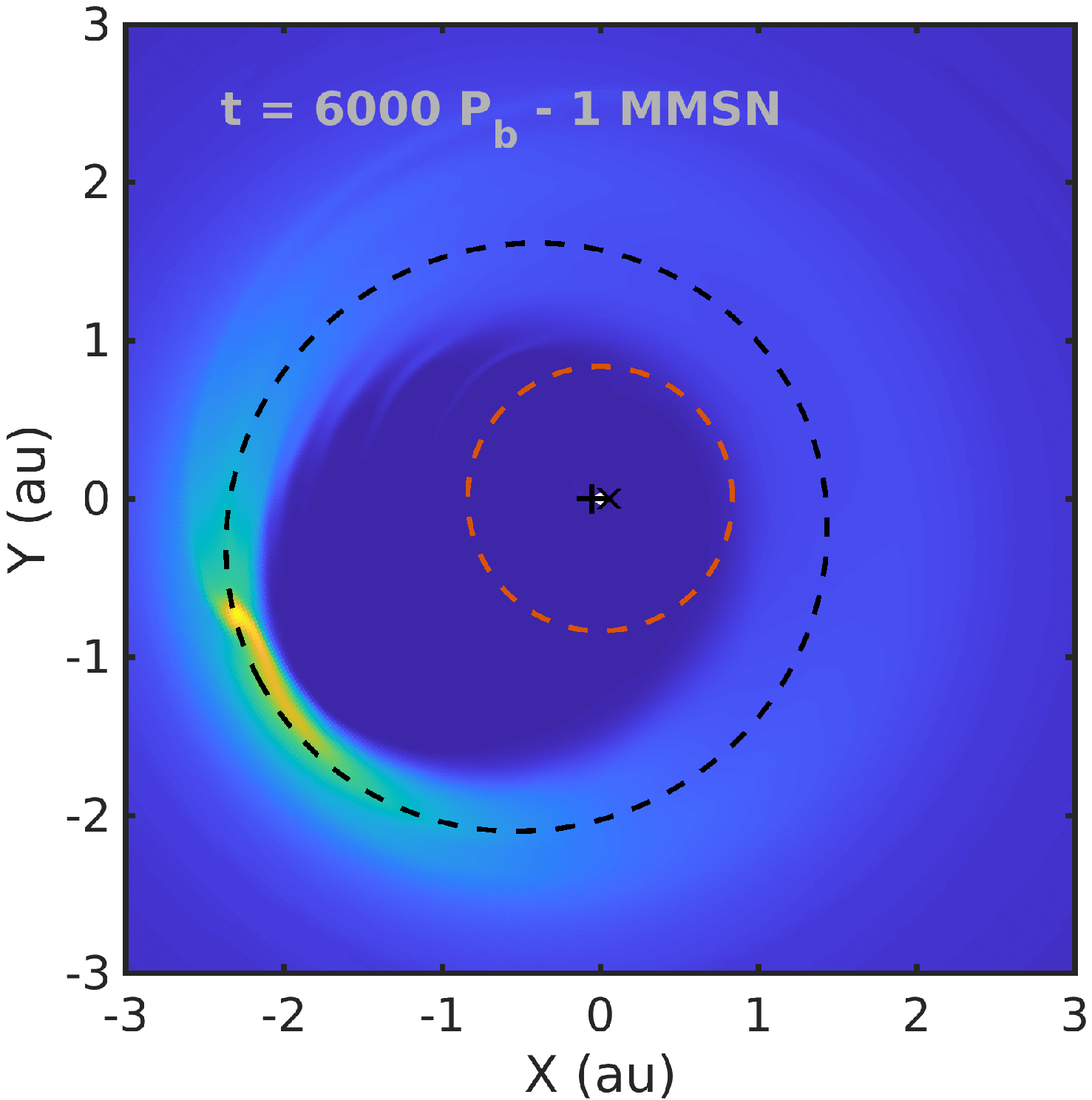}}
\hspace{-2pt}
\subfloat{\includegraphics[width = 0.25\textwidth]{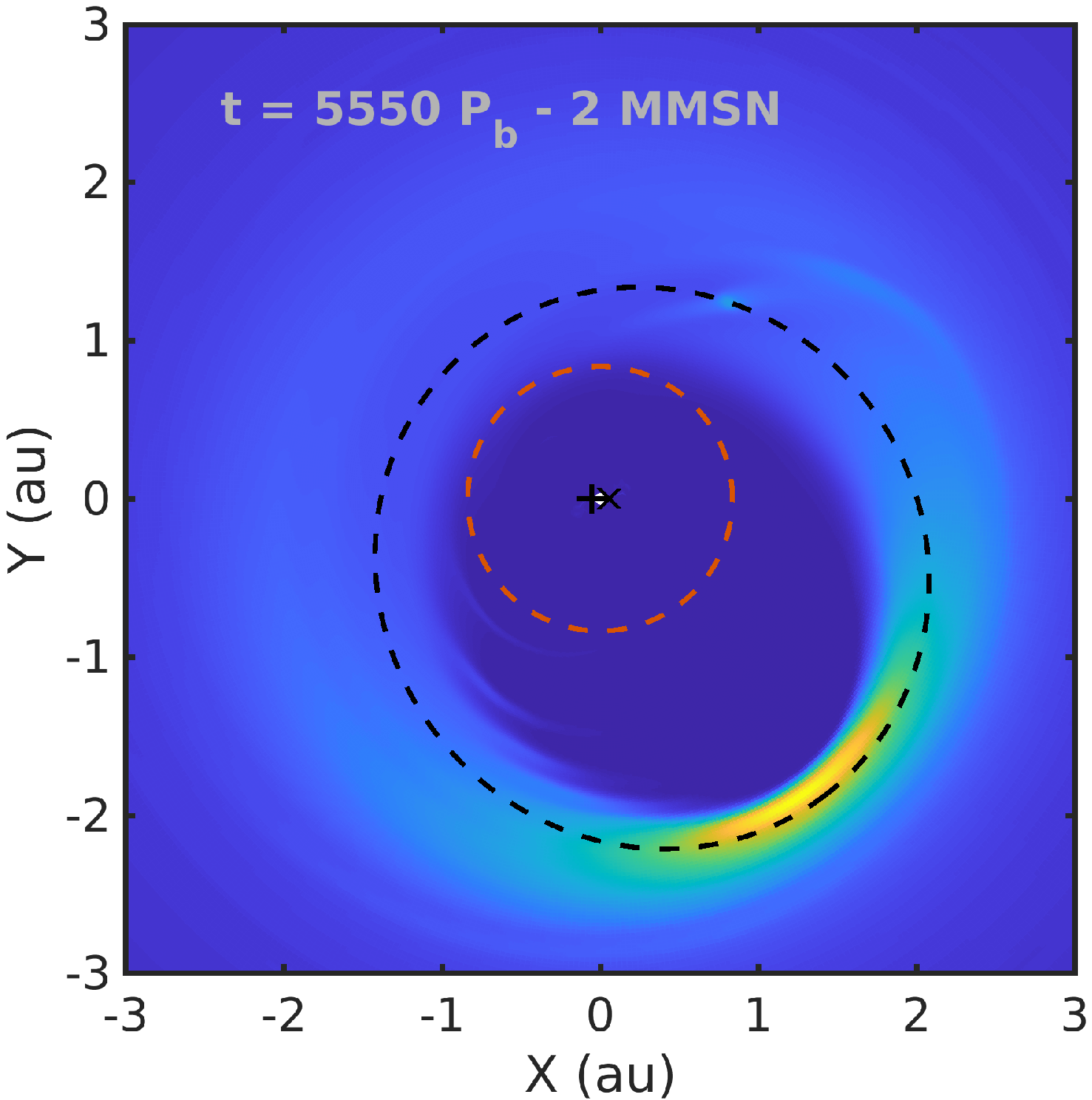}} \\ 
\hspace{-18pt}
\subfloat{\includegraphics[width = 0.25\textwidth]{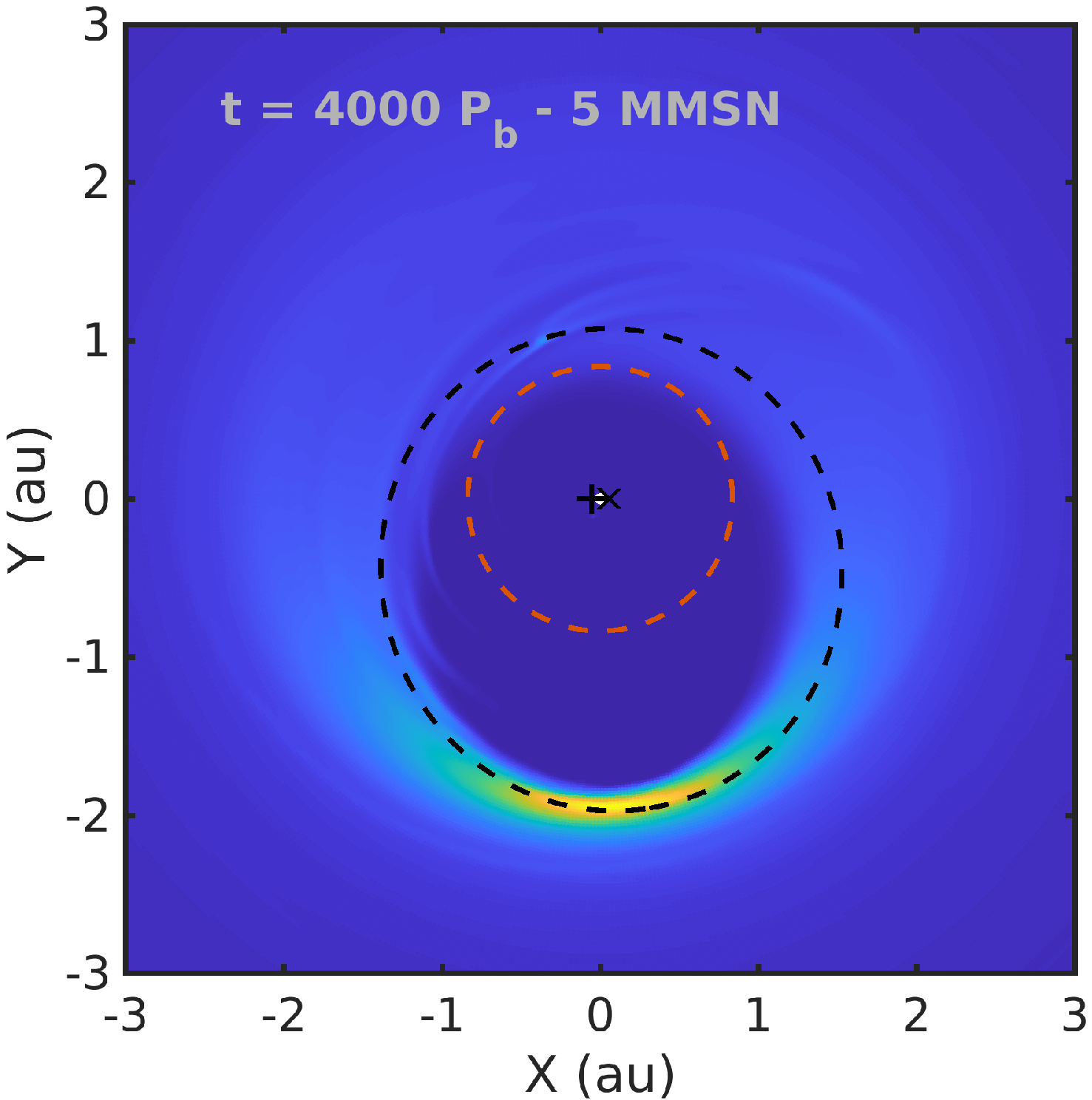}}
\hspace{-2pt}
\subfloat{\includegraphics[width = 0.25\textwidth]{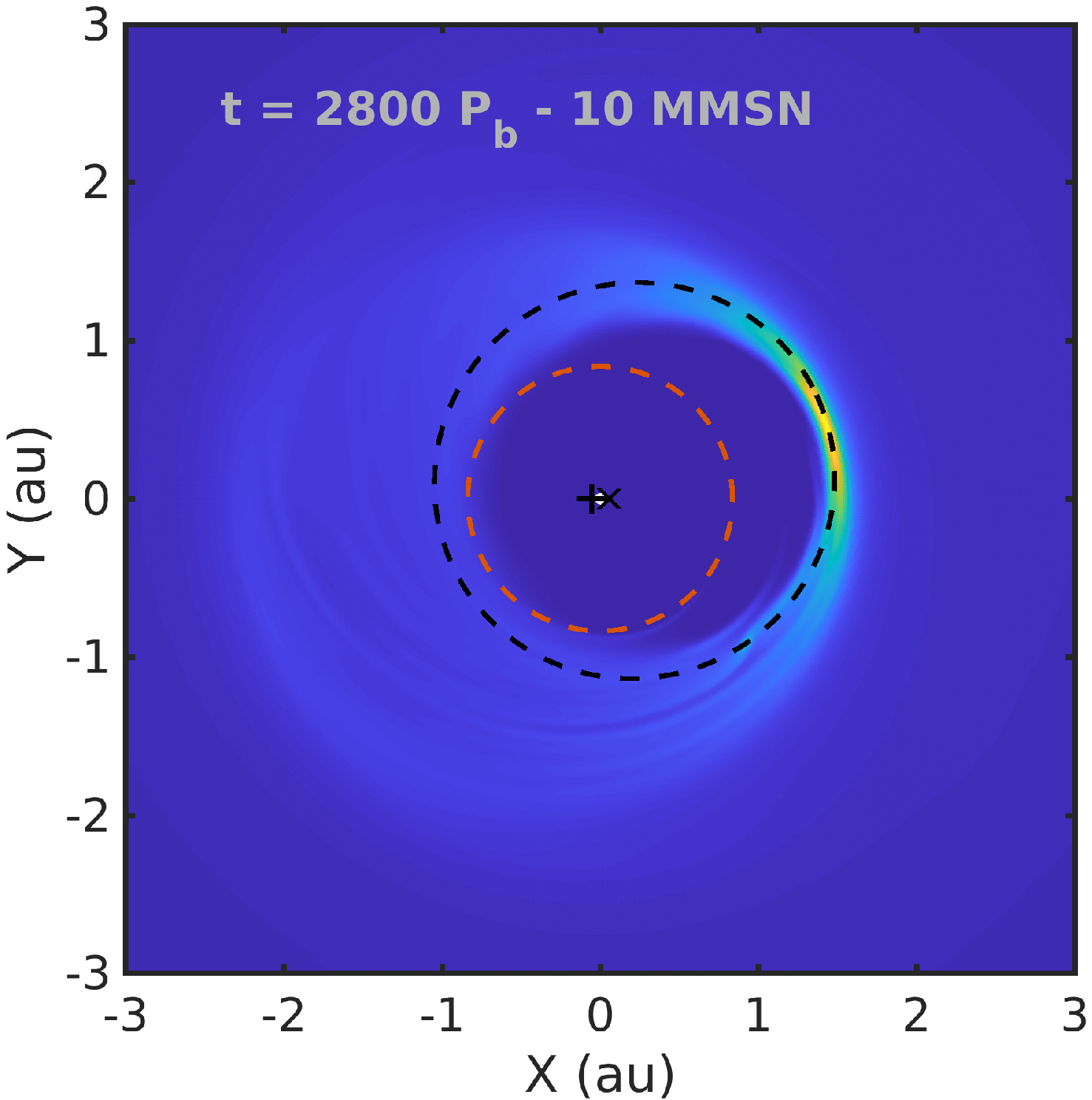}} \\
\hspace{-14pt}
\subfloat{\includegraphics[width = 0.25\textwidth]{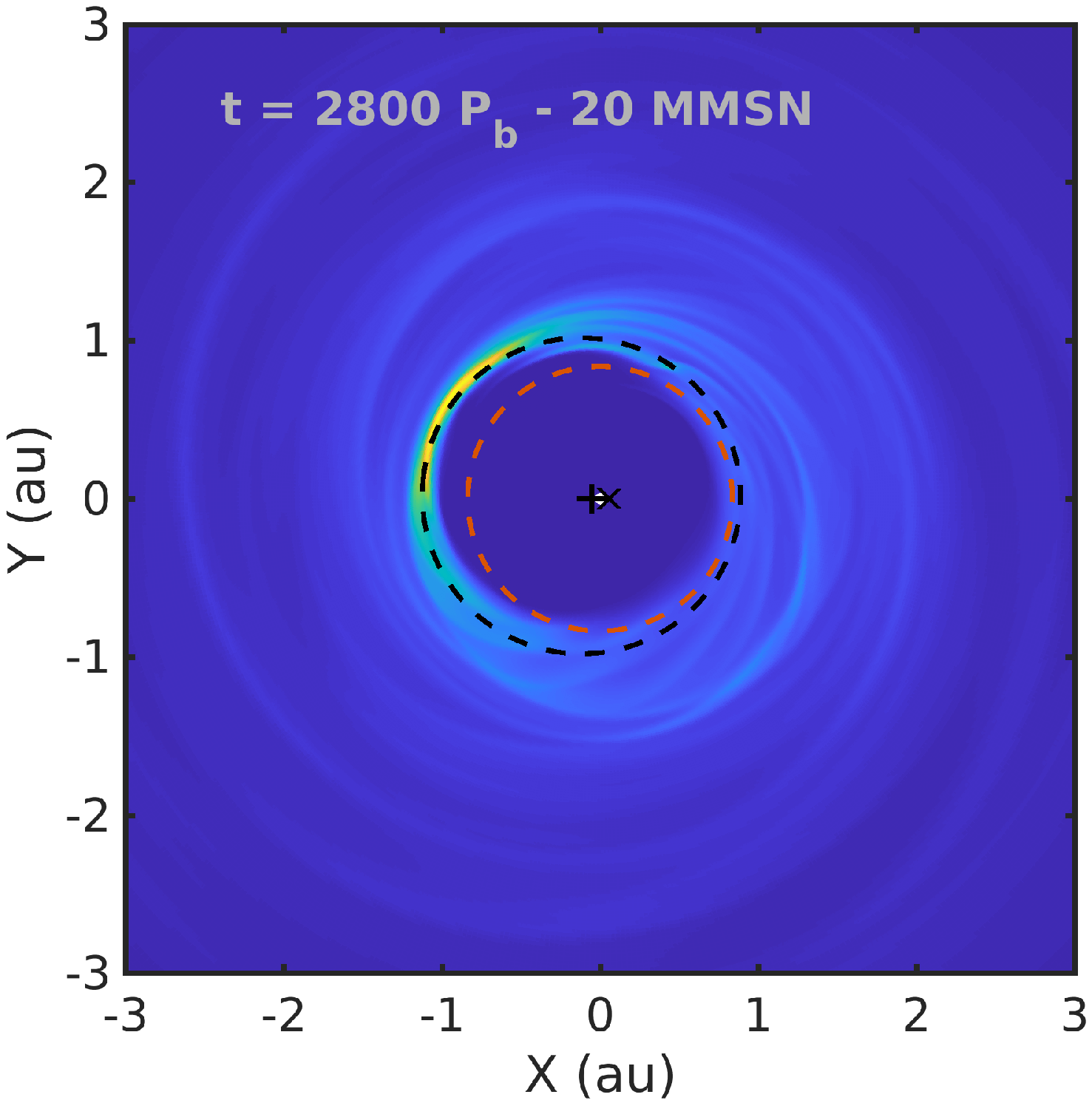}}
\caption{2D surface density profiles showing the structure of the binary-disc-core system once migration has halted for the 1--\mmsn{20} disc models around the Kepler-34 binary. The planetary orbit is shown by the black-dashed line, with the red-dashed line showing the critical stability limit. For each of these models it can be seen that the planet migrates into the inner disc, where it then interacts with the strong eccentric feature. The stopping position can be seen to decrease as the disc mass increases. Clear evidence of pericentre-alignment between the core and the eccentric cavity can be seen}
\label{fig:mig_res_kep34_sdgrid}
\end{figure}

\begin{figure}
\centering
\subfloat[]{\includegraphics[width = 0.47\textwidth]{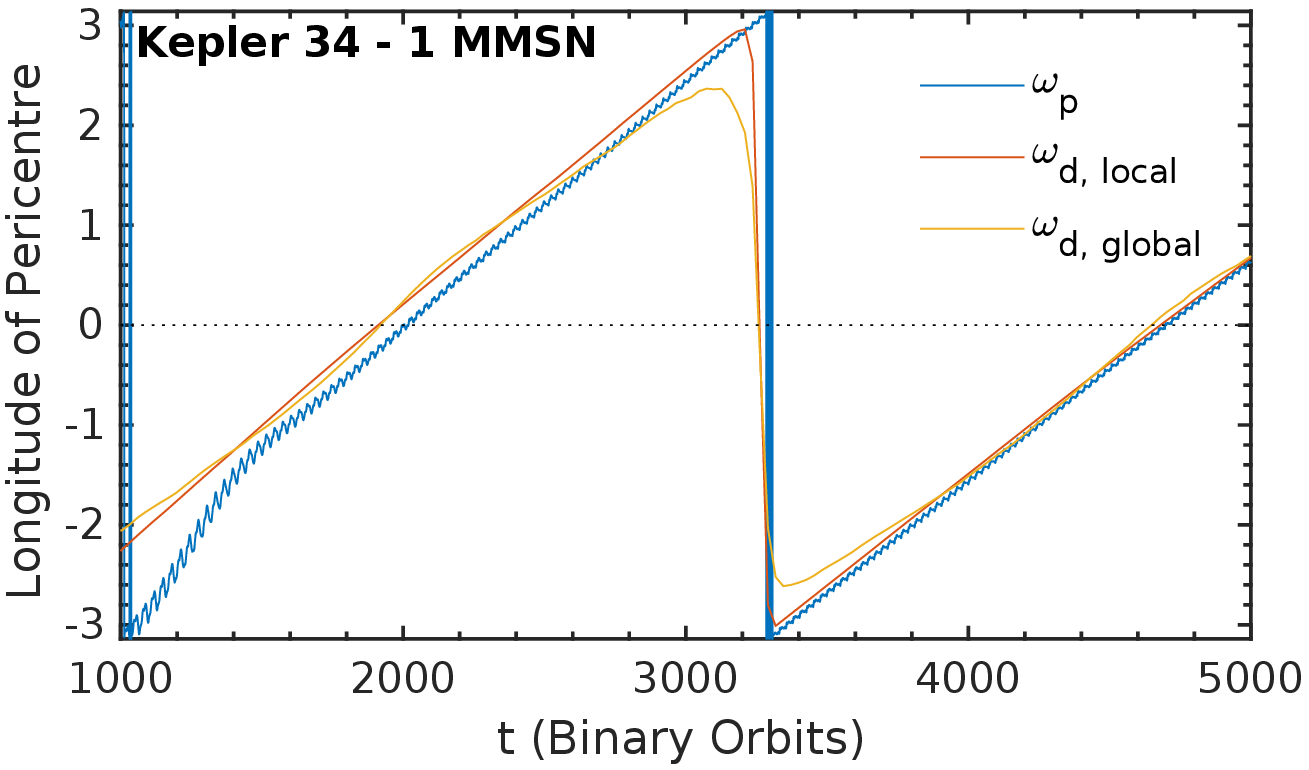}} \\
\vspace{-10pt}
\subfloat[]{\includegraphics[width = 0.47\textwidth]{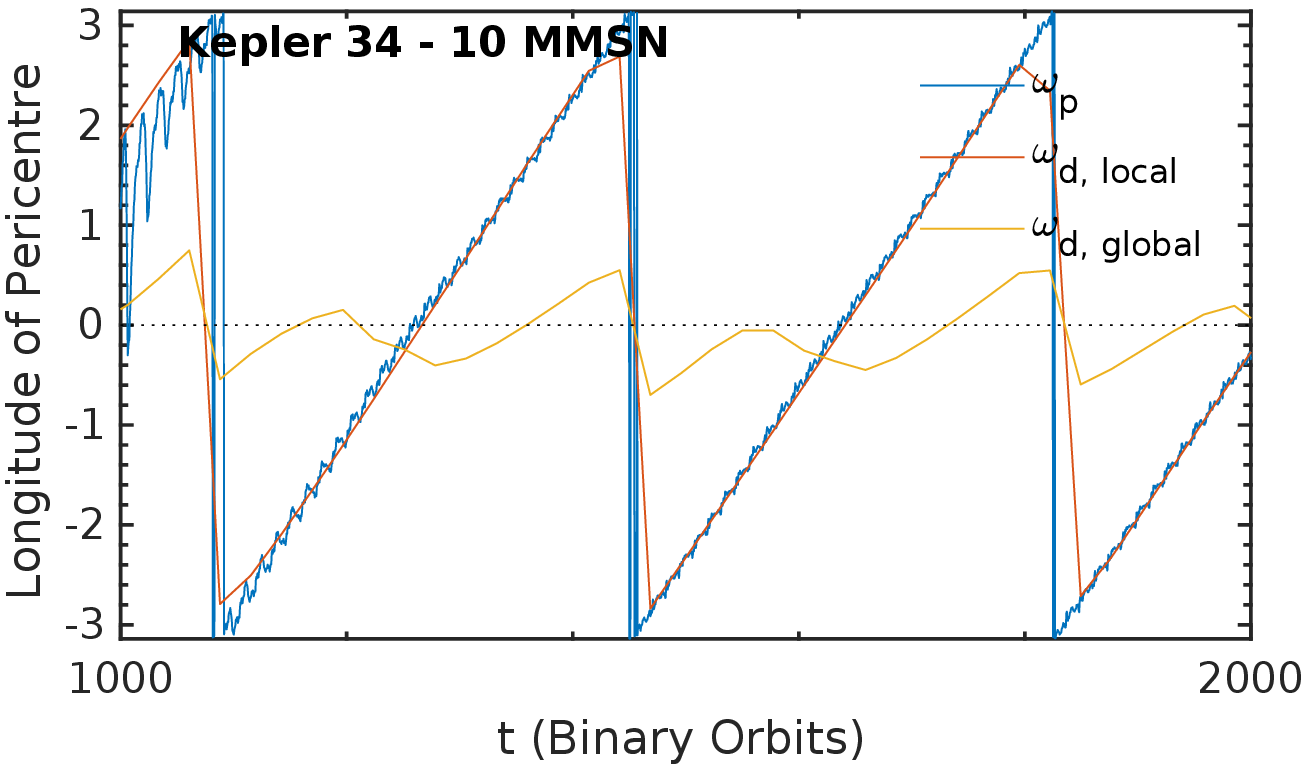}} \\
\vspace{-10pt}
\subfloat[]{\includegraphics[width = 0.47\textwidth]{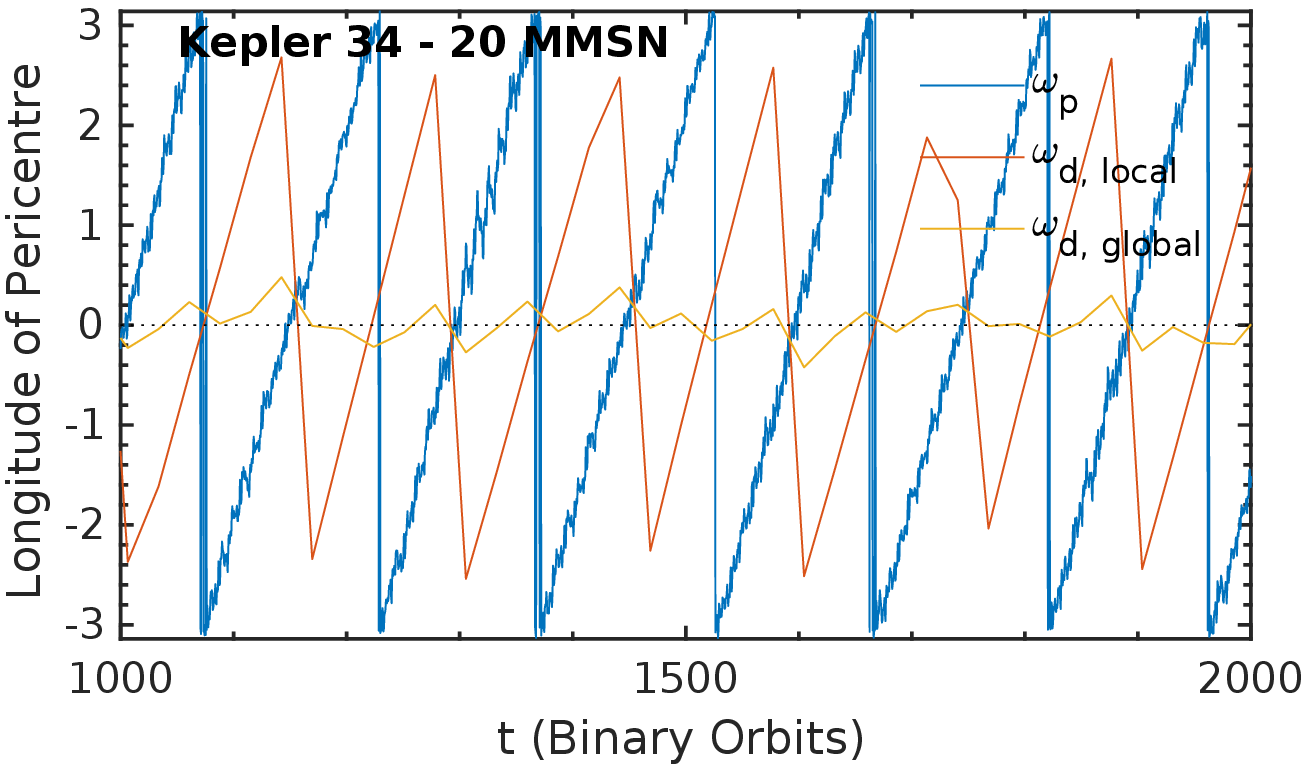}}
\caption{Evolution of longiude of pericentre of the disc-planet system in the 1, 10, and \mmsn{20} disc mass models around Kepler-34. Both the global and local calculations of the disc longitude of pericentre are included (see the text for a description of the differences between these two calculations. In the 1--\mmsn{10} discs the planet evolves into a state where the phase and period of circulation of its precession match that of the inner eccentric cavity of the disc. The two match when the planet has halted its migration at the inner cavity.}
\label{fig:mig_res_kep34_lop}
\end{figure}

As a result of the variety in the evolved disc structures in the Kepler-34 models, we see a large range of results for the migration of protoplanetary cores in these evolved discs. In Paper I we found that as we increase the disc mass from 1 to \mmsn{20}, the size of the initially very eccentric, extended cavity in the least massive disc gradually decreases -- self-gravity acts to compact the scale of the system. As can be seen from Fig. \ref{fig:mig_res_kep34} a similar pattern can be seen in the final stopping positions of the migrating cores, where the halting of migration occurs because \e{p} increases and induces a torque reversal.

In the low-mass discs there is a clear trend for \arm{p} to decrease from \arm{p} $\approx 1.95$ to 1.5 au, and for \e{p} to increase from \e{p} $=0.275$ to 0.3 as the disc mass rises from 1 to \mmsn{5}. Looking at the disc eccentricity distributions in the middle panel of Fig. \ref{fig:sg_disc_edprof}, for a given radius one obtains a smaller value for the average disc cell eccentricity for larger disc masses -- the core therefore has to migrate further through the disc so that \e{p} $\approx$ \e{d}. The large \e{p} seen for the cores in these discs means the strength of the co-rotation torque would be greatly diminished \citep{Fendyke2014}, therefore making torque reversal the dominant mechanism for halting migration. The low-mass discs, especially the \mmsn{1} model, do not match the results obtained in PN13, and show poor agreement with the observed Kepler-34b. The discs in these models tend to have large, highly eccentric cavities compared to the equivalent models in PN13. Our more realistic treatment of the inner disc boundary, allowing for a more accurate capturing of angular momentum flux through the disc due to the binary, is the likely explanation for this.

Whilst the low-mass disc results do not agree well with past results or the observed state of the Kepler-34 planetary system, the \mmsn{10} and \mmsn{20} models agree relatively well with the planetary orbital elements quoted by \citet{Welsh2012b}. The final stopping positions in the \mmsn{10} and \mmsn{20} systems, 1.2 and 1.0 au respectively, bracket the observed value of \arm{p} $\approx 1.1$ au due to the compacting of the system as disc-mass and self-gravity increase. The trend for \e{p} to increase as the disc-mass increases is reversed in the high-mass regime, possibly due to the disc-mass in the vicinity of the planet providing significant damping. We see no evidence of the core being trapped in the outer disc in either model. Examining the last two panels in Fig. \ref{fig:mig_res_kep34_sdgrid}, we can see that although additional eccentric features are present in the outer disc, in both the 10 and \mmsn{20} models, they are far less well-defined than those in Kepler-16. These washed-out features are not strong enough to halt the inwards migration of the cores in this system.

A common feature that can be seen for all the disc models in this system can be observed in Fig. \ref{fig:mig_res_kep34_sdgrid}. One can clearly see that the orbits of the cores in each system are aligned with the precessing eccentric inner cavity. An examination of the evolution of the planet's longitude of pericentre, alongside that of the mean disc longitude of pericentre shows this as well (Fig. \ref{fig:mig_res_kep34_lop}). As the planet migrates into the inner disc, the phase and period of precession both evolve into lockstep with that of the inner disc cavity (which the local calculation of $\omega_{\mathrm{d}}$ traces). The planet and eccentric feature precess with each other, in a pericentre-aligned fashion, a behaviour previously seen for full mass planets in non-self-gravitating discs in the Kepler-34 system \citep{Kley2015}. This is not true in the most massive disc model presented here, \mmsn{20}. In this case, the precession of the planet and disc are half a precession period out of phase such that their eccentric orbits are anti-aligned. In Paper I we give an explanation for the global and local calculations of the disc eccentricity and longitude of pericentre. Whilst the global calculation takes into account all the material in the disc between $R_\mathrm{in}$ and $R_\mathrm{out}$, the local calculation only takes into account material up to and just beyond the position of the surface density peak associated with the inner cavity. This procedure ignores the effect of exterior eccentric, precessing material in the outer disc.

We note that a second separate simulation of a protoplanetary core released at 3 au in the Kepler-34 \mmsn{20} disc model was undertaken to examine migration from a larger radius. This location corresponds to a radius between the second and third additional eccentric features in the disc. Whilst these features are relatively weak, they still alter the surface density profile of the disc. These regions of positive surface density gradient are sufficient to hamper any inwards migration of the planet, but insufficient to excite sufficient eccentricity for it to escape. The forces acting on the planet at this outer position which normally result in inwards migrations are overcome by the small perturbations in surface density -- whilst at the starting position of the first \mmsn{20} core the rate of inwards migration is greater. These weak features could play an important role in the early stages of planet formation, trapping large numbers of planetesimals, boulders or pebbles, but with low eccentricity, providing a reservoir for protoplanetary core creation. For clarity we have not included this simulation in the plots for this section, but we will discuss its further evolution in subsequent sections.

\subsection{Kepler-35}\label{sec:mig_res_35}
\begin{figure}
\centering
\subfloat[]{\includegraphics[width = 0.47\textwidth]{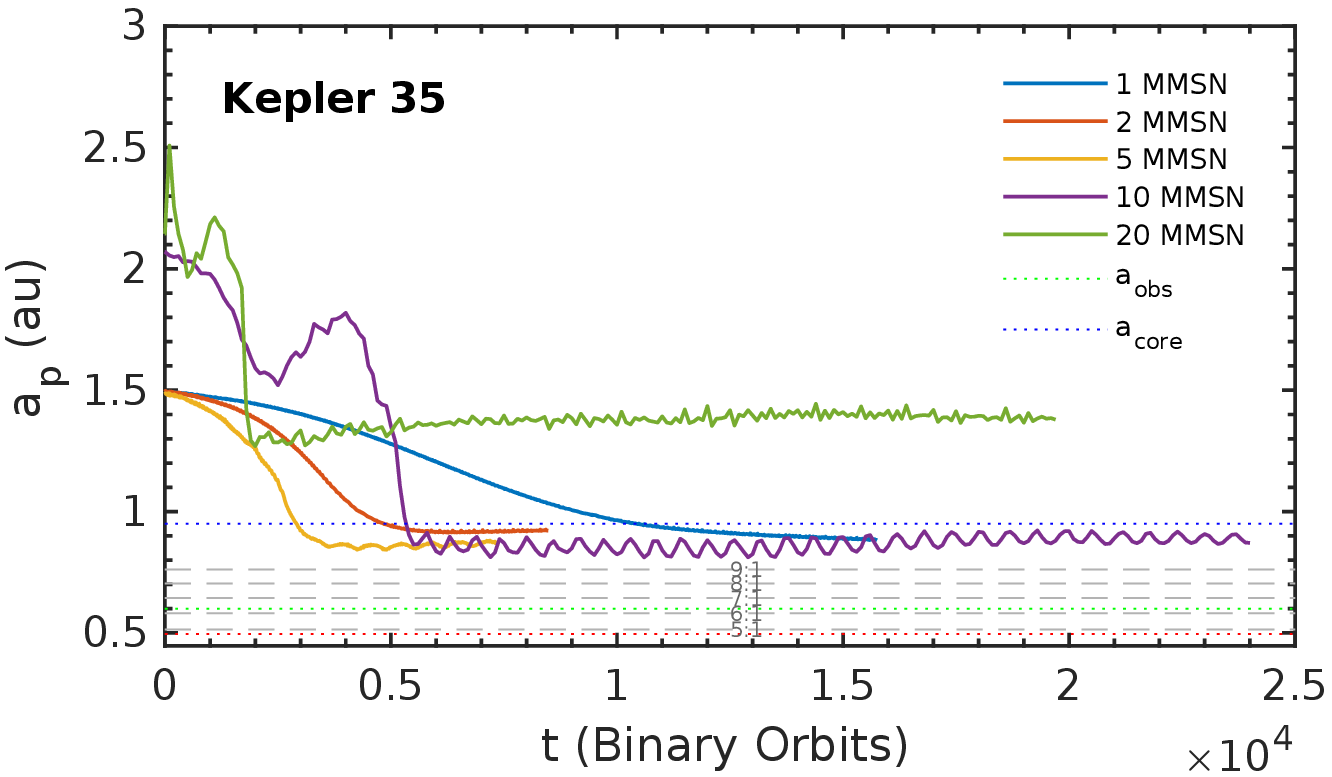}} \\
\vspace{-10pt}
\hspace{-8pt}
\subfloat[]{\includegraphics[width = 0.47\textwidth]{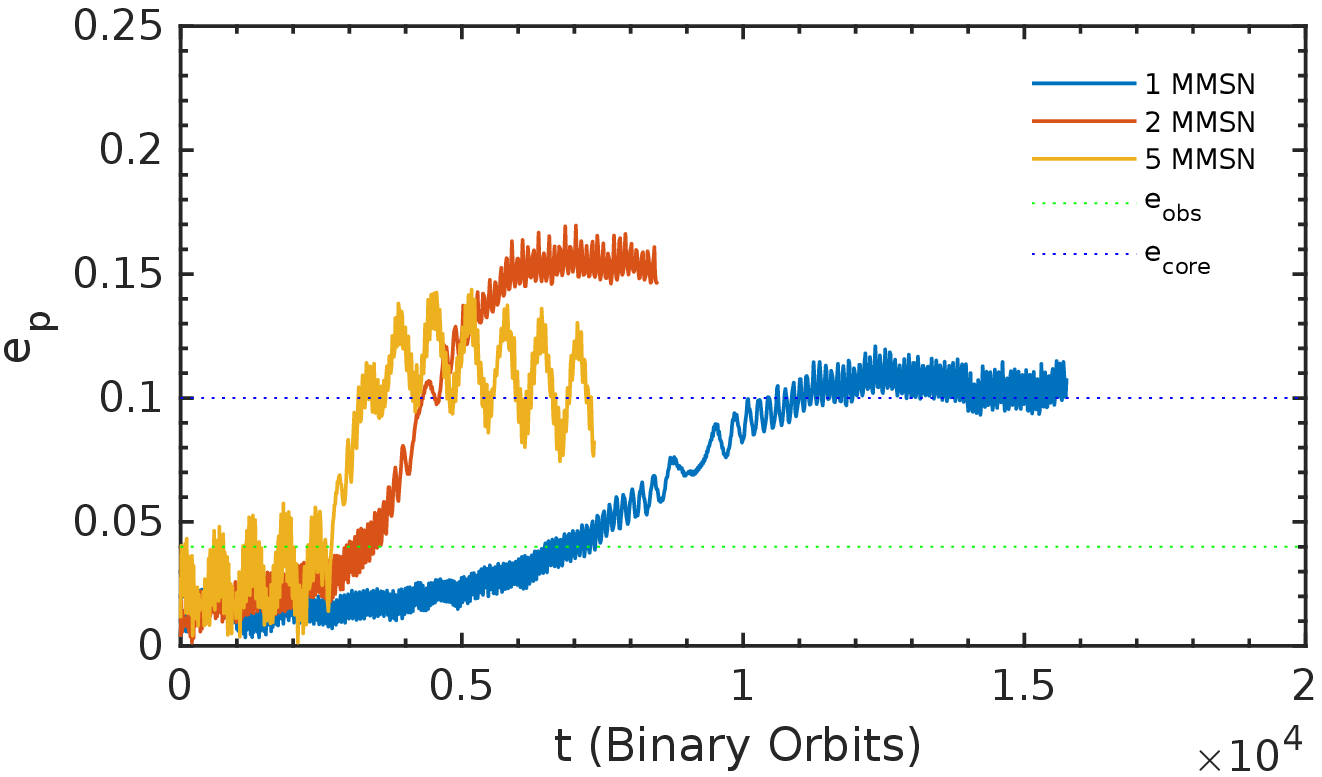}}\\
\vspace{-10pt}
\subfloat[]{\includegraphics[width = 0.47\textwidth]{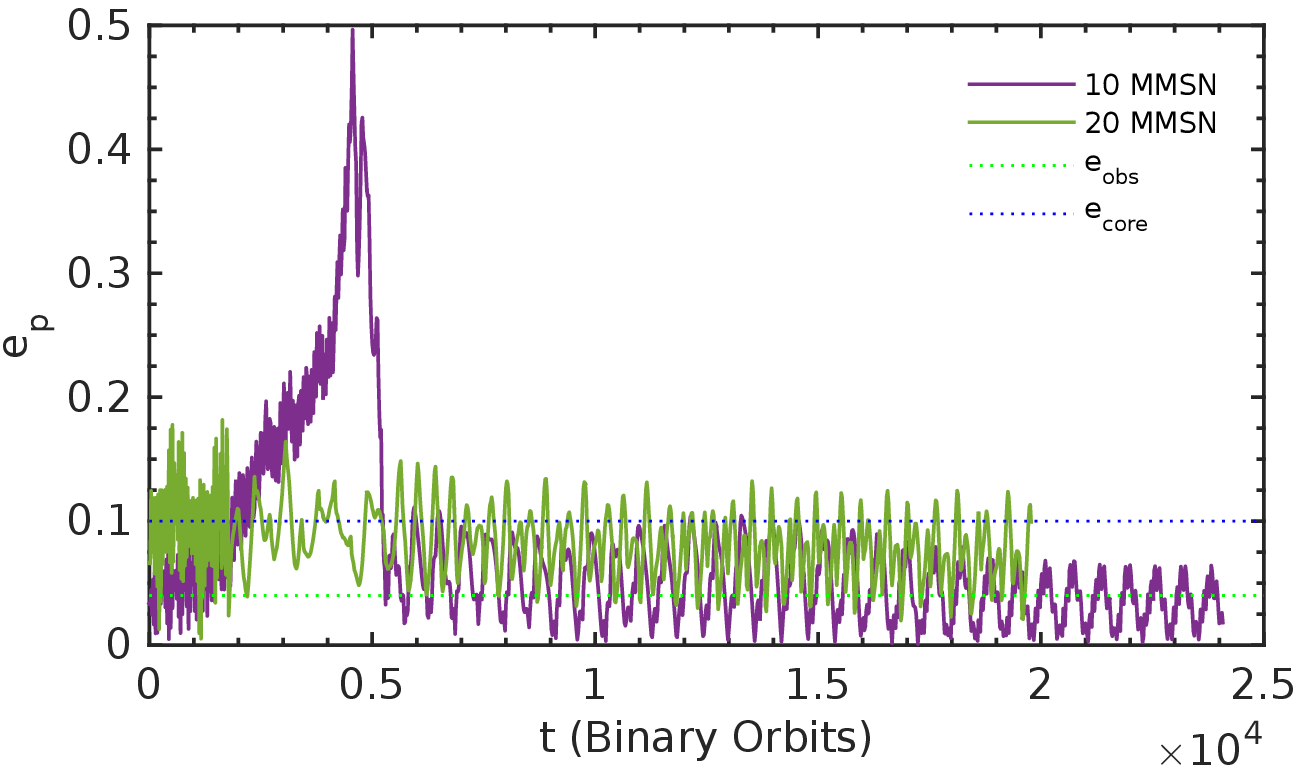}}
\caption{The top panel shows the evolution of $q_{\mathrm{p},0}=6 \times 10^{-5}$ protoplanetary cores' semi-major axes in evolved self-gravitating discs in the Kepler-35 system. The middle and bottom panels shows these cores' eccentricity evolution in the low- and high-mass disc models respectively.}
\label{fig:mig_res_kep35}
\end{figure}

\begin{figure}
\centering
\includegraphics[width = 0.48\textwidth]{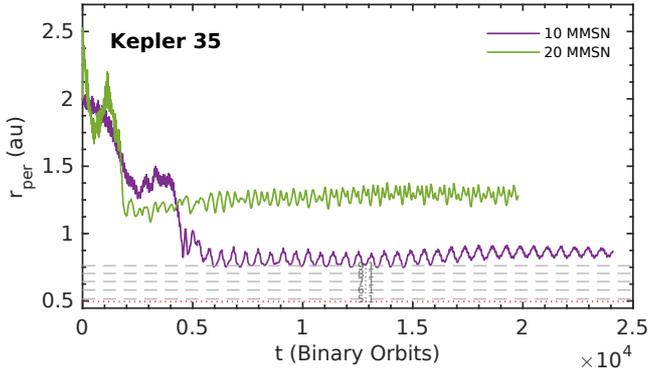}
\caption{Evolution of pericentre distance of the protoplanetary cores in the \mmsn{10} and \mmsn{20} disc models. Of particular note is the decrease between 3000 and \Pb{5000}. During this time the core has been trapped at the location of the first additional eccentric feature. Whilst its eccentricity is being excited, the pericentre distance is decreasing, to the point where it reaches the location of the inner eccentric cavity. Whilst the pericentre distance remains relatively constant, the eccentricity and semi-major axis are damped by this massive feature.}
\label{fig:mig_res_kep35_qpt}
\end{figure}

The results from protoplanetary core migration in the evolved Kepler-35 disc models look very similar to those from the Kepler-16 models, with minor changes caused by the differences in evolved disc structure. This result was expected due to the similarity in evolution and final structure results from Paper I for these low-eccentricity binaries. To recap, in the low-mass discs, the cores migrate inwards through the disc until they reach a location where their eccentricity is excited enough (\e{p} $\approx$ \e{d}(\arm{p})) to induce a torque reversal. This location corresponds to the edge of the eccentric cavity. The location of this edge, which is easily identifiable as the peak in the surface density, lies at a smaller radius in the Kepler-35 system than in Kepler-16, due to its lower binary eccentricity. The final semi-major axis for the cores in the 1--\mmsn{5} models is 0.9 au -- slightly smaller than the previous PN13 result -- and \e{p} oscillates around 0.11 for all three models (see the upper and middle panels of Fig. \ref{fig:mig_res_kep35}). The mean value of \e{p} results from a balance between the highly eccentric disc pumping up the eccentricity and the surrounding material damping the eccentricity. None of the final planetary orbital elements obtained in the low-mass regime are in good agreement with those quoted in \citet{Welsh2012b} for Kepler-35b.

The results from the high-mass models also show the same evolutionary history, with slightly different final values for \arm{p} and \e{p}, as the Kepler-16 high-mass models. The \mmsn{10} model shows evidence of trapping by the $m=1$ eccentric mode at 1.6--1.9 au, where its eccentricity gets rapidly excited to 0.5. This highly eccentric orbit then brings the pericentre close enough to the inner cavity to allow the planet to be captured by the large amount of material skirting the boundary. This material damps the orbit of the planet, decreasing the semi-major axis (\arm{p} $=0.9$ au) and eccentricity(\e{p} $=0.03$), to a near-circular orbit (see upper and lower panels of Fig. \ref{fig:mig_res_kep35}, and Fig. \ref{fig:mig_res_kep35_qpt}).

The \mmsn{20} model, matching the evolution of the core in the Kepler-16 \mmsn{20} disc, migrates inwards through the disc -- keeping a low eccentricity, \e{p} $\approx 0.08$ -- until it is trapped at the first extra eccentric feature, with a final semi-major axis, \arm{p} $=1.4$ au . The eccentricity damping provided by the disc is sufficient that it remains at this location. Both the 10 and \mmsn{20} discs produce cores whose final eccentricity is in good agreement with that of Kepler-35b, but the simulated semi-major axes are too large.

\section{Migration of Accreting Cores}\label{sec:acc_res}
The simulations that have been presented so far in this paper all adopted a fixed mass for the planetary cores, corresponding to a mass ratio between the planet and central binary of $q_{\rm p}=6 \times 10^{-5}$. The actual mass ratios for the observed systems are all larger than this by various factors (see Table~1), and so we now consider what happens to the orbital elements if the planets accrete gas and achieve their observed masses while migrating.
The results shown above indicate that the planets considered so far normally halt their migration at a location that is too far from the binary to provide good agreement with the observations, so we examine whether or not the stopping orbital radii decrease as we increase the planet masses to their observed values. Only the \mmsn{10} and \mmsn{20} cases for the Kepler-34 system produce final stopping radii that agree well with the observations, and this occurs because of the dramatic shrinking of the cavity size in this case for discs where self-gravity is important.

For the simulations presented in this section we undertook accretion scenarios for the evolved binary-disc-planet Kepler-16 and Kepler-34 systems. The initial core mass ratio used throughout this work means that the protoplanetary core in the Kepler-35 models is within $\approx 20\%$ of the observed planet mass, so we didn't simulate gas accretion in this case. The accretion routine of \citet{Kley1999} was used to grow the mass of the protoplanetary cores ($q_{\mathrm{p},0}=6 \times 10^{-5}$) to that of the observed planet mass in the specific system ($q_\mathrm{p}= 3.54\times 10^{-4}$ and $1.01\times 10^{-4}$ in the Kepler-16 and -34 systems respectively). This prescription removes a portion of the gas from the Hill-sphere and adds its mass to that of the planet. The accretion time-scale, i.e. the time in which the Hill-sphere is emptied of gas, is determined as a fraction of the dynamic time-scale of the planet, $t_{acc}=f\,t_{dyn}$. The variable constant $f$ is tuned for the Kepler-16 simulations so that the planet reaches its final mass over \Pb{5000}. We use this approach to inhibit the growth of the planet. A constant value is used ($f=0.01^{-1}$) for all the disc-mass models in the Kepler-34 system, as the final planet mass is relatively low. The issue worth noting with this set-up, in relation to a realistic comparison with the masses of the observed circumbinary planets, is that when accretion is turned on in these simulations the planet finds itself at a location with a wealth of material. Even conservative estimates for the accretion time-scale lead to rapid mass-growth. If planets in circumbinary discs only reach a gas-accretion phase when they are already at the cavity edge, it would be logical to assume that the planet could quickly grow to Jovian mass unless gas accretion is very slow indeed, or occurs at the end of the disc lifetime. Our simulations apply to the former possibility, but it is worth noting that circumbinary systems are self-selecting because too much gas accretion leads to the formation of a Jovian-mass planet, and these tend to be much more unstable due to dynamical interaction with the central binary \citep{Nelson2003}. Even if circumbinary planets grow to be of Jovian-mass close to the cavity edge, we are unlikely to see them as they have a significant probability of being ejected from the system. 

The accretion scenarios that we consider here are run from the point in the simulations from the last section when the planet has reached a pseudo-steady orbit.

\subsection{Kepler-16}\label{sec:acc_res_16}
\begin{figure}
\centering
\subfloat[]{\includegraphics[width = 0.48\textwidth]{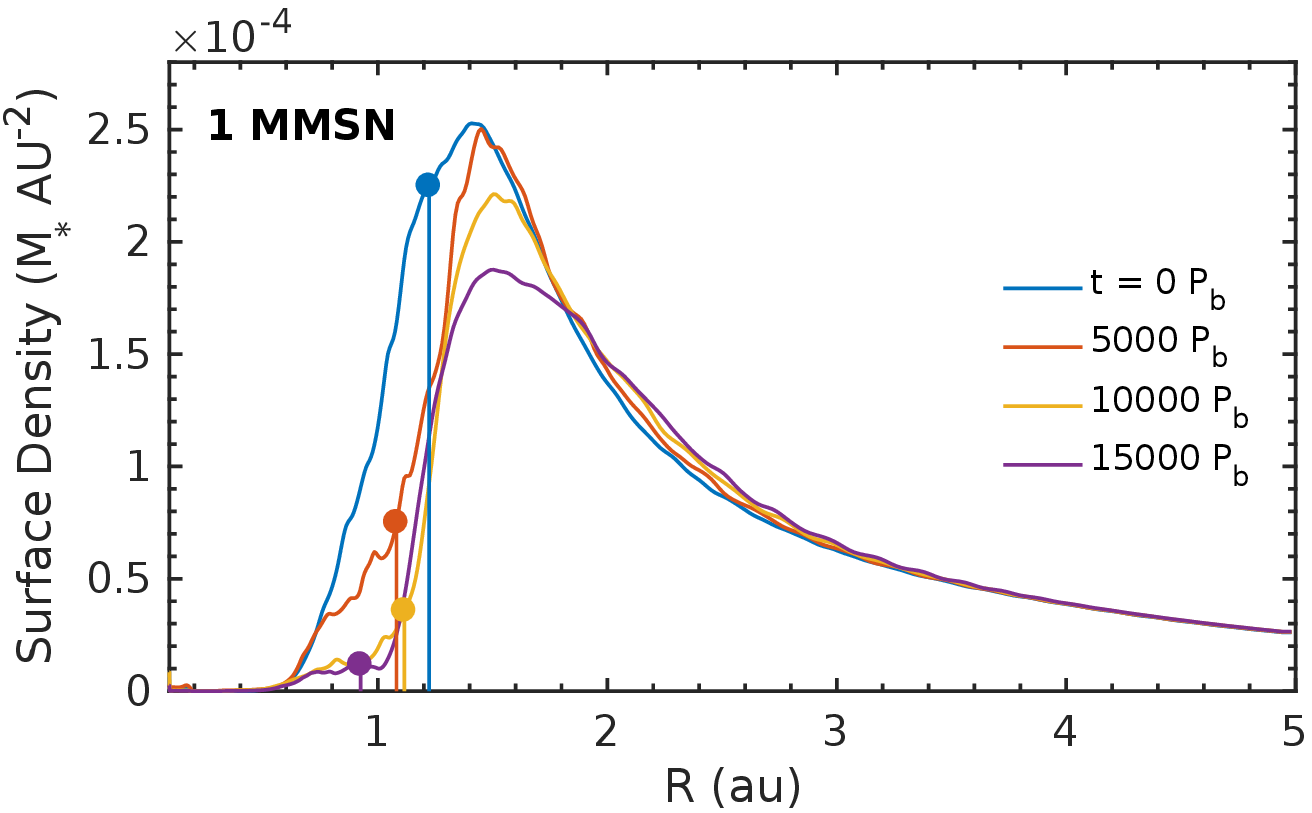}}\\
\vspace{-10pt}
\subfloat[]{\includegraphics[width = 0.47\textwidth]{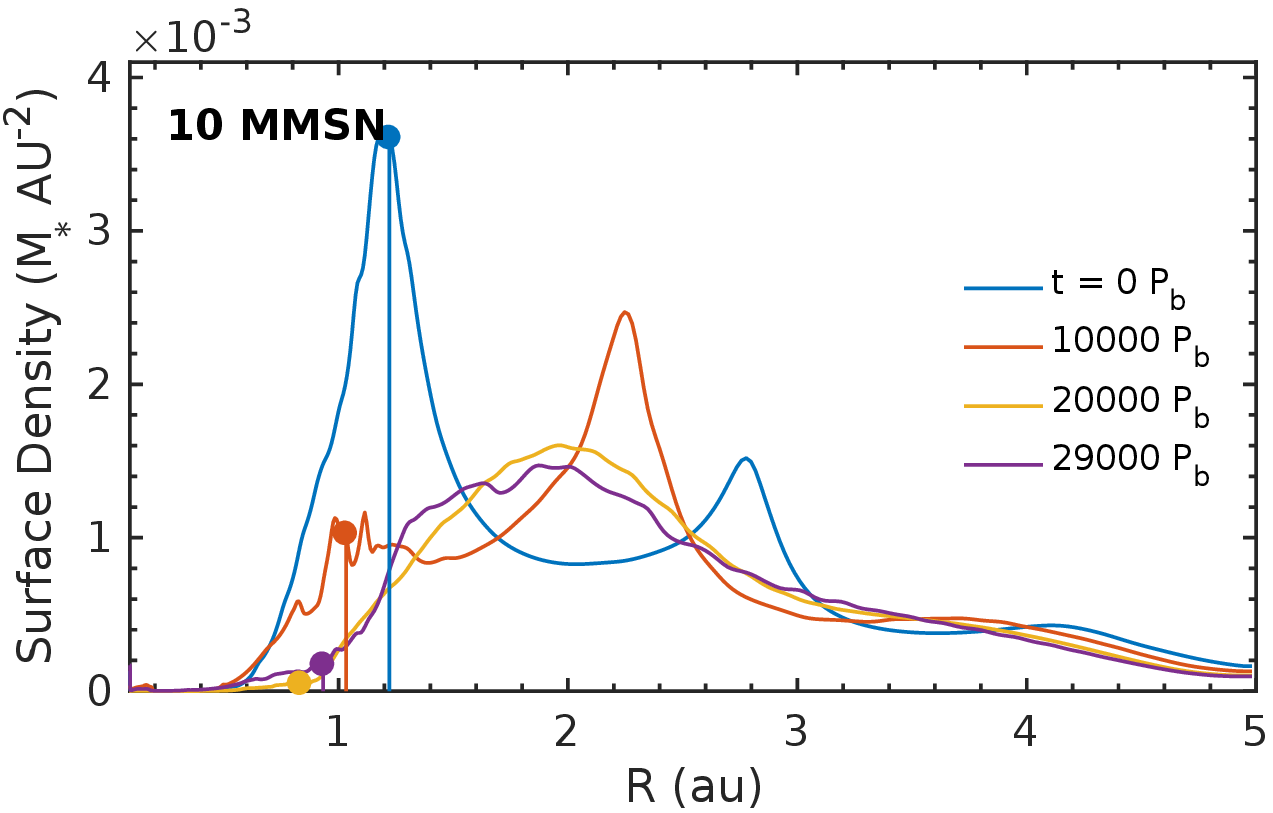}}
\caption{1D snapshots of the surface density profile in the \mmsn{1} (top panel) and \mmsn{10} (bottom panel) models in Kepler-16, once accretion onto the protoplanetary core is allowed. In the low-mass disc the core has grown from its initial mass to the final, observed mass of Kepler-16b between the first and second profiles. It can be seen that whilst the planet has migrated further into the inner disc, it has also carved out a pseudo-gap in the inner edge of the cavity. In the massive disc, whilst the planet doesn't seem to open a gap, it does significantly alter the surface density profile over the whole radial extent.}
\label{fig:acc_res_kep16_sds}
\end{figure}

\begin{figure}
\centering
\subfloat[]{\includegraphics[width = 0.47\textwidth]{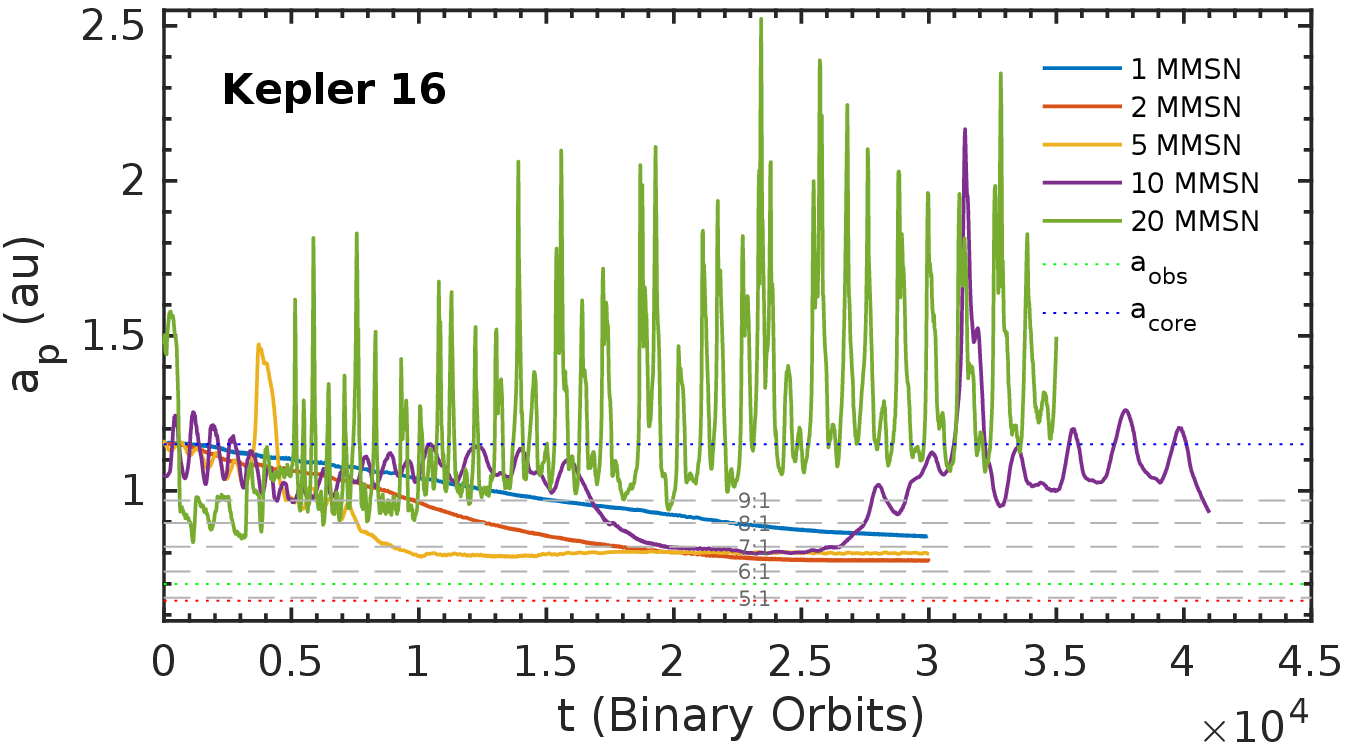}} \\
\vspace{-10pt}
\subfloat[]{\includegraphics[width = 0.47\textwidth]{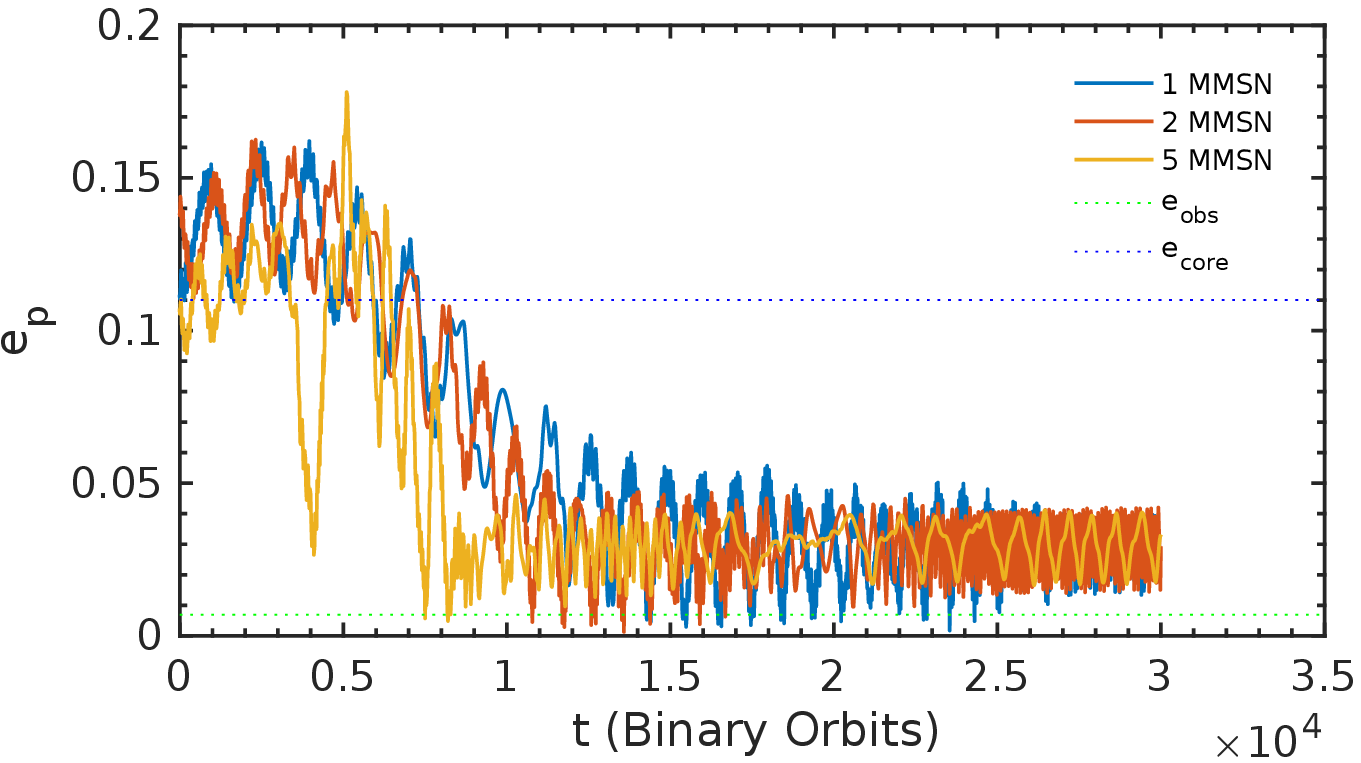}}\\
\vspace{-10pt}
\subfloat[]{\includegraphics[width = 0.47\textwidth]{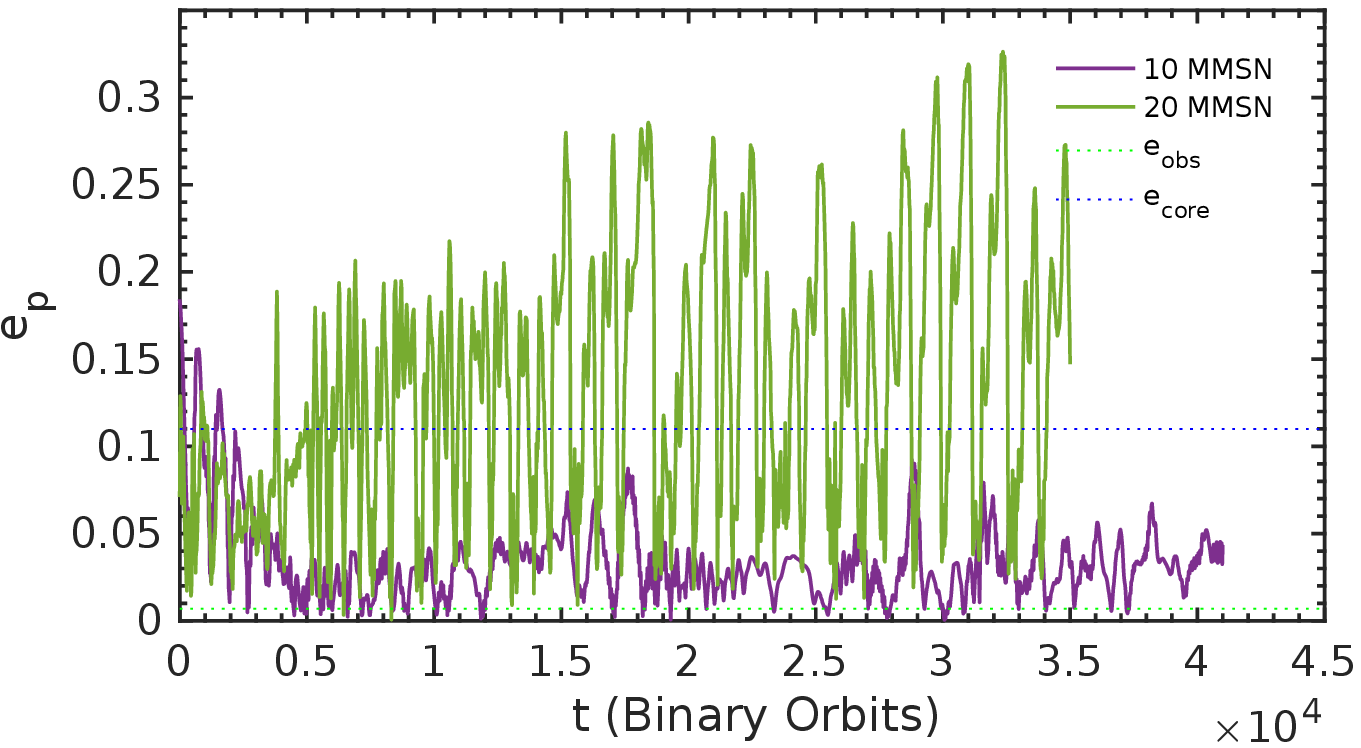}}
\caption{Evolution of accreting protoplanetary cores' semi-major axes in evolved self-gravitating discs in the Kepler-16 system. The middle and bottom panels shows these cores' eccentricity evolution in the low- and high-mass disc models respectively.}
\label{fig:migacc_res_kep16}
\end{figure}

\begin{figure}
\centering
\includegraphics[width = 0.45\textwidth]{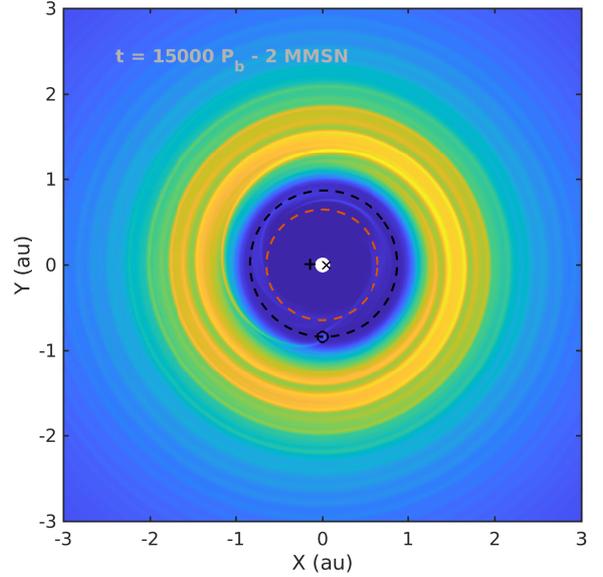}
\caption{Surface density map of the Kepler-16 \mmsn{2} disc model once the core has grown to its observed mass and reached its final semi-major axis. The near-circularity of the planetary orbit is clear, whilst the complex interplay between the density waves launched from the Lindblad resonances with the planet, and those launched by the binary in the self-gravitating disc, has destroyed the eccentric inner cavity. A gap has been opened with a circular cavity interior to it.}
\label{fig:acc_res_kep16_sd2}
\end{figure}

\begin{figure}
\centering
\includegraphics[width = 0.47\textwidth]{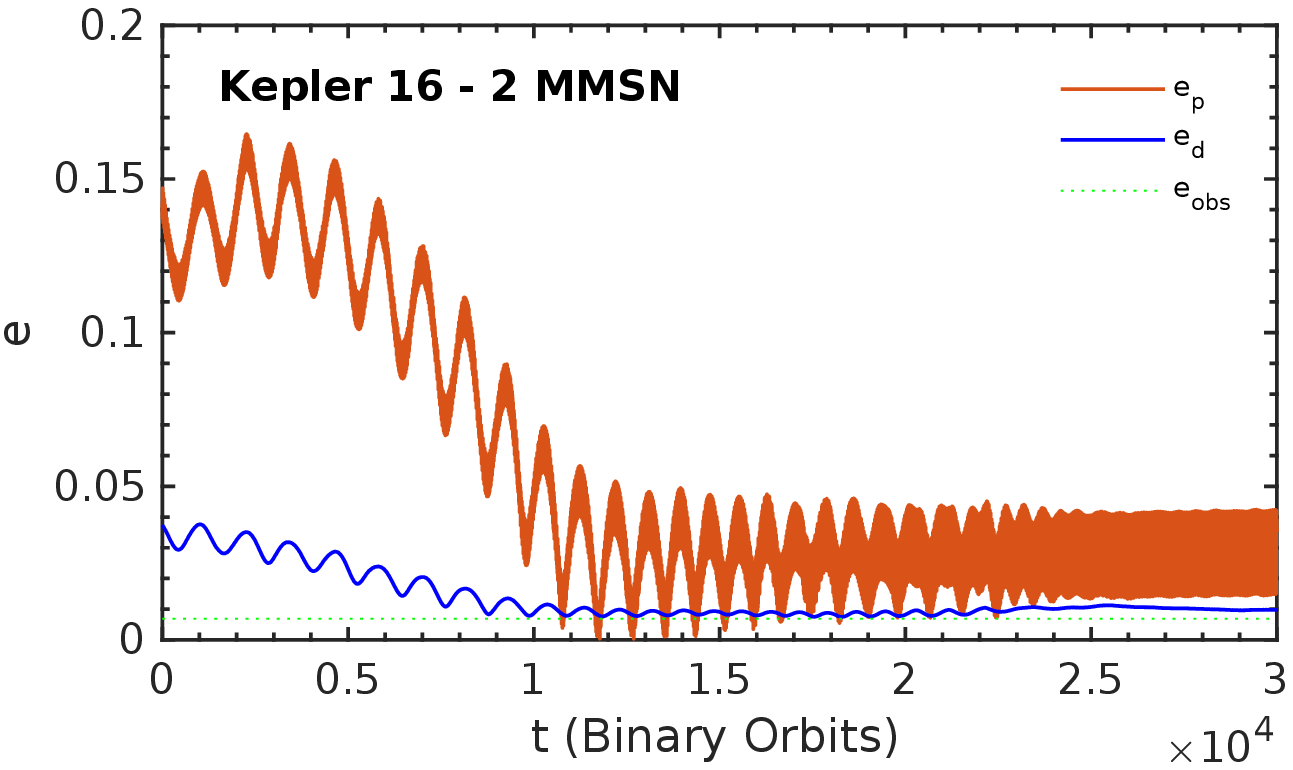} \\
\includegraphics[width = 0.47\textwidth]{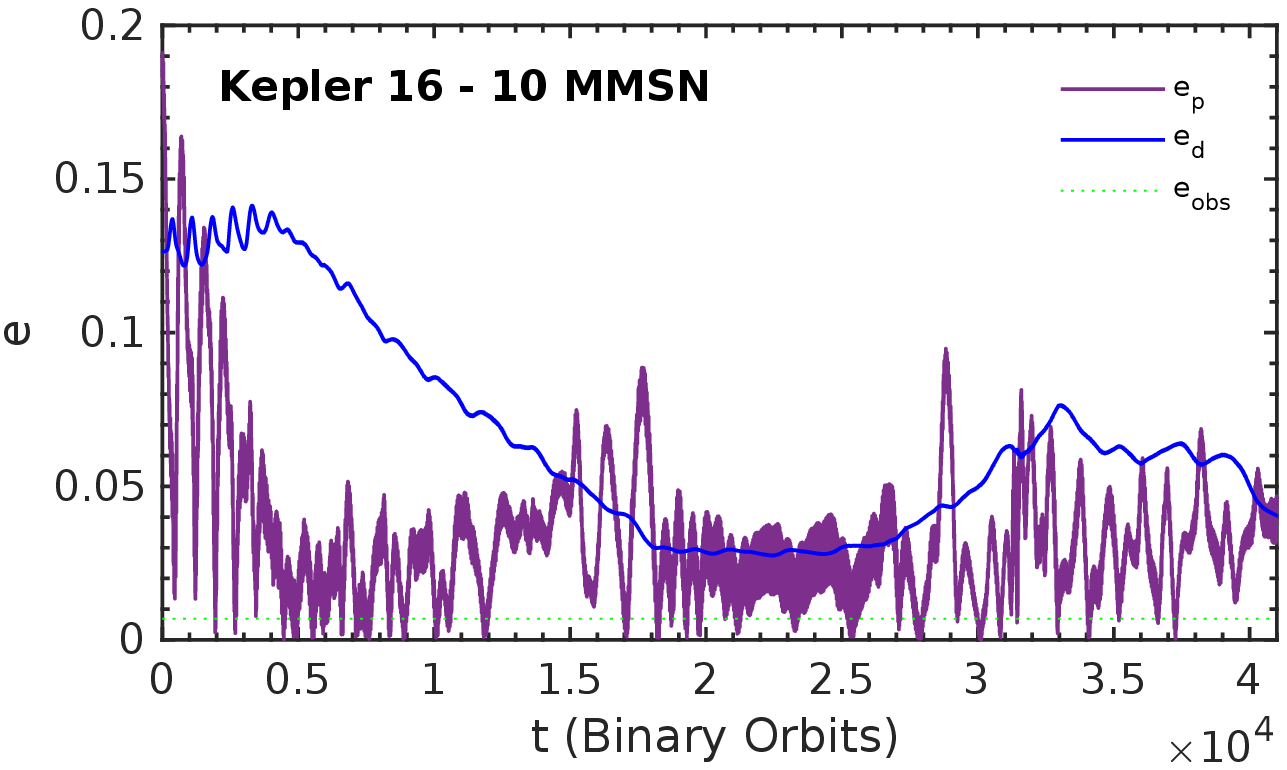}
\caption{Comparison of planetary and disc (global) eccentricity evolution once gas accretion has started (at $t=$\Pb{0}) in the 2 and \mmsn{10} models around the Kepler-16 system. The growth of the core to its observed mass in the first \Pb{5000} of the scenario results in significant alteration of the disc eccentricity profile, as well as the average disc eccentricity. As the planet grows and it approaches the gap-opening regime, a decrease in disc eccentricity can be seen -- in parallel to the decrease in planetary eccentricty. In the \mmsn{10} model the erratic changes in \arm{p} beyond \Pb{2.5 \times 10^4}s are also accompanied by a significant growth in \e{d}.}
\label{fig:acc_res_kep16_epdt2}
\end{figure}

Kepler-16b, with $m_{\rm p}\simeq 0.3\, M_\mathrm{Jup}$ is the most massive of the three circumbinary planets that we consider in this work \citep{Doyle2011}. Using the gap-opening criteria of \citet{Crida2006} which states that for a given set of disc parameters, a planet of mass ratio, $q$ will open a gap if:
\begin{equation}
1.1\left(\frac{q}{h^3}\right)^{-1/3} + \frac{50\alpha h^2}{q} \leq 1,
\label{eq:gap_opening}
\end{equation}
one can see that for the viscous stress parameter and disc aspect ratio used in these simulations, the core will significantly alter the surface density profile of the disc when it approaches its final mass -- as can be seen in Fig. \ref{fig:acc_res_kep16_sds}. In the low-mass discs the core slowly migrates from its initial stopping position at \arm{p} $=1.1$ au further inwards to \arm{p} $=0.75$ au, between the 6:1 and 7:1 MMRs with the binary (top panel of Fig. \ref{fig:migacc_res_kep16}). As the planets migrate into the cavity evacuated by the binary, they carve out this cavity further -- opening one side of a gap. This process destroys the eccentric cavity, as the planet's mass dominates -- resulting in a decrease in eccentricity \e{p} $=0.12\rightarrow0.03$. One anomaly in these results is the rapid outward migration of the core in the \mmsn{5} model at \Pb{4000} (Fig. \ref{fig:migacc_res_kep16}). This is accompanied by a sharp decrease in \e{p}. Examining the evolution of the planet at this epoch, it can be seen that as the planet's mass grows it appears to interact with the 8:1 MMR with the binary. These $n$:1 MMR locations have been shown to be unstable to planetary orbits because they excite the eccentricity \citep{Nelson2000}, and in this case the planet is scattered out. It is not ejected and is able to migrate back into the inner disc, avoiding further scattering events.

The planet in the \mmsn{10} disc alters the surface density similarly to the planets in the low-mass discs. It can be seen in the bottom panel of Fig. \ref{fig:acc_res_kep16_sds} that it doesn't open such a deep gap at the cavity edge. The presence of the planet leads to the destruction of the additional eccentric features in the outer disc. Any planet forming and evolving in the outer disc in a multi-planet formation and migration scenario (see \citep{Kley2015}) would have a very different migration pathway to the first planet. During the accretion phase of the simulation, slow inwards migration occurs to \arm{p} $=1.0$ au, whilst the eccentricity of the core's orbit falls. The next \Pb{10000} is spent at this distance, after which it migrates further into the inner disc, where it reaches \arm{p} $=0.75$ au. For the remainder of the simulation lifetime it has an eccentricity \e{p} $=0.04$, in relatively good agreement with Kepler-16b. However between $2.5 \times 10^4$ and \Pb{3\times10^4} the core seems to undergo a similar scattering event as that seen in the \mmsn{5} disc -- oscillating around the 7:1 MMR and consequently scattering out. This scattering and subsequent inwards migration seems to happen repeatedly over the course of the simulation. The mass of the disc is sufficient to maintain a significant eccentricity of its own, and excite that of the planet.

The evolution of the accreting core in the most massive \mmsn{20} disc is even more disruptive. Initially, when its mass starts to grow, in the first few \Pb{1000}s of the simulation, it escapes the outer planet trap and migrates into the inner disc (\arm{p} $=0.9$ au). During this phase it maintains a significant eccentricity (\e{p} $\approx 0.1$) -- because of this its orbit enters the $n$:1 MMR region. It spends the remainder of the disc undergoing repeated scattering and migration events, where its eccentricity dramatically rises to $\approx 0.3$ and then circularises. If this continues, the core could in principal enter the critical stability limit during one of these events and be ejected from the system -- although we have not yet seen this happen.

Figure \ref{fig:acc_res_kep16_sd2} we can see that the structure of the circumbinary disc has been significantly altered by the growth of the planet to its observed mass. The opening of the gap, as well as strong spiral wakes launched at the Lindblad resonances with the planet act to destroy the eccentric cavity, making it more circular. This is clear in the top panel of Fig. \ref{fig:acc_res_kep16_epdt2}; during the first \Pb{10000} of the simulation, when the planet is accreting mass from the disc and migrating slowly inwards, the eccentricity of the disc decreases to 0.01. In the most massive discs, the growing and migrating planet disrupts the eccentric features in the exterior disc as well as the inner cavity -- leading again to a decrease in \e{d} (bottom panel of Fig. \ref{fig:acc_res_kep16_epdt2}). The erratic changes in the orbit of the planets in the 10 and \mmsn{20} models also leads to corresponding fluctuations in the disc eccentricity. When the planet is on a wider, more eccentric orbit, the eccentricity of the disc can also grow.

In summary, we find that increasing the planet's mass in the low-mass discs leads to further inwards migration, and final orbital elements that are in rather good agreement with the observed values. Gas accretion in the high mass discs, however, leads to repeated interactions with the binary that cause the orbits of the planets to change erratically.

\subsection{Kepler-34}\label{sec:acc_res_34}
\begin{figure}
\centering
\subfloat[]{\includegraphics[width = 0.47\textwidth]{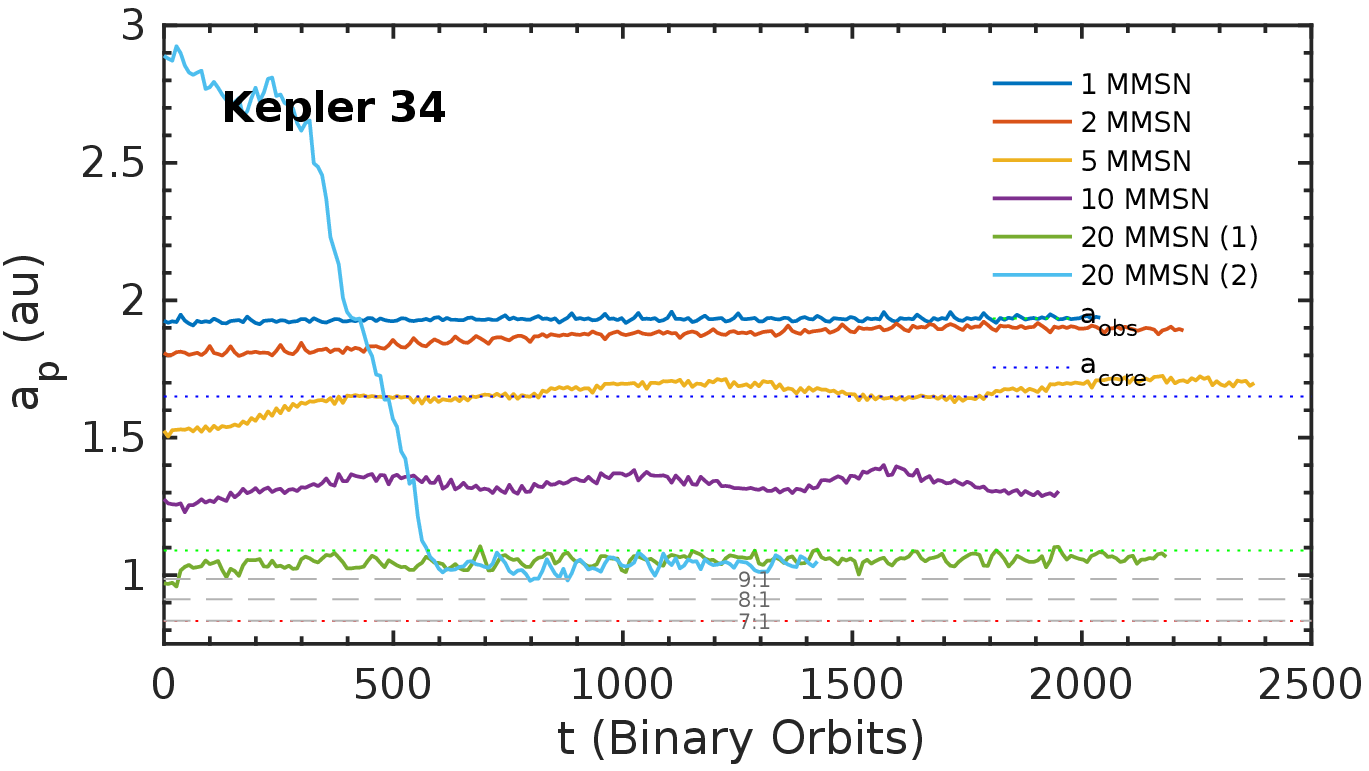}} \\
\vspace{-10pt}
\subfloat[]{\includegraphics[width = 0.47\textwidth]{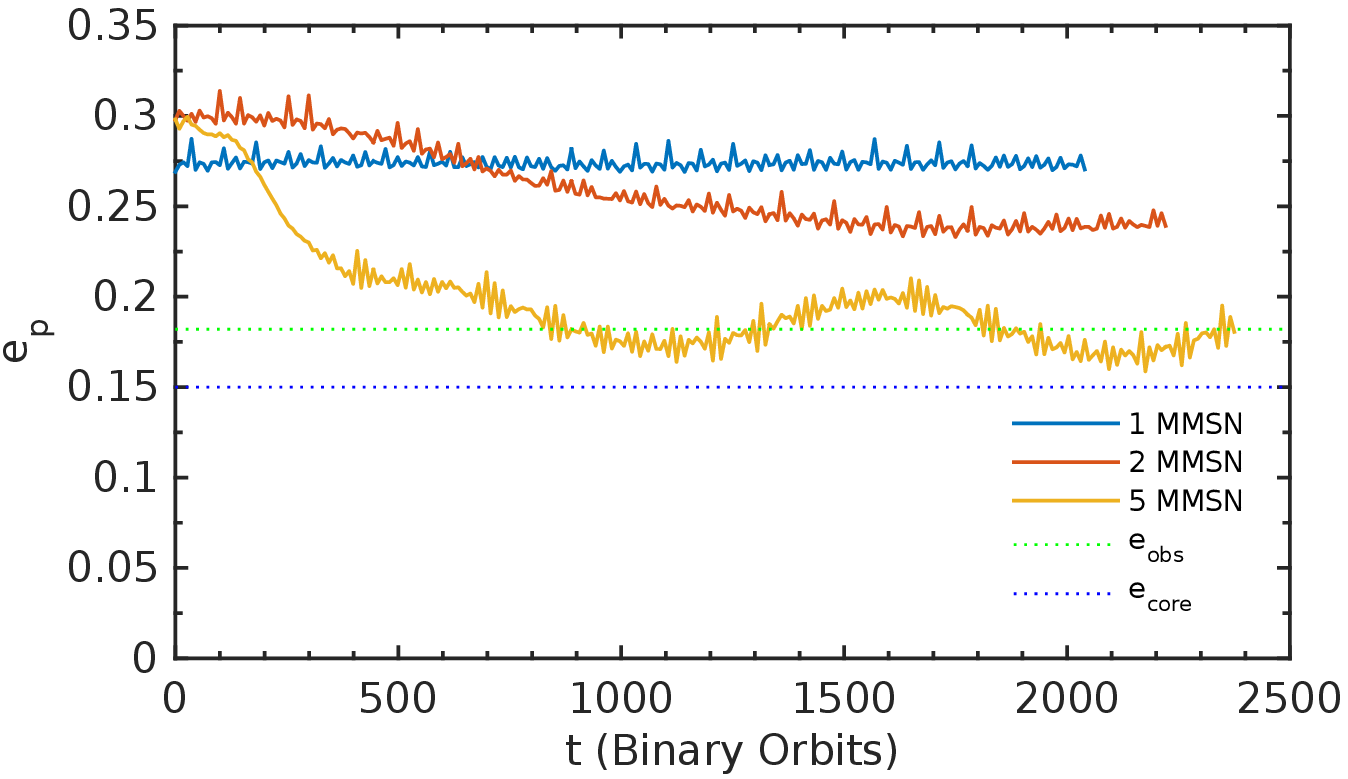}}\\
\vspace{-10pt}
\subfloat[]{\includegraphics[width = 0.47\textwidth]{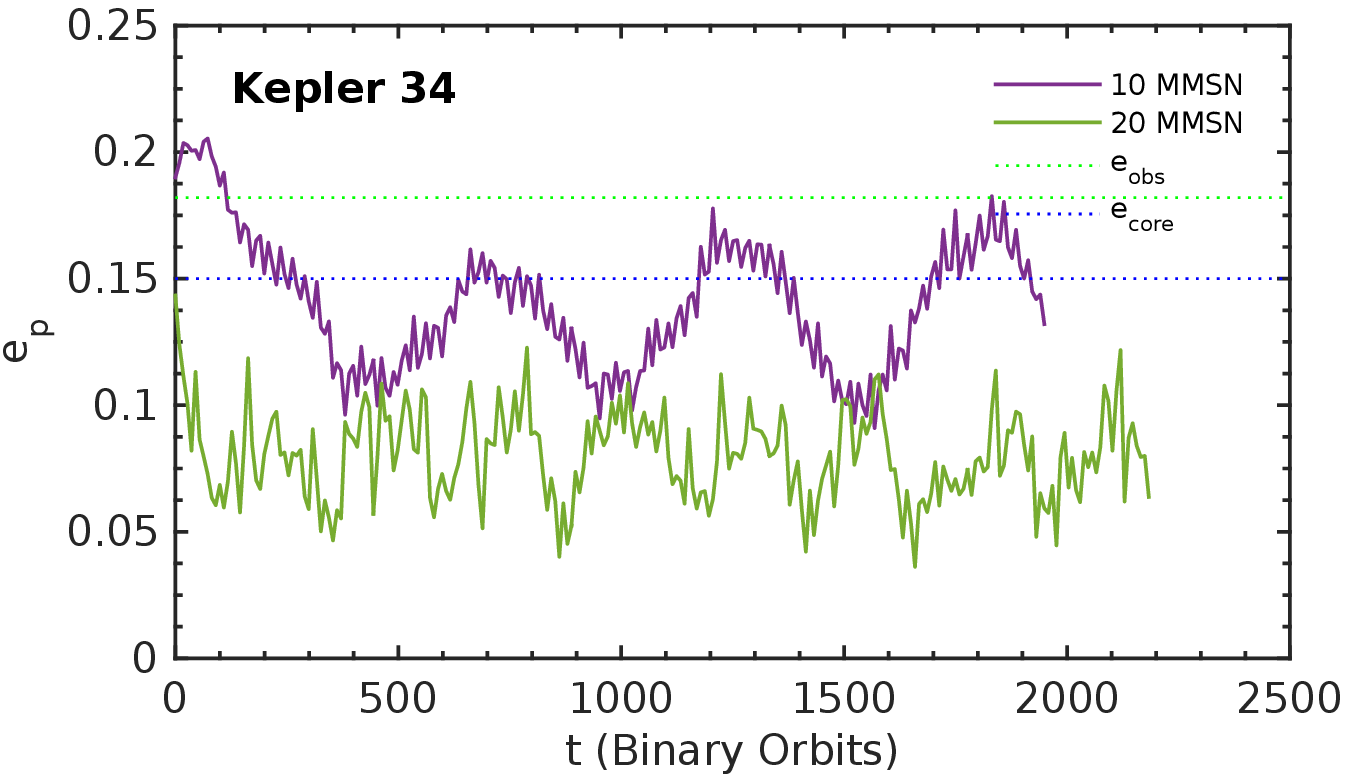}}
\caption{Evolution of accreting protoplanetary cores' semi-major axes in evolved self-gravitating discs in the Kepler-34 system. The middle and bottom panels shows these cores' eccentricity evolution in the low- and high-mass disc models respectively. The difference between the two cores in \mmsn{20} models is the initial starting position. It can be seen that a small increase in planet mass is sufficient to alter the surface density in the outer disc, so that the previously trapped core can migrate inwards.}
\label{fig:migacc_res_kep34}
\end{figure}

Increasing the protoplanet's mass from $q_{\mathrm{p},0}=6 \times 10^{-5}$ to the observed mass of Kepler-34, $q_\mathrm{p}=1\times 10^{-4}$ -- an increase a little over $60\%$ -- results in little change of orbital parameters. This is unsurprising as the core is still in the Type I planet migration regime in our disc models, and according to Eq. \ref{eq:gap_opening}, is not capable of sufficiently disturbing the surface density distribution to open a gap. This lack of significant activity was apparent after a relatively short simulation time ($\approx$ \Pb{2000}), where after a period of relaxation the system reaches a pseudo-steady state. In Fig. \ref{fig:migacc_res_kep34} one can see a slight outward migration of the planets in the \mmsn{2} and \mmsn{5} models, associated with a circularisation of the orbit. A lack of change in the low-mass models mean that there is still poor agreement with the observed configuration of Kepler-34b. The semi-major axes of the cores in this system are too large (\arm{p} $=1.7$--$2$ au), with eccentricities which are too excited (\e{p} $=0.25$--$0.275$) -- although good agreement with the observed eccentricity is obtained for the \mmsn{5} core model where \e{p} is oscillating around 0.18. The 10 and \mmsn{20} model cores also show little change when accretion is switched on, apart from a slight decrease in eccentricity in the \mmsn{10} case due to more efficient damping by the disc.

The second \mmsn{20} model run in the Kepler-34 system which is initially released further out in the disc but is trapped close to its starting position -- mentioned at the end of Section \ref{sec:mig_res_34} -- shows the most dramatic response to accreting mass. The increased core mass is sufficient for it to escape the region of weak eccentric features, created by the self-gravitating disc response to the binary potential, in the outer disc. It quickly migrates through the disc, finally reaching \arm{p} $=1.0$ au, the same as the first \mmsn{20} model presented and in good agreement with the observed value of \arm{p}, although both models have small values of \e{p}. In this system it is especially hard to produce a planet so close-in with a non-negligible eccentricity that matches the observations. Increasing \e{p}, hence lowering the pericentre distance, further increases the risk of destabilising encounters with the $n$:1 MMR region and the chance of a catastrophic ejection event. This, along with post-disc dissipation evolution with the binary, may be the reason why we may yet to observe a very close-in circumbinary planet (like Kepler-16b or -35b) with a significant eccentricity like that of Kepler-34b.

\section{Disc Dissipation Impact on Migrating Cores}\label{sec:dis_res}
From the beginning of this investigation we have been using the disc-mass as a proxy for the age of the circumbinary disc. It is logical to assume that when the disc first forms into a stable entity around the central binary it is at its most massive, and over the course of its lifetime loses mass due to a number of different processes -- accretion onto the central binary, loss from photo-evaporative and/or magnetised winds from the surface of the disc, etc. Whilst we have simulated the disc structure and evolution at different eras throughout its lifetime, we have not investigated the effects of \textit{transitioning} from a high-mass environment to that of a low-mass one. The dichotomy of results from Paper I suggest that the additional eccentric features seen in the outer disc will disperse as the disc-mass and the strength of self-gravity decrease. Without a sustaining action, the viscous forces in the disc will dissipate these eccentric features. As the strength of self-gravity diminishes in the disc we would also expect the compactness of the system to relax back to that seen in the least-massive \mmsn{1} disc -- the eccentric cavity will increase in size, especially those seen in the Kepler-34 system. The surface density profile will alter as the disc relaxes and we would expect the planet to migrate outwards with the cavity. In those discs where the planets are halted by the counteracting of the Lindblad torque by the positive co-rotation torque, the planet may be able to stay at this stable stopping location whilst the disc relaxes. We are not investigating the mechanisms and physics which dictate disc mass-loss and dispersal -- these are topics of ongoing research -- and perhaps deserve their own work in the context of massive self-gravitating discs. Instead, and as a computation time saving exercise, we use a simple exponential decay to dissipate the mass of the disc:
\begin{equation}
	\Sigma^{n+1}_{ij} = \Sigma^{n}_{ij} \exp{\left(-\frac{\Delta t}{\tau}\right)},
	\label{eq:disc_decay}
\end{equation}
where $\Sigma_{ij}$ is the cell surface density value, $\Delta t=t^{n+1} - t^{n}$, is the time between successive time levels $n$ and $n+1$, and $\tau$ is the decay time constant. In each disc this value is chosen so that after \Pb{5000} the total disc mass $m_\mathrm{d}$ will have decreased from its initial value down to the equivalent \mmsn{1} model in that system. For reference, to reach a \mmsn{1} disc from a \mmsn{10} or \mmsn{20} mass disc in the Kepler-16 system, a time constant of $\tau=1520$ or 1175 is used respectively. This decay length is sufficiently large that the dynamical time-scales associated with the disc and planet are much smaller, and can therefore respond to any changes in the disc. Once the disc has reached a total disc-mass equivalent to the initial \mmsn{1} disc mass, the dissipation mechanism is stopped, to allow the disc and planet to reach a pseudo-steady-state on time-scales of a few $\times$ 10,000 binary orbits. This procedure is started in the disc once -- similarly to the previous subsection looking into accretion scenarios -- the initial binary-disc-protoplanet systems from Section \ref{sec:mig_res} have reached quasi-steady state. This allows us to track the response of planets, trapped at the cavity edge or by eccentric rings, to the diminishing disc mass and relaxation or dissipation of eccentric features. During this procedure, and in the post-dissipation evolution of the planets, we consider non-accreting cores.

\subsection{Kepler-16}\label{sec:dis_res_16}
\begin{figure}
\centering
\subfloat[]{\includegraphics[width = 0.47\textwidth]{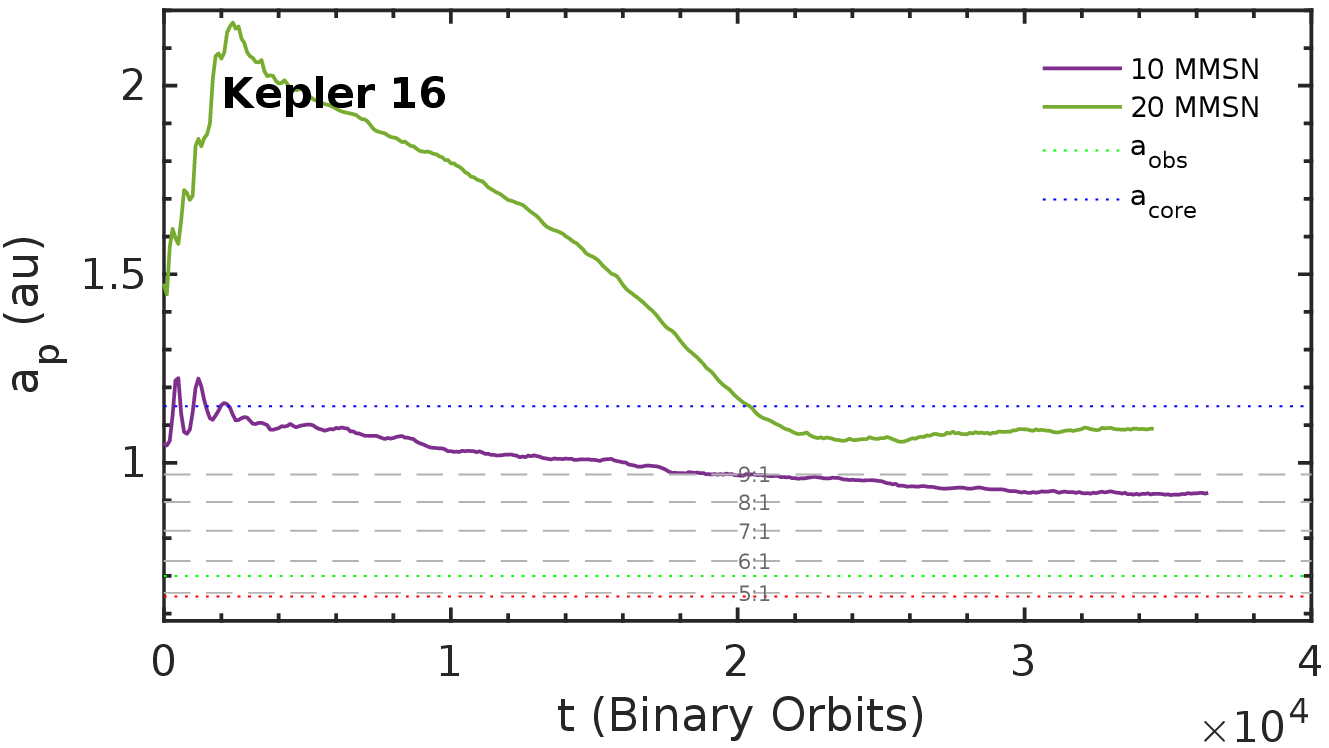}} \\
\vspace{-10pt}
\subfloat[]{\includegraphics[width = 0.47\textwidth]{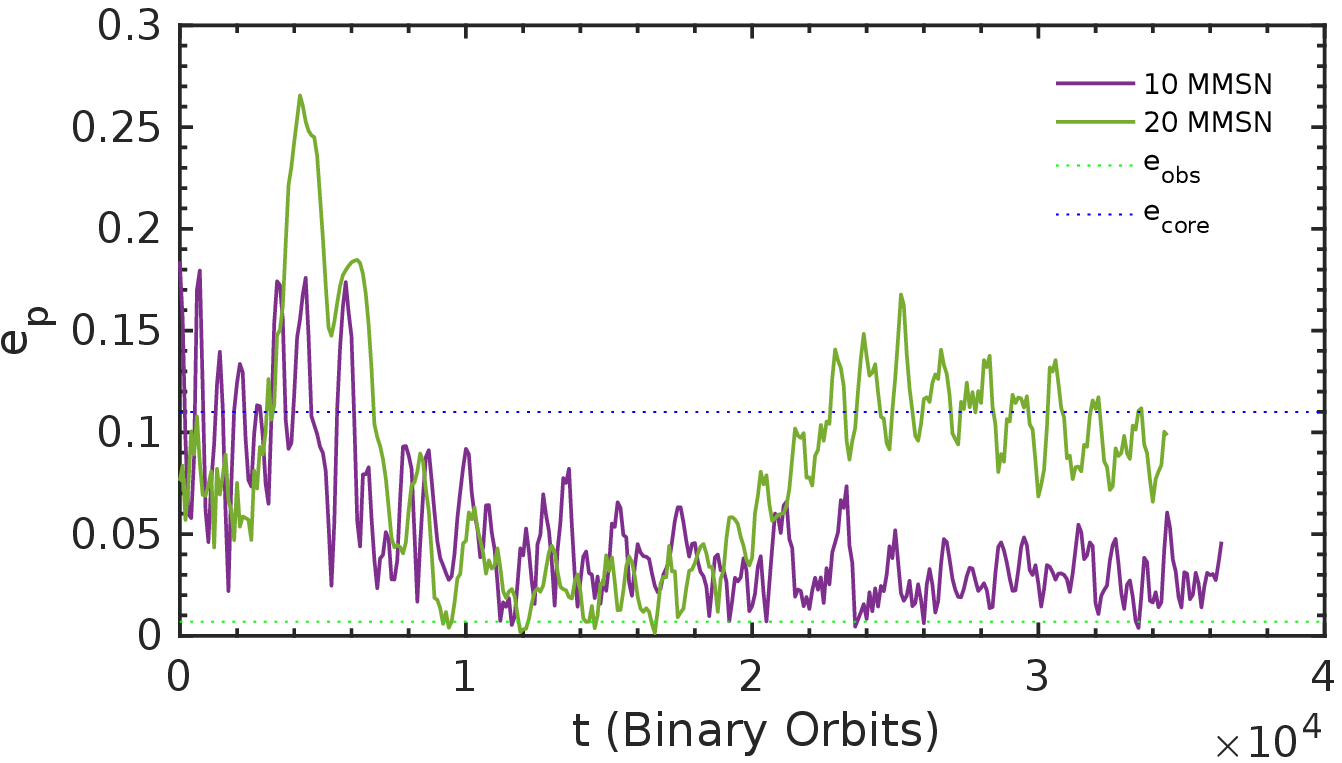}}
\caption{Evolution of protoplanetary cores' semi-major axes in dissipating self-gravitating discs, in the Kepler-16 system (top panel). The bottom panel shows the core eccentricity evolution in the 10MMSN and 20MMSN disc models.}
\label{fig:migdis_res_kep16}
\end{figure}
\begin{figure}
\centering
\includegraphics[width = 0.45\textwidth]{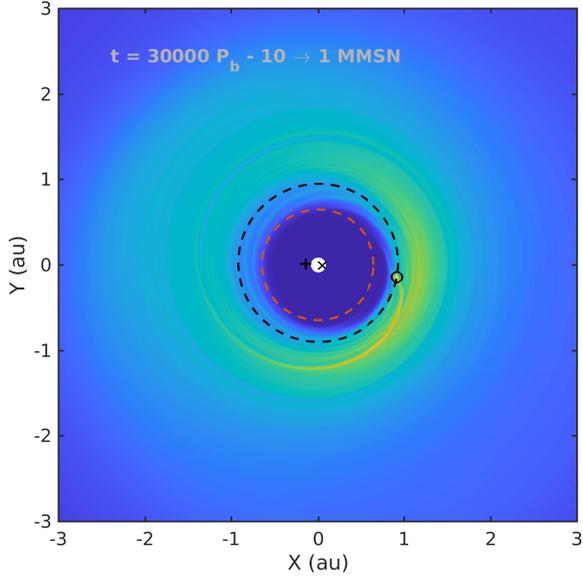}
\caption{Surface density map of the Kepler-16 \mmsn{10} disc model once the disc has undergone mass dissipation, and the core has reached its final orbital position. Whilst the eccentric cavity has relaxed from the very compact initial \mmsn{10} state to a cavity with a larger extent, the core has managed to retain a close-in, circular orbit.}
\label{fig:dis_res_kep16_sd10}
\end{figure}
\begin{figure}
\centering
\hspace{-5pt}
\subfloat[]{\includegraphics[width = 0.472\textwidth]{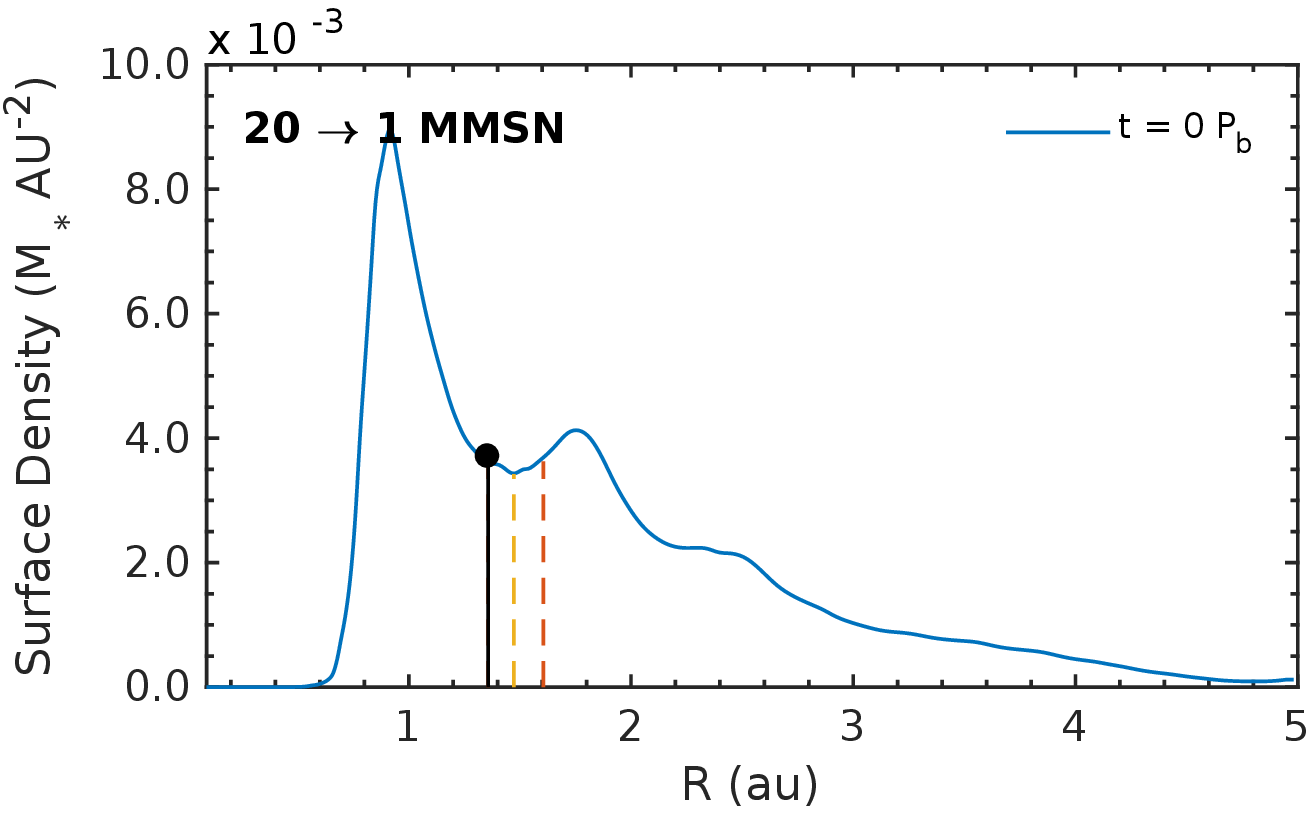}} \\
\vspace{-22pt}
\subfloat[]{\includegraphics[width = 0.47\textwidth]{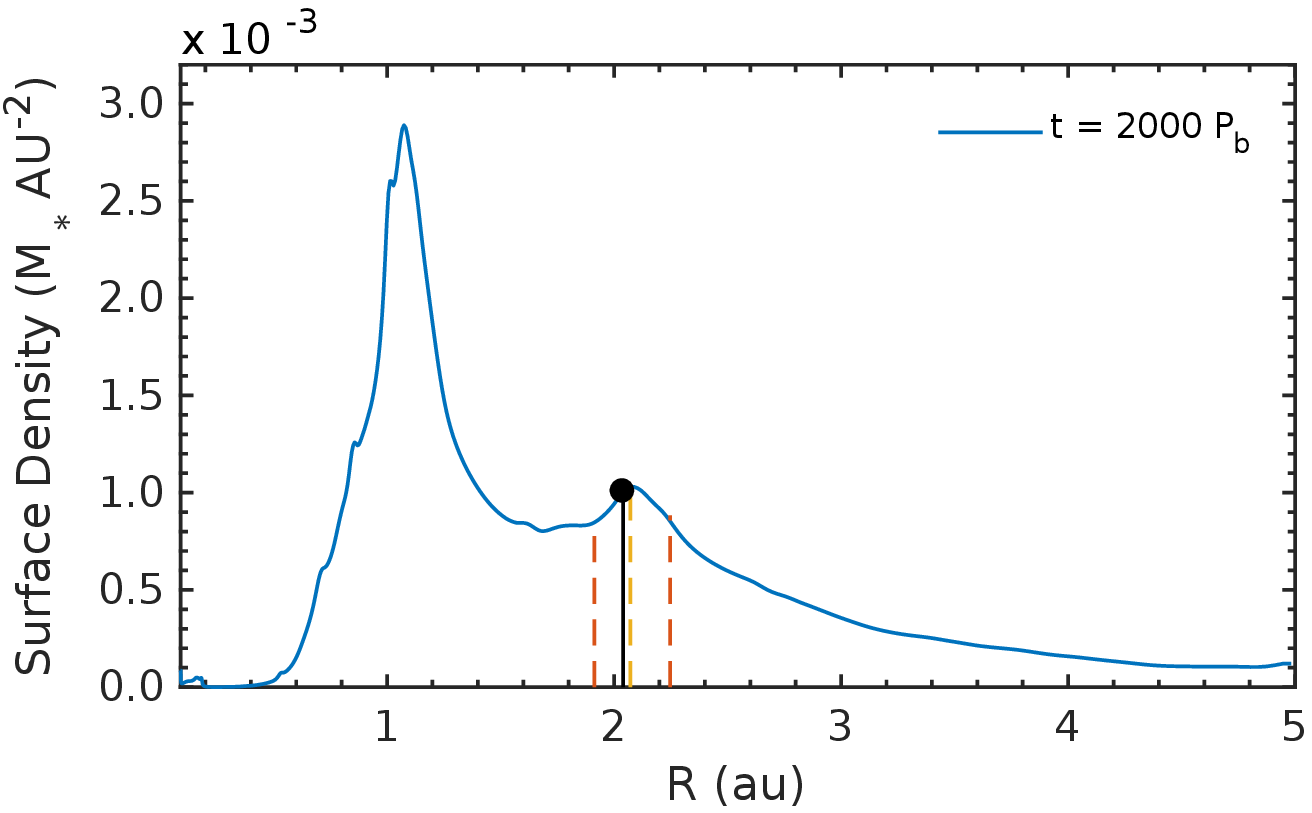}} \\
\vspace{-22pt}
\subfloat[]{\includegraphics[width = 0.47\textwidth]{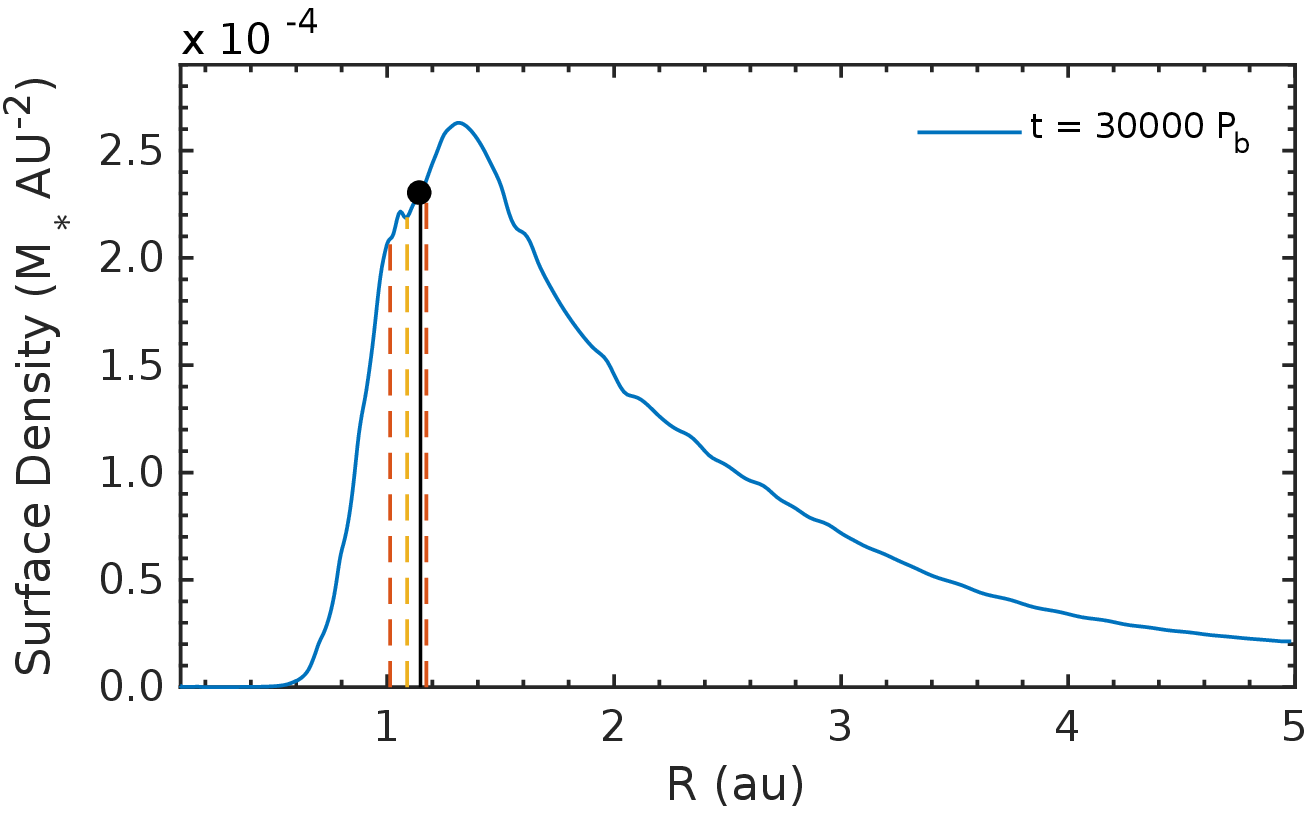}}
\caption{1D surface density profiles are shown here during the evolution of the \mmsn{20} disc, and its embedded core. All line types and colours carry the same meaning as previous plots of this type. The first panel shows the system at the beginning of disc dissipation, the second is \Pb{2000} orbits through this process, and the final subplot shows the system once the disc has relaxed and the planet has reached its final stopping position. The middle panel's surface density distribution looks very similar to those seen from \mmsn{10}, whilst the last panel looks very similar to a \mmsn{1} model. The planet initially migrates outwards as it follows the dissipating eccentric feature. Once this feature has completely dissipated, and the surface density gradient in the outer disc is negative once more, the planet then migrates into the inner disc. It finally stops at the edge of the inner cavity, \arm{p} $=1.1$ au, in good agreement with the results from core migration in the 1--\mmsn{5} models presented earlier.}
\label{fig:dis_res_kep16_sds20}
\end{figure}
The similarity of the results from the 1--\mmsn{5} models seen in Section \ref{sec:mig_res_16} for the Kepler-16 system prompted us to only carry out disc dispersal simulations for the most massive \mmsn{10} and \mmsn{20} models. Figure \ref{fig:migdis_res_kep16} shows the response of the planets' semi-major axes (top panel) and eccentricities (bottom panel) to the disc dispersal, which occurs during the first \Pb{5000} of these plots. A significant amount of post-dissipation evolution of the cores can be seen, especially in the $20\rightarrow$\mmsn{1} model.

As previously detailed, the core in the \mmsn{10} disc is orbiting at the edge of the inner eccentric cavity when migration halts. When dissipation starts to occur, the semi-major axis of the core increases from 1 to 1.2 au -- the position of $R_\mathrm{max}$, or the cavity edge in the least massive \mmsn{1} model. The eccentricity of the core also increases in this period, increasing from around 0.05 to 0.1. This increase in eccentricity, due to reduced damping by the disc, increases the positive torque contribution from the outer disc, even as it relaxes due to dissipation. The balance between reduced eccentricity damping and the diminishing influence of the positive torque from the outer disc, dictates whether the core migrates inwards or outwards as the disc dissipates. After this initial period of outwards migration, this balance inverts. The semi-major axis decreases -- past the initial stopping distance -- further into the inner disc as the eccentricity drops -- reaching a final orbit with \arm{p} $=0.9$ au and \e{p} $\approx 0.025$. This model gives better final agreement with the observed Kepler-16 system than the low-mass models. When dissipation starts, the core's small \arm{p} and non-negligible \e{p} mean its pericentre distance lies around 0.8 au. The core retains this small value during dissipation, and as the eccentricity is damped by the disc. The final semi-major axis corresponds to a location between the 8:1 and 9:1 MMR with the binary -- the core's low eccentricity however keeps it clear of interaction with these destabilising regions. Examining the surface density profile and planetary orbit in Fig. \ref{fig:dis_res_kep16_sd10}, one can see the similarity to the low-mass Kepler-16 discs (Fig. \ref{fig:mig_res_kep16_1sd}). The eccentric features in the outer disc have dissipated and the inner eccentric cavity has relaxed, to a size in good agreement with the 1--\mmsn{5} models. The difference in the shape of the planetary orbit is also clear, a more circular orbit inside, rather than tracing the outside edge of the cavity, is attained.

A very different evolution is seen in the \mmsn{20} model. The core in this model is trapped at the location of the first eccentric feature in the outer disc when the process of dissipation begins. During dissipation the planet migrates outwards from \arm{p} $=1.4$ to 2.2 au. After the first \Pb{2000} of disc dissipation, migration reverses and the core migrates into the inner disc, reaching a final semi-major axis of 1.1 au. Examining the azimuthally averaged surface density profiles in Fig. \ref{fig:dis_res_kep16_sds20} this evolutionary history can be explained. Comparing the three plots several important things can be extracted; the first being that between the first two panels the surface density profile has relaxed to one resembling a \mmsn{10} profile. The tightly wound eccentric features in the \mmsn{20} model dissipate outwards in the disc, but one relatively strong eccentric feature at 2.1 au can still be seen. If dissipation stopped here, we might see a migration scenario much like the \mmsn{10} model from Section \ref{sec:mig_res_16}, where the planet gets trapped, but then subsequently escapes. Dissipation does continue however, and the core is free to immediately migrate into the inner disc, as the eccentric features are destroyed. The last panel shows the planet at its final location of \arm{p} $=1.1$ au. This value and the surface density profile are very similar to the final state of the \mmsn{1} model from Section \ref{sec:mig_res_16}, and its final \e{p} $\approx0.075$ is in good agreement with observations. To summarise, the core -- still trapped by the eccentric feature -- migrates outwards as it follows the dissipating perturbation, until the eccentricity of the core dimishes enough so that the positive torque contribution from the outer disc stops. The net negative torque migrates the core inwards towards the central cavity, where the eccentricity increases again, inducing another torque reversal, halting migration in the inner disc.

\subsection{Kepler-34}\label{sec:dis_res_34}
\begin{figure}
\centering
\subfloat[]{\includegraphics[width = 0.47\textwidth]{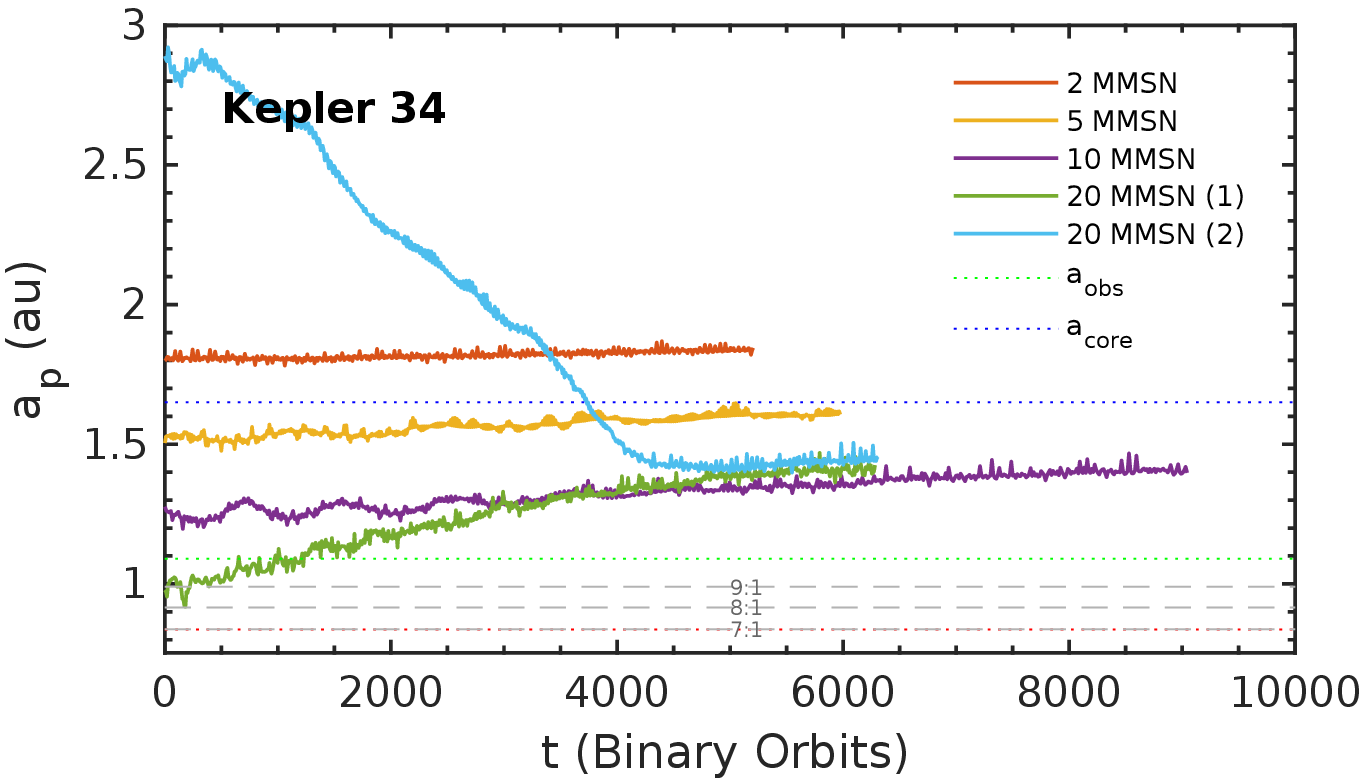}} \\
\vspace{-10pt}
\subfloat[]{\includegraphics[width = 0.47\textwidth]{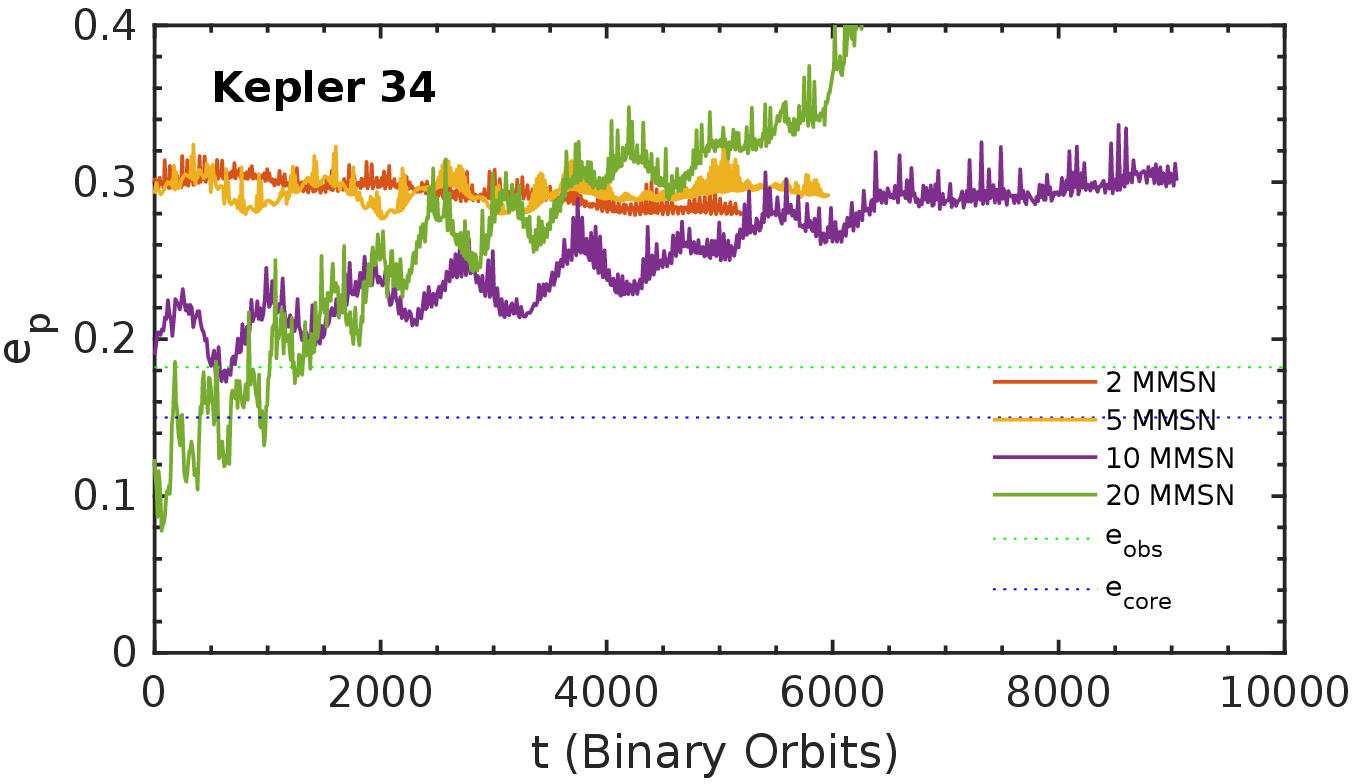}}
\caption{Evolution of protoplanetary cores' semi-major axes in dissipating self-gravitating discs in the Kepler-34 system. The middle and bottom panels shows these cores' eccentricity evolution in the low- and high-mass disc models respectively.}
\label{fig:migdis_res_kep34}
\end{figure}

\begin{figure}
\hspace{-15pt}
\subfloat{\includegraphics[width = 0.25\textwidth]{./figures/surf_p/34/SG1p_sd600.eps}}
\hspace{-2pt}
\subfloat{\includegraphics[width = 0.25\textwidth]{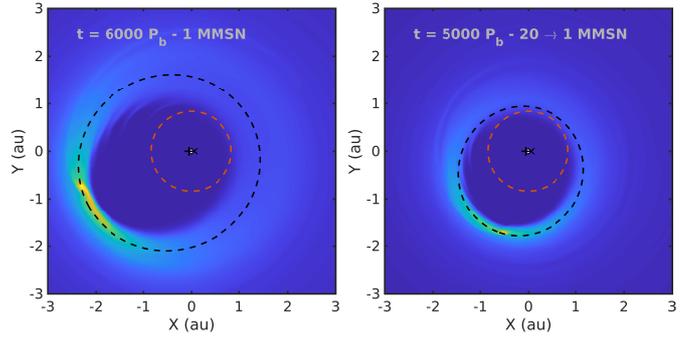}}
\caption{Surface density maps of the Kepler-34 \mmsn{1} system once the planet has reached its final orbit (left), and the final state of the 20$\rightarrow$\mmsn{1} dissipation model (right), once the disc has relaxed and the planet has migrated outwards to the orbit seen here. Despite the mass being equal in these discs, the cavity in the dissipated case is still much smaller and less eccentric.}
\label{fig:dis_res_kep34_sds}
\end{figure}
In contrast to Kepler-16, in the Kepler-34 system we undertook disc dissipation in the 2--\mmsn{20} models, reducing their disc mass to \mmsn{1}. With these models, we would expect the core semi-major axes and eccentricities to converge on the values reached by the core in the \mmsn{1} model, as the disc mass dissipates. Examining the top panel of Fig. \ref{fig:migdis_res_kep34}, the evolution of \arm{p}, we see this is not the case. Whilst there is evidence of slight outwards migration as a result of the disc relaxing, they do not migrate significantly to \arm{p} $\approx 2$ au -- the location of the \mmsn{1} model core. The \mmsn{2} core shows little change, \mmsn{5} migrates outwards to 1.6 au where it halts, and the high-mass discs all converge to 1.4 au. As the eccentric features in the outer disc dissipate, the core in the second \mmsn{20} model is able to escape the outer disc and migrate into the inner disc, where it halts at 1.4 au, close to the stopping radius of the other \mmsn{20} run with reducing disc mass, and the corresponding \mmsn{10} case. Comparing the surface density maps in Fig. \ref{fig:dis_res_kep34_sds}, the lack of agreement between the basic \mmsn{1} migration scenario and the disc-dissipated 20$\rightarrow$\mmsn{1} models is clear. Whilst there is some evidence of the disc relaxing during its dissipation, the cavity in the latter model is still more tightly bound around the central binary, as a result the planet is in a much closer orbit than expected. It appears that the presence of the planet in the inner cavity interferes with the relaxation of the disc and prevents it from relaxing to the configuration expected from the \mmsn{1} run. It is for this reason that we achieve a smaller stopping radius for the planets when the disc mass transitions from high to low mass, and indicates that the history of the system influences the final stopping location of the planet.

\subsection{Kepler-35}\label{sec:dis_res_35}
\begin{figure}
\centering
\subfloat[]{\includegraphics[width = 0.47\textwidth]{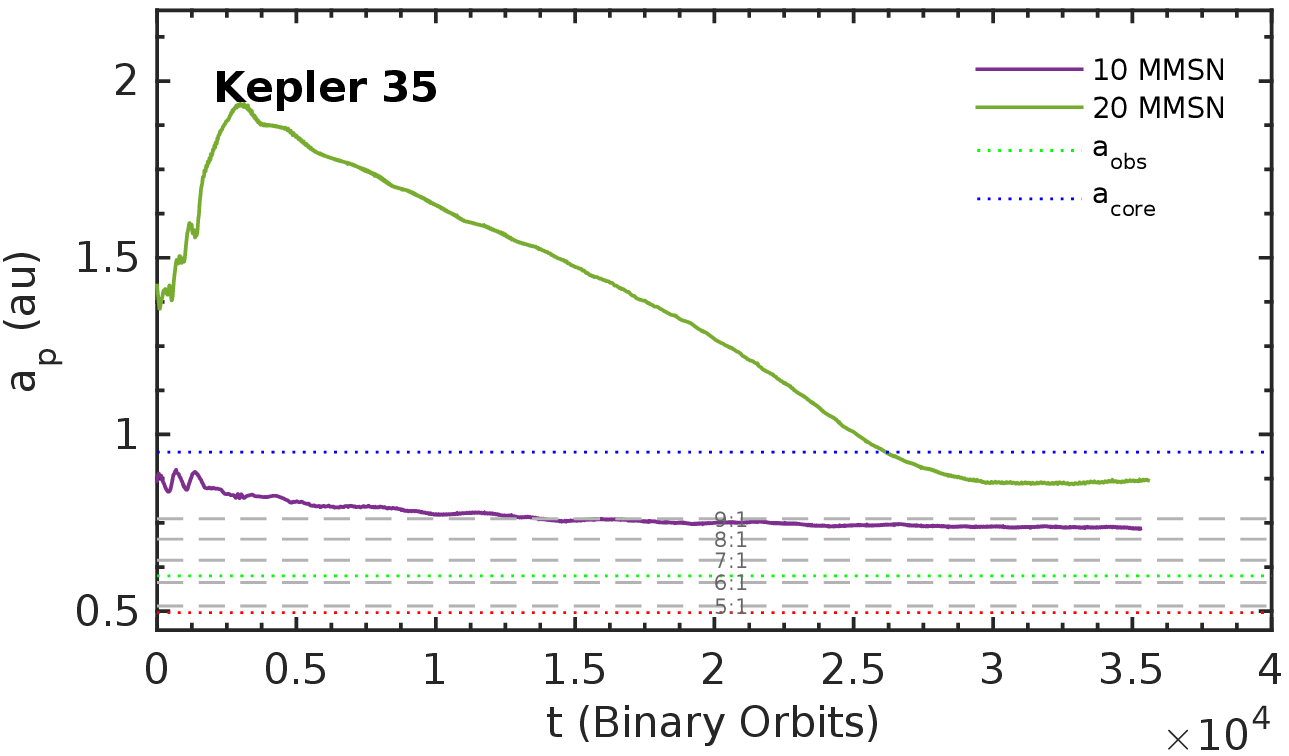}} \\
\subfloat[]{\includegraphics[width = 0.47\textwidth]{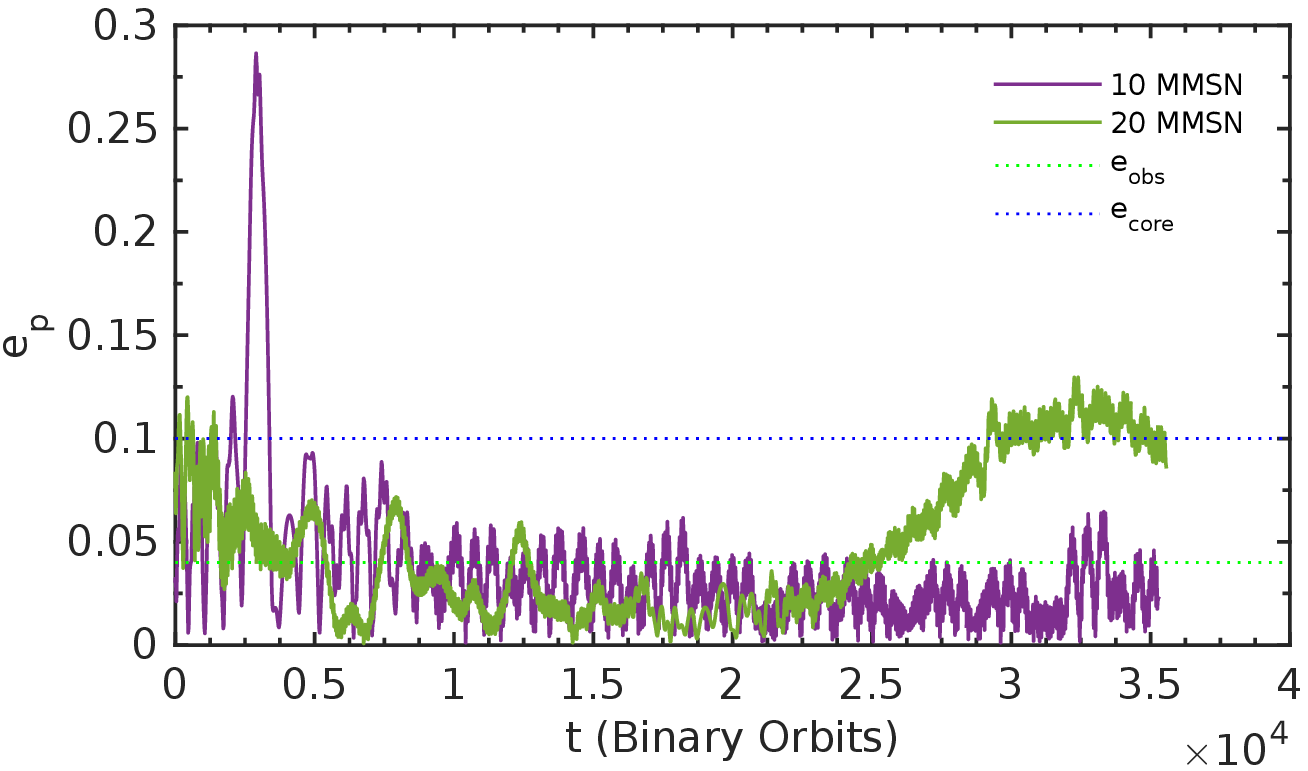}}\\
\caption{Evolution of protoplanetary cores' semi-major axes in dissipating self-gravitating discs in the Kepler-35 system. The bottom panels show these cores' eccentricity evolution in the high-mass disc models.}
\label{fig:migdis_res_kep35}
\end{figure}
Similar results to Kepler-16 in Section \ref{sec:mig_res_16} prompted a similar approach for running disc dissipation scenarios in the Kepler-35 systems; ignoring the low-mass models which show consistent results and focusing on the high-mass models which show the most variation, both with each other and the low-mass cases. The cores in the \mmsn{10} and \mmsn{20} models start in much the same positions as those in the Kepler-16 models; the \mmsn{10} core on a close-in orbit (\arm{p} $=0.9$ au) with a low eccentricity, and the \mmsn{20} core trapped in the outer disc by the first additional eccentric feature. Examining Fig. \ref{fig:migdis_res_kep35} the similarity continues -- the cores follow the same migration pathway as their counterpart cores in the Section \ref{sec:dis_res_16} simulations. The replication of the same evolutionary scenarios in different mass-ratio binary systems, suggest that the mechanisms observed in the above sections are relatively robust. For low-mass, isothermal discs, with the same structure, the zero torque location should be the same -- hence planets in these discs halt migration at the same location. The core in the dissipating \mmsn{20} model reaches the same semi-major axis as the core in the \mmsn{1} model, in our first set of simulations, because when it reaches the inner disc the structure is the same because the disc has already relaxed. On the other hand, the core in \mmsn{10} has already reached the inner disc, with a different disc structure. As the disc dissipates, the planet also has an impact on the final disc structure, which shifts the zero torque location -- inwards in this case. The tendency for the cores in the \mmsn{10} models to converge onto shorter period orbits between the 8:1 and 9:1 MMRs, resulting in better agreement with the observations than the other models when the planet mass is kept constant, is likely a fingerprint of the high-mass disc structure at \mmsn{10}, which when it disperses allows the planets to achieve shorter period orbits with low eccentricities.

\section{Summary and Discussion}\label{sec:sumdis}
This is the second paper in a series that examines the influence of disc self-gravity on the evolution of gaseous circumbinary discs, and on the evolution of planets that are embedded in those discs. The focus of Paper I was on the evolution of the discs alone. Several disc masses, ranging between 1--\mmsn{20} equivalent discs, were used to probe the evolution of disc structure throughout the lifetime of a circumbinary disc under the influence of self-gravity. The main results to emerge from this study were that self-gravity leads to two important effects: i) the size of the tidally truncated, eccentric inner disc cavity that forms tends to be smaller for larger disc masses as self-gravity compacts the system scale; ii) additional precessing eccentric modes emerge at large orbital radii in discs where self-gravity is important. 

In the present paper we use the end-points of the simulations from Paper I as initial conditions for simulations that examine the orbital evolution of embedded planets, with the binary parameters having been chosen to correspond to the Kepler-16, -34 and -35 planet-hosting systems. Most of the simulations that we present assume that the planet-binary mass ratio is fixed at $q=6 \times 10^{-5}$. The aim of this work is to examine whether or not self-gravity can improve the level of agreement between the migration stopping locations of planets in the simulations and their currently observed orbital radii. In addition to examining the influence of disc mass, we also examined how the results changed when allowing planets to accrete gas so that they reached their observed masses while migrating, and the influence of allowing the disc mass to decrease with time such that high-mass discs transition to become low-mass discs after the planets have migrated to the their stopping radii. We summarise and discuss the results for the different binary systems below.

\subsection{Kepler-16}
We found that the cavity size in this case only changes significantly when the disc mass exceeds \mmsn{10}. The migration of planetary cores of fixed mass in the lower mass discs resulted in them stopping close to the edge of the cavity, but with semi-major axes and eccentricities that were too large compared to the observations (\arm{p} $\sim1.15$~au and \e{p} $\sim 0.11$ versus observed values of 0.705~au and 0.007, respectively). The stopping location, however, was found to agree well with our previous work presented in PN13 that adopted different boundary conditions. Migration in the \mmsn{20} disc resulted in the planet being halted by one of the additional eccentric features further out in the disc that acted as a planet trap, so in spite of the disc cavity being significantly smaller in this case, the planet was unable to reach the cavity such that it could park closer to the central binary as required by the observations.

Allowing the planets to accrete gas so that they reach the mass inferred from observations (this requires the mass ratio to grow from $6 \times 10^{-5}$ to $3.54\times 10^{-4}$) resulted in much better agreement with observations for the low mass discs. Here, the planet grows in excess of the gap forming mass, and this allows it to push deeper into the tidally truncated cavity. Furthermore, the growth of the planet causes the eccentricity of the central cavity (and the other eccentric features) to diminish significantly, and this leads to the eccentricity of the planet orbits reducing significantly. For the lowest mass disc we obtain \arm{p}=0.78~au and \e{p}=0.03, which agrees rather well with the observed values for Kepler-16b, giving us confidence that the formation and evolution scenario that we are exploring in this work is probably the correct one. The evolution of the accreting planets in the higher-mass \mmsn{10} and \mmsn{20} discs did not result in such good agreement with observations. Here, the planets have their eccentricities excited by the eccentric disc modes to values that cause them to interact more strongly with the central binary, leading to a sequence of scattering events that send them out into the disc and then back again over the full run times of the simulations.

Finally, allowing the disc mass to decrease for the heavy \mmsn{10} and \mmsn{20} discs, while keeping the planet-binary mass ratio $=6 \times 10^{-5}$ caused the planets to end up orbiting closer to the star than when the disc masses were at their initial values. In particular, the reduction in disc mass causes the additional eccentric features in the disc to dissipate, and this allows the planet in the \mmsn{20} case to migrate inwards. The level of agreement with observations in these cases, however, is not as good as that obtained by allowing the planet masses to increase to their observed values. We conclude that for the Kepler-16 system, self-gravity of the disc does not provide a positive contribution to obtaining agreement between observations and theoretical predictions. 

\subsection{Kepler-34}
The binary system in Kepler-34 has an eccentric orbit, and this leads to the formation of a wide and highly eccentric cavity when the disc mass is low. For large disc masses, however, self-gravity causes the disc cavity to shrink substantially, and this has a strong influence on the orbital evolution of embedded planets.

In the low mass discs the planets migrate inwards and stop at the edge of the cavity, which is too far from the binary for the stopping location to agree with the observations (\arm{p}$\sim 1.8$~au and \e{p}$\sim 0.28$ versus the observed values 1.09~au and 0.182, respectively). In the high mass cases, however, we find that the additional eccentric features that form in the disc are somewhat weaker than in the Kepler-16 run described above, and consequently the planets can normally migrate all the way to the central cavity in these cases. We find that the semi-major axes and eccentricities of the planets in the \mmsn{10} and \mmsn{20} discs straddle the observed values for Kepler-34b, indicating that self-gravity in this case provides the possibility of obtaining much better agreement with the observations.

Switching on gas accretion makes very little difference to the results of these simulations because the final mass of Kepler-34b is only 60\% larger than the initial mass that we start with. Allowing the disc mass to decrease inevitably leads to the cavity sizes of the most massive discs increasing as the influence of self-gravity is diminished. Interestingly, however, we find that the presence of the planet prevents the cavity from relaxing to the size expected for a lower mass disc, and instead the amount of expansion observed is relatively modest. (The time to establish the approximately steady state cavity configuration, in the absence of planets, is typically $\sim 3500$ binary orbits. We have run our simulations for longer than this to ensure that we have achieved a quasi-steady state.) Although the planets in these more massive discs no longer show such good agreement with observations once the disc mass has diminished, they provide much better agreement than those planets that form and migrate in low mass discs. This leads us to conclude that self-gravity can have a positive impact on obtaining agreement between simulations and the observations of Kepler-34b because of the rather dramatic influence that it has on the cavity size, and also because the system retains memory of its larger initial disc mass when the mass of the disc is slowly decreased. Formation of a planet in a heavy disc, followed by its migration and then rapid disc removal would seem to provide one way in which agreement with observations could be obtained for this system.
 
\subsection{Kepler-35} 
As mentioned previously in this paper, the similar eccentricity of the Kepler-16 and -35 binaries leads to very similar outcomes both in terms of disc structure and orbital evolution. Mass growth of Kepler-35b from the initial planet-binary mass ratio of $6 \times 10^{-5}$ was not considered in this paper because the final mass is only 20\% larger than the initial mass. One consequence of this lower planet mass is that growth to a gap forming object that can push further into the inner cavity is difficult to invoke so that good agreement between the simulation outcomes and the observations of Kepler-35b can be obtained, in contrast to the situation with Kepler-16b. The final orbital radii of Kepler-35b analogues were always too large by a factor of 1.5 compared to the observed values. Allowing the planet to be in the partial gap forming regime, such that it might push deeper into the cavity, can probably only be achieved by a significant reduction in the disc pressure scale height. Given that the two stars in Kepler-35 are more massive and hotter than in the Kepler-16, it is not immediately obvious why the disc should be cooler in this case. Fitting this system using simulations therefore remains an unsolved problem, and will require a more sophisticated treatment of the disc thermodynamics to examine whether or not Kepler-35b could have been in the gap forming regime when the protoplanetary disc was present.

Assuming that the scenario we have explored in this paper, namely that planets form at large orbital radii in circumbinary discs, and then migrate inwards to be stopped near the cavity edge, is the correct one, then we can use the Kepler-16, -34 and -35 systems to constrain models of planet formation and protoplanetary disc dynamics. The simplicity of our models means that there is still a large amount of work that needs to be done to achieve this goal.
In this work, gas accretion onto planets to their final masses and disc dissipation scenarios have been carried out separately. Combining the two, akin to the method in \citep{Pierens2013}, in self-gravitating discs might help to fit the orbital properties of the planets but also shed light on the era in which the planets may have accreted their masses. Whilst disc-mass, and the influence of self-gravity can significantly alter disc structure, other physics will also play important roles. The inclusion of an adiabatic equation of state with radiative physics has been investigated in circumbinary systems, along with the effect on planet migration \citep{Kley2014, Kley2015}. The non-uniform, time dependent, radiation field produced by the two stars, however, has not yet been explored in combination with a more realistic thermal treatment of the disc. 3D effects are also likely to be important. These include, but are not limited to: disc warping when the disc and orbit plane of the central binary are misaligned \citep{LarwoodPapaloizou1997}, which can in turn lead to the development of a parametric instability in the disc that may be a source of hydrodynamic turbulence \citep{OgilvieLatter2013}; the development of eccentric modes in the disc leading to parametric instability and hydrodynamic turbulence \citep{Papaloizou2005, BarkerOgilvie2014}; and the formation of Spiral Wave Instabilities existing in a disc that is tidally forced by a binary system, can lead to a parametric instability and hydrodynamic turbulence \citep{BaeNelsonHartmann2016}. Equally importantly will be the inclusion of MHD, since the underlying angular momentum transport mechanism operating in circumbinary discs is likely to be of magnetic origin \citep[e.g.]{Balbus1991, Bai2013}. Simulations carried out in 3-D will allow us to examine these and other effects. Finally, both theory and observations indicate that planets do not normally form in isolation, so the evolution of multi-planet systems \citep{Kley2015} may provide better agreement with at least a subset of circumbinary planet observations.
 
The authors thank the referee for the comments and suggestions in their helpful report. Computer time for the simulations performed with {\small GENESIS} was provided by HPC resources of Cines under the allocation A0010406957 made by GENCI (Grand Equipement National de Calcul Intensif). Those performed with {\small FARGO} utilised: Queen Mary's MidPlus computational facilities, supported by QMUL Research-IT and funded by EPSRC grant EP/K000128/1; and the DiRAC Complexity system, operated by the University of Leicester IT Services, which forms part of the STFC DiRAC HPC Facility (\url{www.dirac.ac.uk}). This equipment is funded by BIS National E-Infrastructure capital grant ST/K000373/1 and STFC DiRAC Operations grant ST/K0003259/1. DiRAC is part of the National E-Infrastructure. This research was also supported in part by the National Science Foundation under Grant No. NSF PHY-1125915.



\begingroup
\raggedright
\bibliographystyle{mnras}
\bibliography{sgcbIIrevised}
\endgroup


\bsp	
\label{lastpage}
\end{document}